\pdfoutput=1 
\documentclass[
  aps,%
  prx,
 twocolumn,%
 groupedaddress,%
 superscriptaddress,%
  showpacs,%
 letterpaper,%
 amsfonts,%
 footinbib,%
  10pt,%
 floatfix,%
]{revtex4-2}
 
\usepackage{dcolumn}
\usepackage{bm}

\usepackage{hyperref}
\hypersetup{
    colorlinks=true,
    citecolor=blue,
    linkcolor=blue , 
    filecolor=cyan,      
    urlcolor=magenta,
    pdfcreator = {\LaTeX\ and \flqq hyperref\frqq},
}

\RequirePackage{graphicx,amsmath,amssymb,bm}
\usepackage{bbm}
\usepackage{float}
\usepackage{framed} 
\usepackage{amsthm}
\usepackage{caption}
\usepackage{subcaption}
\usepackage[normalem]{ulem}
\usepackage{comment}
\usepackage{slashed} 
\usepackage{upgreek}
\usepackage{xcolor}
\usepackage{enumitem}

\newcommand{\bb}{\mathbb}

\newcommand{\Ha}{{\cal{H}}}  
\newcommand{\Aa}{{\cal{A}}}

\newcommand{\1}{\mathbbm{1}}

\newcommand{\bbz}{\bb{Z}}

\def\be{\begin{equation}}
\def\ee{\end{equation}} 
\def\bsh{\begin{shaded}}
\def\esh{\end{shaded}} 
\def\bpm{\begin{pmatrix}}
\def\epm{\end{pmatrix}}

\begin{document} 

\title{Bosonization and Kramers-Wannier dualities in general dimensions}
\author{Lei Su}
\affiliation{Department of Physics, University of Chicago, Chicago, Illinois 60637, USA}
\author{Ivar Martin}
\affiliation{Materials Science Division, Argonne National Laboratory, Lemont, Illinois 60439, USA}
\affiliation{Department of Physics, University of Chicago, Chicago, Illinois 60637, USA}

\begin{abstract}  
It is well known that the noninteracting Majorana chain is dual to the one-dimensional transverse-field Ising model, either through the Jordan--Wigner transformation or by gauging fermion parity. In this correspondence, the minimal translation of the Majorana chain maps to the celebrated Kramers--Wannier (KW) duality of the spin model, with the critical point mapped to the self-dual point. In this work, we generalize this mapping to two and higher dimensions by constructing a unitary equivalence between the parity-gauged fermionic system and a spin system defined on arbitrary polyhedral decompositions of space. Imposing the flatness condition on the gauge field yields a bosonization duality between the original (ungauged) fermionic system and a gauged spin system obeying a Gauss law. The dependence of the Gauss law in the spin system on the Kasteleyn orientation (and the discrete spin structure) of the fermionic system is made explicit. Applying this bosonization to one or two copies of Majorana fermions on translationally invariant lattices, we derive higher-dimensional analogs of KW (self-)dualities in spin systems arising from fermionic minimal translations. The KW (self-)dualities are non-invertible due to projections onto eigenspaces of higher-form symmetries in the associated symmetry operators. The bosonization framework we present is intuitive, general, and systematic, encompassing other known exact bosonization methods while offering a novel approach to establish new connections  between fermionic and spin systems in arbitrary dimensions.
\end{abstract} 
\maketitle

\tableofcontents

\section{Introduction}
Bosonization in 1D space is well-developed and commonly used in both condensed matter physics and high energy physics. It represents discrete or continuous fermionic systems by equivalent bosonic systems. For example, a Luttinger liquid of interacting electrons can be described by bosonic fields \cite{mattis1965exact}. Also, the Thirring model of Dirac fermions is transformed into the quantum sine-Gordon theory \cite{coleman1975}. A key characteristic to preserve is the anticommutativity of fermions at spacelike separation, in contrast to the commutativity of bosons. One famous exact bosonization strategy to account for this difference is the Jordan-Wigner transformation that maps a 1D fermionic system to a spin model where a local fermionic creation/annihilation operator is represented by a local spin operator with a string of Pauli operators attached to it \cite{jordan1928}. This nonlocal string of operators accounts for the anticommutativity of fermionic operators. In general, bosonization (and fermionization) is a powerful technique that has led to many theoretical developments.

There have been many attempts to generalize \textit{exact} bosonization on a lattice to higher dimensions \cite{srednickiHidden1980a, fradkinFermion1980,  wosiek1981local, szczerba1985spins, fradkin1989jordan, bravyi2002fermionic, kitaev2006anyons, verstraete2005mapping, ball2005fermions, Nussinov2012arbitrary, seeley2012bravyi, havlicek2017operator, zohar2018eliminating, chen2018exact, chen2019bosonization,  setia2019superfast, Jiang2019majorana, steudtner2019quantum, chen2020exact, tanti2020jordan, shirleyFractonic2020, shukla2020tensor, bochniak2020bosonization, derby2021compact, borla2022quantum, li2022higher,  Chiew2023discoveringoptimal, obrien2024ultrafast, seifertWegners2024, obrien2025local}. In particular, a series of recent works, starting from the analysis of spin-TQFT and fermionic topological matter \cite{gaiotto2016spin}, map a parity-even fermionic system to a gauged spin system in arbitrary dimensions when the space is triangulated \cite{chen2018exact, chen2019bosonization, chen2020exact}, though this method is equivalent to many earlier approaches \cite{chen2023equivalence}. In this work, we revisit the generalization of bosonization to general dimensions for a general polyhedral or, up to continuous deformations, regular cellular decomposition of space, with an objective of generalizing the celebrated Kramers-Wannier (KW) (self-)duality \cite{kramers1941} in 1D quantum spin chains to higher dimensions. 

The KW duality has been studied for many decades and the self-duality has been applied to determine the quantum critical point of the 1D transverse-field Ising model (TFIM) (and the thermal critical point of the 2D classical Ising model). It can be interpreted as an isomorphism of operator (bond) algebras of the spin model \cite{cobanera2011bond, lootens2023dualities}. A more modern perspective was obtained by deriving the KW duality by gauging a $\bbz_2$ global symmetry of the quantum system, leading to a novel  interpretation of general KW dualities as noninvertible symmetries \cite{schafer2024ictp, shao2023}, that is, (topological) symmetry operators without an inverse. Yet another direction of progress was driven by generalizations of ordinary (0-form) symmetries to higher-form symmetries \cite{gaiotto2015generalized}. The charge operator of a $p$-form symmetry on a $(d{+}1)$D spacetime manifold is supported on a $(d{-}p)$D submanifold. Gauging a $p$-form global symmetry in $(d{+}1)$D spacetime by coupling the system to a flat $(p{+}1)$-form gauge field induces a dual $(d{-}p)$-form symmetry in the gauged system. Conversely, gauging the $(d{-}p)$-form symmetry of the dual system can recover the original $p$-form symmetry. Consequently, the celebrated KW duality can be made self-dual in 1D space by gauging a 0-form $\bbz_2$ symmetry. To obtain a KW self-duality in 3D, one needs to gauge 1-form discrete symmetries \cite{kaidi2022kramers, choi2022noninvertible}. In 2D, since gauging a 0-form (1-form) yields a 1-form (0-form) dual symmetry, to obtain a KW self-duality, one needs to consider a combined system with both a 0-form symmetry and a 1-form symmetry.

\begin{figure}[tb]
    \centering \includegraphics[width=1.0\linewidth]{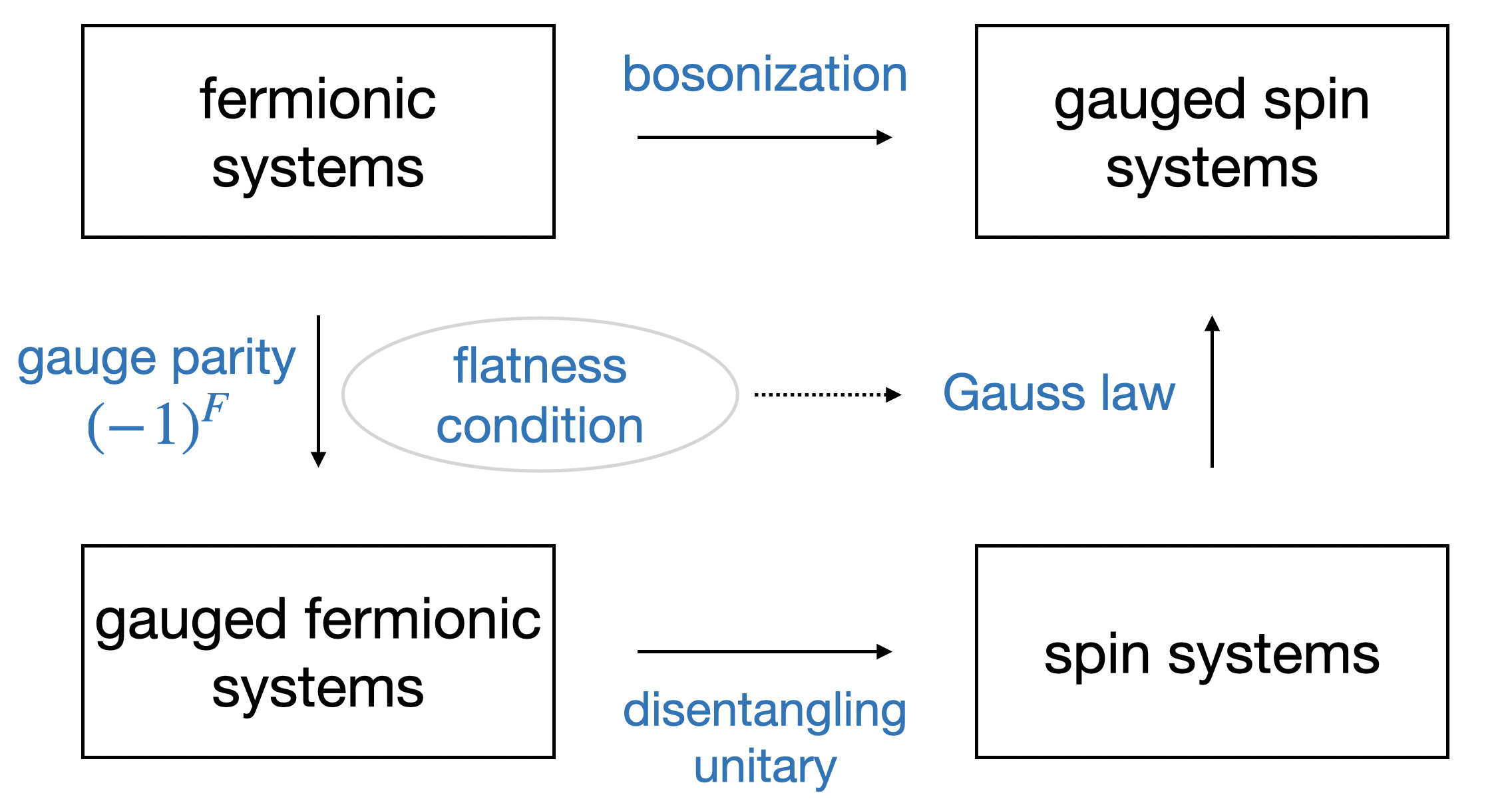}
    \caption{Schematic of the relations between fermionic systems and spin systems. A parity-gauged fermionic system is shown in the main text to be equivalent to a spin system via a disentangling unitary. Imposing the flatness condition on the gauge spins establishes a duality between the ungauged (parity-even) fermionic system and the gauged fermionic system. On the spin side, this flatness condition becomes a Gauss-law constraint. The combination of parity gauging (under the flatness condition) and the disentangling unitary transformation defines a bosonization map from a general fermionic system to a gauged spin system.}
    \label{fig_schematic}
\end{figure}

In this work, however, we take a different route to generalize KW (self-)dualities to two or higher dimensions. It is known that a translationally invariant Majorana chain is mapped to the self-dual TFIM and that the minimal translation of the Majorana chain is mapped to the KW self-duality by the Jordan-Wigner transformation \cite{seiberg2024majorana, seiberg2024non}. The KW self-duality essentially stems from the flexibility in choosing a unit cell of two Majorana fermions (forming a complex fermion) for the Majorana chain. Alternatively, one can show that gauging the fermion parity of the Majorana chain also yields the TFIM by a unitary transformation that disentangles fermions from ``gauge" spins. From this perspective, the KW self-duality originates from two different ways to gauge the fermion parity of the same system. The dual paths toward the TFIM from either the (ungauged) fermion system or the parity-gauged fermionic system are not surprising because in 1D gauging a discrete global symmetry of a system can be reversed by gauging the dual symmetry. More physically, the gauging process preserves the information of the system (up to boundary conditions and degeneracies) and thus can be undone. To generalize the picture directly to higher dimensions, we will emphasize the equivalence between a parity-gauged fermionic system and a spin system, which has been studied both abstractly and  concretely in the literature. For example, an explicit map between a parity-gauged fermionic system and a spin system in a higher dimension was briefly discussed in Ref.~\cite{chen2019bosonization} from an operator algebraic point of view. In our work, we consolidate this equivalence by constructing explicit disentangling unitaries from parity-gauged fermionic systems to spin systems (see Fig.~\ref{fig_schematic}). Different from the 1D case, to preserve the duality between the ungauged (parity-even) fermionic system and the parity-gauged fermionic system, an extra flatness condition on the ``gauge field" is needed on a lattice. This condition ensures that there is no gauge flux in the fermionic system, which is automatically guaranteed in 1D. Imposing the flatness condition on the ``gauge field" induces a Gauss law in the spin system (Fig.~\ref{fig_schematic}). The corresponding spin system with the Gauss law then is the bosonized dual incarnation of the ungauged fermionic system.

In our work, we present a prescription for constructing an explicit disentangling unitary on an arbitrary embedded surface graph in 2D and on an arbitrary polyhedral decomposition in higher dimensions. On a 2D surface graph, where a complex fermion (or two Majorana fermions) is placed on each face of the graph, the construction of the disentangling unitary relies crucially on an assignment of a Majorana bilinear $i\gamma_{f_1(e)}\gamma_{f_2(e)}$ between two neighboring faces $f_1$ and $f_2$ to the shared edge $e$, a process that can be carried out by partitioning the edges of each face into two sets and assigning them to the two Majorana fermions associated with the same face. Each edge is then assigned to two Majorana fermions in the two neighboring faces sharing the edge. The disentangling unitary can then be written down as a product of controlled local parity transformations, a fermionic analog of controlled-$Z$ gates.  The flexibility in this assignment procedure can be employed to derive many equivalent bosonization schemes, including the methods by Bravyi--Kitaev\cite{bravyi2002fermionic} and Chen--Kapustin \cite{chen2019bosonization}. In the process of the construction, one also needs to specify the order within each Majorana bilinear appearing in the unitary, i.e., $i\gamma_{f_1(e)}\gamma_{f_2(e)}$ versus $i\gamma_{f_2(e)}\gamma_{f_1(e)}$. This step is somewhat arbitrary -- it simply modifies the signs in front of the Gauss-law constraints associated with each vertex on the spin side (see Fig.~\ref{fig_schematic}) when the flatness condition on the gauge spins is imposed on the fermionic side. We can choose an assignment scheme such that the Majorana bilinears associated with each edge, together with  the Majorana bilinear of the local fermion parity $(-1)^{F_f} = i \gamma'_f \gamma_f$ on  each face, form a dual surface graph equipped with an inherent dimer covering configuration. Assigning an orientation to each edge of this graph specifies the ordering within each Majorana bilinear. By further choosing a particular Kasteleyn orientation -- such that each dual face has an odd number of edges oriented clockwise -- we fix the sign  of the product of concatenated Majorana bilinears around closed loops. This, in turn, determines the disentangling unitary and the signs in front of the Gauss-law constraints. Since equivalent Kasteleyn orientations are in one-to-one correspondence with the different spin structures of the system in 2D, this choice of edge orientations (and the associated bilinear ordering) reveals the dependence of bosonization on the discrete spin structure. This spin-structure dependence has also been discussed in Ref.~\cite{chen2019bosonization} in the context of triangulations with branching structures. Our alternative geometric construction applies to general surface graphs; fermions are not required to reside in top-dimensional simplices of a triangulated space. In other words, the bosonization scheme is more general. Most importantly, we provide an intuitive, general, and unifying framework that can be extended in a straightforward manner to higher-dimensional bosonization on arbitrary polyhedral decompositions.

Equipped with higher-dimensional bosonization, we can now apply it to the special case of  Majorana lattices with translational invariance. Generalizing the  relation between minimal Majorana translation of the Majorana chain and the KW (self-)duality of the 1D TFIM, we map the minimal translation of the Majorana fermion on a higher-dimensional lattice to the KW (self-)duality in the gauged spin system under bosonization. Similar to the explicit form of the (noninvertible) symmetry generator of the KW duality, which can be derived from the fermionic system \cite{seiberg2024majorana, seiberg2024non}, the corresponding symmetry generator of the KW duality in higher dimensions can also be derived from the fermionic system.  Analogous to the 1D case, the self-duality originates from the flexibility in choosing a unit cell of two Majorana fermions (or a complex fermion), as induced by a minimal Majorana translation. In 1D, there is only one translational direction. In higher dimensions, however, there are multiple translational directions, and choosing a unit cell of two Majorana fermions can break the discrete rotational isotropy. This results in an anisotropic KW (self-)duality on the square lattice of Majorana fermions with nearest-neighbor couplings (i.e., hoppings). However, if the Majorana couplings are not restricted to nearest neighbors, multidirectional KW (self-)dualities in the bosonized spin system can be obtained. For example, when the Majorana system consists of two decoupled copies of 2D Majorana lattices, KW (self-)dualities can exist along multiple directions specified by the primitive lattice vectors. The minimal translation is defined as an exchange between all $\gamma$ and $\gamma'$ followed by a one-unit-cell translation of $\gamma'$. The Majorana lattice is not limited to the square lattice; for example, it can also take the form of the honeycomb lattice. At the critical point, the Majorana system possesses multiple $U(1)$ symmetries that are related by the minimal Majorana translation. After bosonization, this translation is mapped to the KW (self-)dualities between the corresponding  $U(1)$ charges. This directly generalizes the relation between the two $U(1)$ charges in a free Dirac chain and those in the $XX$ model and its variants \cite{chatterjee2025quantized, pace2024lattice, su2025z2}.

The paper is organized as follows. In Sec.~\ref{sec:review1d}, we briefly review the bosonization of the Majorana chain to obtain the 1D TFIM by gauging fermion parity followed by a disentangling unitary transformation. The minimal (half-unit-cell) Majorana translation is mapped to the KW duality under bosonization. We describe in Sec.~\ref{sec:bos2d} the main generalization procedure, parity-gauging and unitary transformation, on the 2D square lattice. In Sec.~\ref{sec:disentanling},  we construct the disentangling unitary that maps a parity-gauged fermionic system to a spin system. As a warm-up, we elaborate on two different ways to carry out the assignment step and construct the unitaries on the square lattice, before generalizing to general surface graphs. We also highlight the connection to the Bravyi--Kitaev superfast transformation. In Sec.~\ref{sec:flatness}, we impose the flatness condition on the gauge spins to enforce the duality between the ungauged fermionic system  and the parity-gauged fermionic system, with the condition mapped to the Gauss law in the spin system, completing the bridge of bosonization between a (ungauged) fermionic system and a gauged spin system. We elaborate on the results on the square lattice in Sec.~\ref{sec:bos_square} which reproduces the bosonization method in Ref.~\cite{chen2018exact}. The dependence of the product of concatenated fermionic bilinears along a closed loop on the equivalence class of Kasteleyn orientations is presented in Sec.~\ref{sec:Kasteleyn}. In  Sec.~\ref{sec:triangulation}, we explore the correspondence between the Kasteleyn orientation and the discrete spin structure on triangulated graphs and more general surface graphs. In Sec.~\ref{sec:KW}, we apply the bosonization method to translationally invariant Majorana systems in order to generalize the KW (self-)duality to 2D. We begin with an intuitive look at the duality beyond 1D, followed by a detailed discussion of KW dualities in bosonized spin systems derived from free Majorana fermions on the square lattice. In Sec.~\ref{sec:noninv}, we write down an explicit expression for the noninvertible KW duality transformation in terms of spin operators, revealing the role of higher-form symmetries. Then we analyze KW dualities derived from two copies of decoupled Majorana fermions on the square and the honeycomb lattice in Sec.~\ref{sec:twocopies}, where self-dual KW dualities also interchange $U(1)$ charges at the self-dual point. In Sec.~\ref{sec:general}, we extend the bosonization framework --- gauging fermion parity via minimal coupling, imposing the flatness condition, and constructing the disentangling unitary --- and KW dualities to general polyhedral decompositions in arbitrary dimensions, with a special focus on the 3D case. We discuss the dependence on generalized Kasteleyn orientations and spin structures. Finally, we conclude with a discussion and outlook in the last section. Additional technical details are provided in the Appendices. This work is accompanied by a complementary study in Ref.~\cite{su$mathbbZ_2$2025a}.

\section{Bosonization in 2D}
\label{sec:bosonization2d}
\subsection{From Majorana chain to 1D TFIM: a brief review}
\label{sec:review1d}
\begin{figure}[bt]
    \centering
    \includegraphics[width=0.8\linewidth]{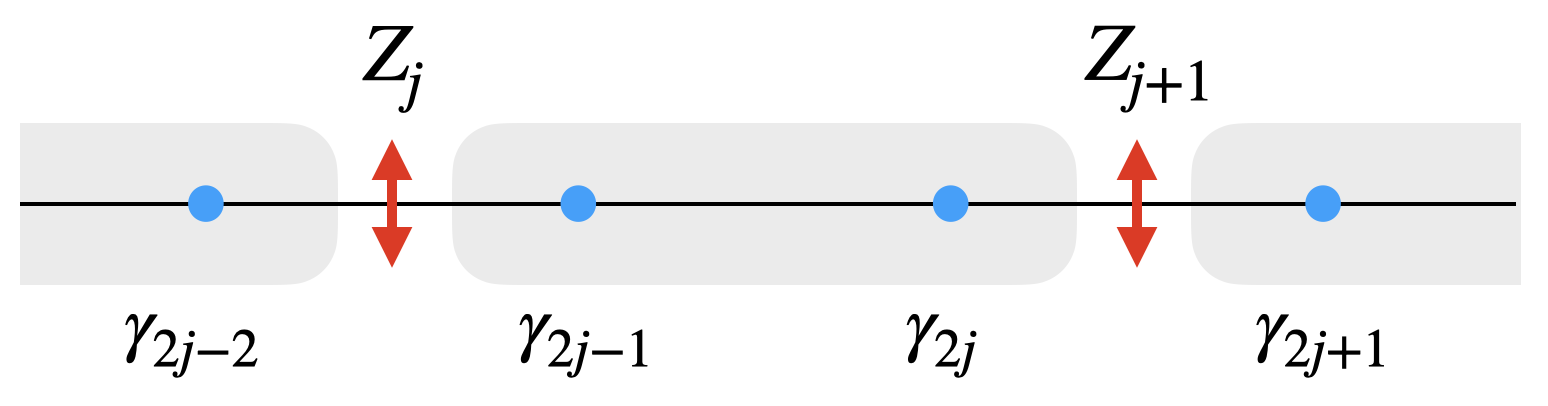}
    \caption{Parity-gauging of the Majorana chain. We place one ``gauge" 1/2-spin on each bond between unit cells of two Majorana fermions and impose the Gauss law $G_j \equiv Z_{j-1}(i \gamma_{2j-1}\gamma_{2j}) Z_j = 1$ for all $j$.}
    \label{fig:maj}
\end{figure}

Before discussing bosonization in higher dimensions, let us briefly review the well-known correspondence between the free Majorana chain and the one-dimensional TFIM. Provided that the Hamiltonian is even under fermion parity, there are at least two ways to bosonize fermionic systems on a 1D lattice: the Jordan-Wigner transformation and gauging fermion parity. In the former approach, string operators are attached to spin ladder operators in order to account for fermionic anticommutation relations between different sites. Remarkably, these nonlocal strings cancel out in the resulting Ising couplings, leading to a local TFIM Hamiltonian. A naive extension of this ``string attachment" construction to higher dimensions, however, typically requires introducing an ordered ``snake" path that covers the entire lattice, leading to highly nonlocal interactions unless special techniques are employed to eliminate the strings, as in, e.g., Ref.~\cite{verstraete2005mapping}. With the goal of extending bosonization to more general settings, we will instead focus on the latter approach: gauging fermion parity.

Given a spinless fermionic system on a 1D chain, we can define Majorana fermions in terms of the complex fermions:
\be 
\gamma_{2j-1} = c_j +c_j^{\dagger},\quad \gamma_{2j} = i (c_j -c_j^{\dagger}).
\ee 
They satisfy the anticommutation relation: $\{\gamma_k, \gamma_l\} = 2\delta_{kl}$. The local fermion parity is defined as 
\be 
(-1)^{F_j} = (-1)^{c^{\dagger}_j c_j}= i \gamma_{2j-1}\gamma_{2j}.
\ee 
Suppose we are given a Hamiltonian that commutes with the total fermion parity $(-1)^F = \prod_j (-1)^{F_j}$. We can gauge the fermion parity $(-1)^F$. Although the gauging process is similar for all spinless fermionic systems, for simplicity, let us focus on a free Majorana fermion chain:
\be 
H_{\text{Maj}} = -i \sum_j t_1  \gamma_{2j-1}\gamma_{2j} + t_2  \gamma_{2j}\gamma_{2j+1}.
\label{eq:H_Maj1d}
\ee 
We either work on an infinite chain or use the periodic boundary condition for the complex fermions in this section. 

Now let us follow the standard gauging procedure.  We first place one 1/2-spin or qubit on the bond between unit cells that consist of two Majorana fermions, $\gamma_{2j-1}$ and $\gamma_{2j}$. See Fig.~\ref{fig:maj}. Denoting the bond between $2j$ and $2j+1$ as $\langle 2j, 2j+1\rangle$, we will use the short-hand notation $Z_j = Z_{\langle 2j, 2j+1\rangle}$, $X_j = X_{\langle 2j, 2j+1\rangle}$, and so on. Here, $X$ and $Z$ are the Pauli operators. We now impose the Gauss-law constraints:
\be 
G_j \equiv Z_{j-1}(i \gamma_{2j-1}\gamma_{2j}) Z_j = 1.
\label{eq:gausslaw}
\ee 
The Hamiltonian is minimally coupled to the ``gauge" spins:
\be 
\tilde{H}_{\text{Maj}} = -i \sum_j t_1 \gamma_{2j-1}\gamma_{2j} + t_2 \gamma_{2j}X_j\gamma_{2j+1}.
\ee 
It is easy to check that $\tilde{H}_{\text{Maj}}$ is gauge-invariant because $[\tilde{H}_{\text{Maj}}, G_j] =0$ for all $j$.

To obtain the TFIM, we now construct a unitary transformation $U$ that disentangles 
the fermions (matter) from gauge spins. It should satisfy the following requirement 
\be 
G_j^U \equiv U G_j U^{-1} = i \gamma_{2j-1}\gamma_{2j}.
\ee
Since a unitary transformation preserves gauge invariance, it is clear that the transformed $\tilde{H}_{\text{Maj}}$ contains only fermion bilinears generated by $G_j^U =i \gamma_{2j-1}\gamma_{2j}$. For example, a term consisting of a fermion bilinear $i \gamma_{2j}\gamma_{2j+1}$ anticommutes with $G_j^U$ and is thus not gauge invariant. Therefore, once we enforce the transformed Gauss-law constraints $G_j^U =i \gamma_{2j-1}\gamma_{2j} =1$, we manage to disentangle fermions from spins. Physically, this is a gauging-fixing process. Let us construct such a unitary $U$ explicitly by requiring the following conditions to be satisfied:
\be 
U \gamma_{2j-1} U^{-1} = Z_{j-1} \gamma_{2j-1}, \quad U \gamma_{2j} U^{-1} = Z_j \gamma_{2j}
\ee 
for all $j$. 
By observing how the Majorana operators transform in the eigenbasis of all $Z_k$, we can immediately construct the following operator 
\be 
U = \prod_j(  P_{j}^+ + P_j^- i\gamma_{2j}\gamma_{2j+1}),
\label{eq:U1}
\ee 
where we have defined projection operators $P_j^{\pm}\equiv (I\pm Z_{j})/2$. It is easy to check that $U =  U^{\dagger} = U^{-1}$ and is therefore unitary. The reader familiar with the language of quantum information can easily notice that each factor in this unitary transformation $U$ is a direct analog of the controlled-$Z$ gate: it is a local fermion-parity transformation conditioned on the eigenvalue of $Z_j$. Also note that each factor in $U$ contains only fermion bilinears (as opposed to odd-fermion-parity operators), and therefore it acts only locally on the lattice. 

Let us now see how $\tilde{H}_{\text{Maj}} $ is transformed under this unitary transformation $U$:
\begin{align}
  U\tilde{H}_{\text{Maj}} U^{-1} &= -\sum_j t_1 Z_{j-1} (i \gamma_{2j-1}\gamma_{2j}) Z_j + t_2 X_j \nonumber\\
  & = -\sum_j t_1 Z_{j-1} Z_j + t_2 X_j.
\end{align} 
In the second line, we have plugged in the condition $i \gamma_{2j-1}\gamma_{2j} =1$. Thus we have successfully disentangled fermions from spins, and mapped the parity-gauged Majorana chain to the TFIM. Note that when the chain is periodic with $2L$ Majorana fermions, a product of concatenated fermion bilinears over the entire chain $\prod_{j=1}^{2L} (i \gamma_j \gamma_{j+1})$ is mapped to $\prod_{j=1}^L X_j$ up to a phase that depends on the boundary condition of the fermions and the total size $2L$. Thus the bosonization in this case is  only an isomorphism up to different boundary conditions and charge sectors. Nevertheless, the bosonization process can be reversed by coupling the spin model to the Majorana chain and gauging the diagonal $\bbz_2$ symmetry or by fermionic gauging \cite{aksoyLiebSchultzMattis2024, su$mathbbZ_2$2025a}.

It is important to notice that there are two ways to choose a unit cell in the Majorana chain that contains two Majorana fermions, and, equivalently, two ways to factorize the parity operator $(-1)^F = \prod_j i\gamma_{2j-1}\gamma_{2j}$. In our derivation above, we imposed the Gauss-law constraints on $i\gamma_{2j-1}\gamma_{2j}$, but we can also impose the Gauss-law constraints as 
\be 
G'_j \equiv Z_{j-\frac{1}{2}}(i \gamma_{2j}\gamma_{2j+1}) Z_{j+\frac{1}{2}} = 1.
\label{eq:gausslaw2}
\ee 
The gauged Hamiltonian is then
\be 
\tilde{H}'_{\text{Maj}} = -i \sum_j t_1 \gamma_{2j-1}X_{j-\frac{1}{2}}\gamma_{2j} + t_2 \gamma_{2j}\gamma_{2j+1}.
\ee 
Going through the disentangling process again, we obtain
\begin{align}
  U'\tilde{H}'_{\text{Maj}} U'^{-1} &= -\sum_j t_1 X_{j-\frac{1}{2}}  +t_2 Z_{j-\frac{1}{2}} (i \gamma_{2j}\gamma_{2j+1}) Z_{j+\frac{1}{2}} \nonumber\\
  & = -\sum_j t_1 X_{j-\frac{1}{2}}  +t_2 Z_{j-\frac{1}{2}} Z_{j+\frac{1}{2}}.
\end{align} 
Shifting this Hamiltonian by $j\to j+1/2$, we obtain the dual TFIM with $t_1$ and $t_2$ exchanged. More explicitly, $Z_{j-1} Z_{j} \to X_j \to Z_j Z_{j+1}$. This is the famous KW duality: $H_{\text{TFIM}}(t_1, t_2) = H_{\text{TFIM}}(t_2, t_1)$. In particular, at $t_1= t_2$, the TFIM is self-dual under this process. Thus, the self-duality can be traced back to the minimal (half-unit-cell) translation invariance of the Majorana chain.

\subsection{Brief introduction of generalization to 2D}
\label{sec:bos2d}
Having reviewed the 1D bosonization by gauging fermion parity $(-1)^F$ followed by a disentangling unitary $U$, we now extend it to 2D. To be concrete, in this section, we use the square lattice as an example to elaborate our main idea.

\begin{figure}[bt]
    \centering
    \includegraphics[width=1.0\linewidth]{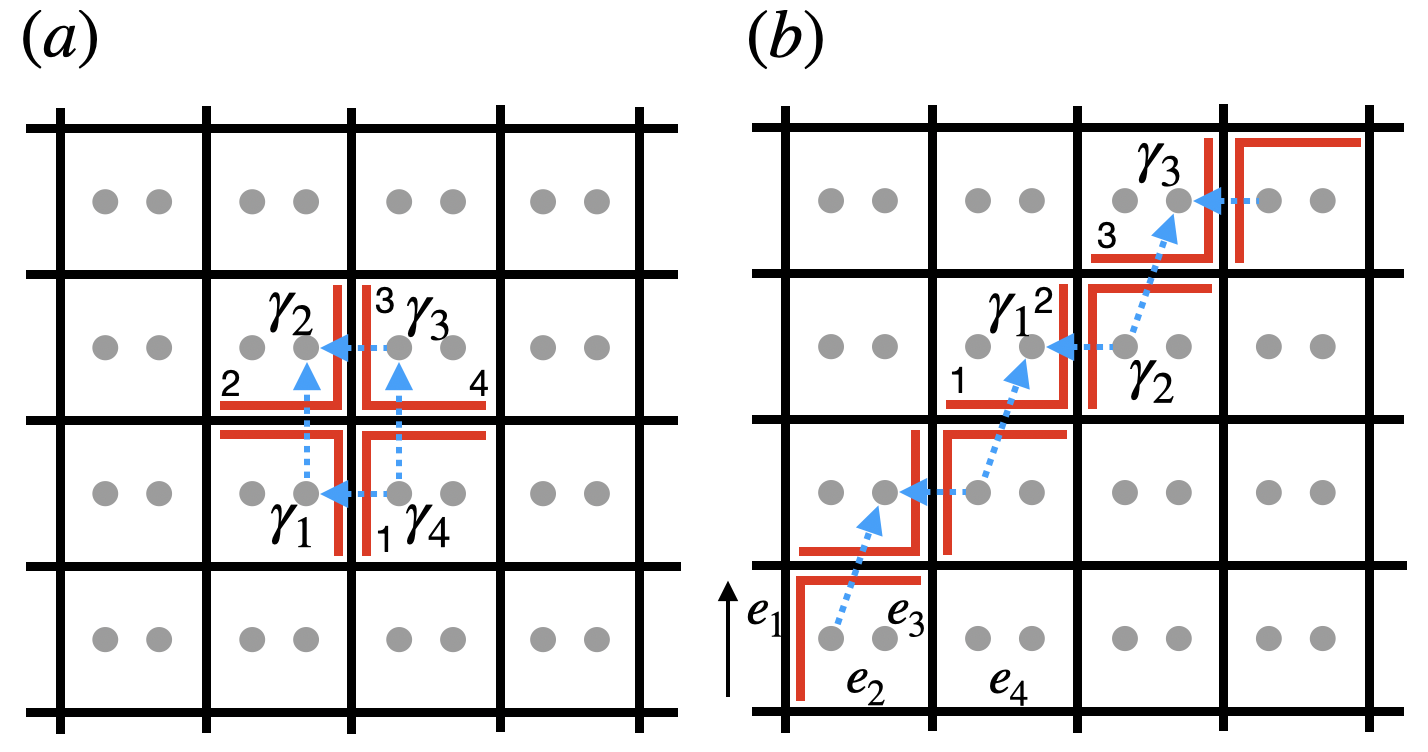}
    \caption{Quartet (a) and zigzag (b) assignment schemes of edges on the square lattice. In both schemes, two edges (red) in a face are assigned to a Majorana fermion on the face. The dashed blue arrows indicate the ordering of fermions in the bilinear $S_e$: an arrow from $\gamma_{f_2(e)}$ to $\gamma_{f_1(e)}$ fixes $S_e = i \gamma_{f_1(e)} \gamma_{f_2(e)}$. The global order of edges $e_{n-1}<e_n$ used in Eq.~(\ref{eq:square_U}) are indicated by the labels on the lower left corner of (b). The black arrow shows the increasing direction once ordering in a row is done. In other words, the edges are ordered from left to right and then from bottom to top. }
    \label{fig:uni1}
\end{figure}

We place one spinless complex fermion or two Majorana fermions $\gamma_f$ and $\gamma'_f$ on each face $f$ of the square lattice (see Fig.~\ref{fig:uni1}). $\gamma$ and $\gamma'$ are slightly displaced from the center and placed to the right side and the left side of each square, respectively, so the system can be regarded as a Majorana lattice. Here, $f = (m,n)$ can also be viewed as the coordinate of face $f$. Let us consider two types of fermion bilinears: intracell coupling $(-1)^{F_f} = i \gamma'_f \gamma_f$ and intercell coupling $i\gamma_f \gamma'_{f+a_{x/y}}$. Here we use $a_x$ ($a_y$) to denote a horizontal (vertical) translation to the right (top) by one unit cell. By making use of the (anti-)commutation relations of the fermion operators, we can see that these bilinears commute with each other unless they have one and only one overlapping fermion operator. It is easy to see that any operator consisting of an even number of fermions can be factored as the product of bilinears of this form. In other words, the parity-even algebra is generated by these two types of bilinears. From an operator algebra point of view, to construct a bosonization scheme, we need to map these bilinears to bosonic operators such that all the (anti-)commutation relations are preserved. A more mathematically rigorous description is presented in Appendix \ref{app:algebra}.  In our work, we follow the intuitive ``gauging + disentangling" procedure discussed earlier in the 1D case and generalize it properly to 2D.  

To gauge the fermion parity $(-1)^F = \prod_f (-1)^{F_f} =\prod_f (i \gamma'_f \gamma_f)$, we place one gauge $\bbz_2$-spin on each edge of the square lattice and impose the Gauss-law constraints
\be 
G_f \equiv (i \gamma'_f \gamma_f) \prod_{e\subset f} Z_e = 1.
\label{eq:gausslaw2d}
\ee  
To preserve gauge invariance, the intercell couplings across edge $e$ need to be dressed by $X_e$. For example, $i \gamma_{f} \gamma'_{f'}$, where $f$ and $f'$ are nearest neighboring faces with shared edge $e$, becomes
$i \gamma_{f} \gamma'_{f'} X_e$.   The intracell couplings $i\gamma'_f \gamma_f$ remain intact. Even though we do not need to specify the Hamiltonian, a simple gauged Hamiltonian $\tilde{H}$ is a sum of all these bilinear terms with coupling constants, such as 
\be 
\begin{split}
 \tilde{H} =& -i\sum_{f} (t_1 \gamma'_{f} \gamma_{f} +t_2 \gamma_{f} \gamma'_{f+a_x} X_{e^f_x}) \\
 & + (t'_1 \gamma_{f} \gamma_{f+a_y} X_{e^f_y} +t'_2 \gamma'_{f} \gamma'_{f+a_y}X_{e^f_y}).   
\end{split} 
\ee
Here $e^f_{x/y}$ is the edge on the right/top of face $f$.  It is straightforward to check that $[\tilde{H}, G_f] =0$ for all $f$.

After finishing the gauging process, let us try to construct a unitary transformation that disentangles fermions from spins, which is a straightforward generalization of the 1D case. We want to construct a unitary $U$ that maps $\gamma'_f$ and $\gamma_f$ in a way that cancels the plaquette operator $\prod_{e\subset f} Z_e$ in $G_f$. In other words, we want 
\be 
U G_f U^{-1} =   i \gamma'_f \gamma_f,
\label{eq:Gf}
\ee  
so that we can put $(-1)^{F_f} =i \gamma'_f \gamma_f =1$ in the end. We are guaranteed that the transformed Hamiltonian $U \tilde{H} U^{-1}$ contains only bilinears of the form $(-1)^{F_f} = i \gamma'_f \gamma_f$. Indeed, the unitary transformation $U$ should preserve gauge invariance: $[\tilde{H}, G_f] =0$ for all $f$, i.e., 
\be [U \tilde{H}U^{-1}, (-1)^{F_f}] =0\ee
for all $f$. Since any fermion bilinear with just one overlapping fermion operator with $(-1)^{F_f} = i \gamma'_f \gamma_f$ anticommutes with it, each term in $U \tilde{H}U^{-1}$ contains either zero or two overlapping fermions with $(-1)^{F_f}$, i.e., it should be factored out in $(-1)^{F_f}$ for a subset of faces. Therefore, substituting $(-1)^{F_f} =1$ into $U \tilde{H}U^{-1}$ yields a Hamiltonian with only spin degrees of freedom. 

The above procedure can be extended to a general surface graph very easily. In the following section, we explicitly construct such unitaries.  

\subsection{Disentangling unitaries}
\label{sec:disentanling}
\subsubsection{Square lattice: Example I}
Note that in the 1D case, the local unitary factors $U_j$ in the unitary transformation $U$ commute with each other because they act on different degrees of freedom. It is not hard to see that this condition must be generalized in 2D. As mentioned earlier in Eq.~(\ref{eq:Gf}), we want $U$ to map $\gamma'_f$ and $\gamma_f$ in a way that cancels the plaquette operator $\prod_{e\subset f} Z_e$ in $G_f$. To do that, we partition the Majorana fermion into groups of four. A quartet of Majorana fermions is shown in Fig.~\ref{fig:uni1}(a), and the rest are shifted diagonally by one unit cell so that the quartets are non-overlapping. The four edges of each face are partitioned into two sets, each of which contains two edges and is assigned to the two Majorana fermions associated with the face. Also, each edge is assigned to exactly two Majorana fermions on the neighboring faces.    Labeling the four fermions and four edges around a vertex $v$ as in Fig.~\ref{fig:uni1}(a), we want to construct a $U_v$ such that 
\be
\begin{split}
  U_v \gamma_1 U_v^{-1} & = Z_1 Z_2 \gamma_1, \quad
    U_v \gamma_2 U_v^{-1}  = Z_2 Z_3 \gamma_2,  \\
    U_v \gamma_3 U_v^{-1} & = Z_3 Z_4 \gamma_3, \quad
    U_v \gamma_4 U_v^{-1}  = Z_4 Z_1 \gamma_4,  
\end{split} 
\label{eq:Urequirement}
\ee
so the cancellation occurs. We also require that $U_v$ should only contain parity-even operators, for otherwise it acts nontrivially on fermions outside of this quartet.

Analogously to the 1D case, we focus on the eigenbasis of $Z_e$'s where they take values $\pm 1$. Upon closer inspection of  Eq.~(\ref{eq:Urequirement}), we see that each $Z_e$ for $e = 1, 2, 3, 4$ shows up twice on the right-hand side. Flipping its value gives a sign change for exactly two fermions. For example, suppose that we are in the subspace where $(Z_1, Z_2, Z_3, Z_4) = (+,+,+,+)$, then $U_v$ is the identity operation. If we now switch to the $(-,+,+,+)$ subspace, i.e., $Z_1$ takes $-1$, then the effect of $U_v$ is to flip the sign of $\gamma_1$ and $\gamma_4$, the two neighboring fermions that share edge $e =1$. This can be realized by the unitary $ i\gamma_4 \gamma_1$ up to a phase degree of freedom. For example, $ -i\gamma_4 \gamma_1$ also achieves the desired effect. For each edge $e$ and its two associated Majorana fermions at $f_1(e)$ and $f_2(e)$, let us fix the ordering in a bilinear by choosing an orientation from $\gamma_{f_2(e)}$ to $\gamma_{f_1(e)}$. We place the fermion at the starting point $\gamma_{f_2(e)}$ on the right and the fermion at the end point $\gamma_{f_1(e)}$ on the left in the bilinear. Thus, the bilinear associated with $e$ is defined to be $S_e = i\gamma_{f_1(e)} \gamma_{f_2(e)}$. These orientations are indicated by dashed blue arrows in Fig.~\ref{fig:uni1}(a). It is easy to see that flipping the sign of $Z_1$ induces an independent action and thus can be factored out. Therefore, the corresponding factor in the unitary can be chosen to be $P^+_1 + P^-_1 (i\gamma_1 \gamma_4)$. Again, $P_e^{\pm}\equiv (I\pm Z_{e})/2$. If we include all four edges, then local unitary $U_v$ can be written as 
\be 
U_v = \stackrel{\leftarrow}{\prod}_{e =1, 2, 3, 4} [P^+_e + P^-_e S_e].
\label{eq:Uv2}
\ee 
Note that since the factors do not always commute, we have chosen an ordering to specify $U_v$: $\stackrel{\leftarrow}{\prod}_e O_e = O_4 O_3 O_2 O_1$. Clearly, $U_v = U_v^{\dagger} = U_v^{-1}$ is unitary and contains only parity-even operators. It is also easy to check that $U_v$ satisfies all the conditions in Eq.~(\ref{eq:Urequirement}). Thus, $U = \prod_v U_v$ is such a disentangling unitary that achieves our goal. Note that each $v$ belongs to only one sublattice.   One can check that all bilinears crossing edge $e$ dressed by $X_e$, $X_e S_e$ being one example, involve only factors of $(-1)^{F_f}$ after the unitary transformation.

\subsubsection{Square lattice: Example II}
\label{sec:sq2}
So far we have constructed a $U$ that factorizes into local unitaries $U_v$ acting on a group of four Majorana fermions around a vertex $v$. However, for more general surface graphs or lattices, such a partition of edges and fermions into separate groups is not always possible. To construct a disentangling unitary for a general case, such a restriction must be lifted. Before we discuss the general case, let us revisit the construction on the square lattice. In this and the next section, we will use the more precise notation $\gamma_{v_1(f)} =\gamma'_f$ and $\gamma_{v_2(f)} = \gamma_{f}$ whenever there is any potential notational confusion. $v_1(f)$ and $v_2(f)$ represent the respective (dual) vertices of $\gamma'_f$ and $\gamma_f$. Since the effect of $U$ on each pair of Majorana fermions on the same face, $\gamma_{v_1(f)}$ and $\gamma_{v_2(f)}$, is to produce a plaquette operator $\prod_{e\subset f} Z_e$ to cancel that in $G_f$, there is some flexibility in choosing how each Majorana should be transformed. We can assign $Z_e$ factors for $e \subset f$ differently to $\gamma_{v_1(f)}$ and $\gamma_{v_2(f)}$. We now use a different assignment than in  Fig.~\ref{fig:uni1}(a), as shown in Fig.~\ref{fig:uni1}(b).
As in the first assignment, each Majorana fermion is assigned two edges (indicated by red edges) and each edge is shared by two Majorana fermions. With this new assignment, Majorana fermions on the zigzag path are connected. The orientations on the edges are also indicated by the blue arrows. Repeating the analysis above, we can construct 
\be 
U = \stackrel{\leftarrow}{\prod}_{e} [P^+_e + P^-_e S_e]
\label{eq:square_U}
\ee 
to implement the desired transformation. Note that $U$ still factorizes along the zigzag paths, but each factor contains extensively many Majorana fermions and spins. Also note that there is flexibility in the ordering in the product, and for an arbitrary global ordering of edges we have defined $\stackrel{\leftarrow}{\prod}_e$ by placing operators on larger $e$ to the left of operators on smaller $e$, i.e., $\stackrel{\leftarrow}{\prod}_{e} O_e = \cdots O_{e_n} O_{e_{n-1}} \cdots $ for $e_{n-1}<e_n$. We can choose one ordering $e_{n-1}<e_n$ as shown in Fig.~\ref{fig:uni1}(b). The black arrow to the left indicates the increasing direction after a row is ordered from smaller edges to larger edges.  By moving $X_e$ to the left using $X_e P_e^{\pm} = P_e^{\mp} X_e$ and (anti-)commutation relations of Majorana bilinears,  we obtain 
\begin{align}
 U X_e U^{-1} & = S_e X_e \prod_{e' \in E(e), e'>e} (P^+_{e'} -P^-_{e'})    \nonumber\\
 &= S_e X_e \prod_{e' \in E(e), e'>e}  Z_{e'}. 
 \label{eq:square1}
\end{align} 
Here, $E(e)$ is the set that includes $e$ and all the other edges assigned to $\gamma_{f_1(e)}$ and $\gamma_{f_2(e)}$. For $e' \in E(e)$ and $e'\neq e$, $S_{e'}$ anticommutes with $S_e$, resulting in the minus sign in front of $P^-_{e'}$ for $e' >e$. The unitary transformation maps 
\begin{align}
X_e S_e   & \to  X_e \prod_{e' \in E(e), e'>e}  Z_{e'}   \prod_{e_1 \in E(f_1(e))} Z_{e_1} \prod_{e_2 \in E(f_2(e))}Z_{e_2}  \nonumber\\
& = X_e \prod_{e' \in E(e), e' <e} Z_{e'}.
 \label{eq:square2}
\end{align}  
In the first line, $E(f_1(e))$ and $E(f_2(e))$ are the edges assigned to $\gamma_{f_1(e)}$ and $\gamma_{f_2(e)}$, respectively. By definition, $E(e) = E(f_1(e))\cup E(f_2(e))$. The factors $\prod_{e_1 \in E(f_1(e))} Z_{e_1}$ and $\prod_{e_2 \in E(f_2(e))}Z_{e_2} $ are designed to cancel the plaquette terms.  Note that extra factors in $Z_{e'}$ in Eq.~(\ref{eq:square1}) and Eq.~(\ref{eq:square2}) are important to preserve the (anti-)commutation relations on the fermionic side. In particular, for a vertical edge $e =2$ and a horizontal edge $e =3$ shown in Fig.~\ref{fig:uni1}(b), $S_2 = i \gamma_2\gamma_1 $ and $S_3 = i \gamma_3\gamma_2 $. Then 
\be X_2 S_2 \to X_2 Z_1,\quad X_3 S_3 \to X_3 Z_2.
\label{eq:sq2}
\ee 
This is directly reminiscent of the exact bosonization map on the square lattice in Ref.~\cite{chen2018exact}. We will return to this point later in Sec.~\ref{sec:flatness},  when we discuss the flatness condition on the gauge spins.

\subsubsection{General case}
Having learned from the last example, we now outline the general procedure to construct a disentangling unitary, which is applicable to all polygonal decompositions. We can partition any two-dimensional manifold without a boundary into side-sharing polygons to form a surface graph. Triangulation or quadrangulation is special examples, but in general the polygons do not have to have the same number of edges. The edges need not be directed or oriented. 

Now we put a complex fermion, or equivalently two Majorana fermions $\gamma_{v_1(f)}$ and $\gamma_{v_2(f)}$, on each polygonal face $f$. This face can also be interpreted as the dual vertex using Poincar\'e duality \cite{hatcher2002algebraic}. For each $f$, there is a local fermion parity operator defined as $(-1)^{F_f}=i \gamma_{v_1(f)}\gamma_{v_2(f)}$. To gauge the total fermion parity, we place an Ising spin or qubit on each edge, and impose the Gauss law $G_f = i \gamma_{v_1(f)}\gamma_{v_2(f)} \prod_{e \in \partial f} Z_{e} =1$ for each polygon $f$.

For each polygon $f$, we partition its edges into two subsets and assign them, respectively, to the two Majorana fermions on it: $E(v_1(f)) \sqcup E(v_2(f)) = E(f)$ where $\sqcup$ is a disjoint union and  $E(f)$ is the set of the edges $e \subset f$. We often assume that the edges in both sets $E(v_1(f))$ and $E(v_2(f))$ are connected.  The examples discussed above are two ways to partition the edges of a quadrangle, but more generally, the partition does not have to be even. For example, in a quadrangle, we can assign all four edges to one Majorana fermion and none to the other. We carry out the assignment procedure for all polygons. Since every edge is shared by two polygons, every edge is assigned to two Majorana fermions at $f_1(e)$ and $f_2(e)$  in neighboring polygons. In more mathematical terms, each edge $e$ corresponds to two end points of the dual edge $e^*$ that crosses the edge $e$: $\partial e^* = f_1(e) + f_2(e)$ where coefficients are defined in $\bbz_2$ so choosing an orientation is not necessary. 

The partition and assignment of the edges of the polygon determine a way to transform the Majorana fermions $\gamma_{v_1(f)}$ and $\gamma_{v_2(f)}$:
\be 
\gamma_{v_1(f)} \to \prod_{e \in E(v_1 (f))}Z_e \gamma_{v_1(f)}, \  \gamma_{v_2(f)} \to \prod_{e \in E(v_2(f))}Z_e \gamma_{v_2(f)}.
\label{eq:general_Plaq}
\ee 
For each edge $e$, there is a gauge-invariant bilinear $X_e S_e$ with $S_e = i\gamma_{f_1(e)} \gamma_{f_2(e)}$. As on the square lattice, the ordering of fermions can be fixed by choosing an orientation for the bilinear.  Also, a unitary realizing the actions in Eq.~(\ref{eq:general_Plaq}) disentangles the fermions and spins, and such a disentangling unitary is given by 
\be 
U= \stackrel{\leftarrow}{\prod}_{e}[P_e^+ +P_e^- S_e],
\label{eq:general_U}
\ee 
where a global ordering of all edges is implied. If we write 
\be 
\tilde{P}_e^+ = P_e^+,\quad \tilde{P}_e^- = P_e^- S_e,
\ee 
then $U$ can also be expressed as 
\be 
U= \sum_{s_e = \pm}\quad \stackrel{\leftarrow}{\prod}_{e\in E} \tilde{P}_e^{s_e}.
\ee   
It is easy to check that $U$ maps all $G_f$ to $(-1)^{F_f}$.  As in the last section, $U$ maps 
\be 
  X_e    \to  S_e X_e \prod_{e' \in E(e), e' >e} Z_{e'}  ,
  \label{eq:general1}
\ee
where $E(e)\equiv E(f_1(e))\cup E(f_2(e))$, and 
\be 
X_e S_e    \to   X_e \prod_{e' \in E(e), e' <e} Z_{e'}.
\label{eq:general2}
\ee
This provides a very general expression for the mapping.

\subsubsection{Relation to Bravyi--Kitaev method}
\label{sec:BK}
As an example, we relate the analysis above to the Bravyi--Kitaev superfast scheme \cite{bravyi2002fermionic} by considering an arbitrary surface graph with a total ordering of all the edges $E$. The scheme is summarized in Appendix~\ref{app:BK}. For each polygon $f$, suppose the fermion parity operator is given by $(-1)^{F_f}=i \gamma_{v_1(f)}\gamma_{v_2(f)}$. Then we assign all the edges of this polygon to the first or second Majorana fermions. To be specific, we assign them all to $v_2(f)$ for all $f$, i.e., $E(v_2(f)) = E(f)$ and $E(v_1(f)) =\O$. Then, for each edge $e$, $f_1(e) =v_2(f_1)$ and $f_2(e) =v_2(f_2)$. Also, $E(e) = E(f_1) \cup E(f_2)$ and $S_e = i \gamma_{v_2(f_1(e))}\gamma_{v_2(f_2(e))}$. Thus, the disentangling unitary given by Eq.~(\ref{eq:general_U})
leaves $\gamma_{v_1(f)}$ invariant and maps $\gamma_{v_2(f)} \to  \gamma_{v_2(f)}  \prod_{e\in E(f)} Z_e $  
and $X_e S_e\to X_e \prod_{e' \in E(e), e' <e} Z_{e'} $. To compare with the Bravyi--Kitaev method, we also order the edges of each polygon $f$ by $<_f$. Such local orderings may or may not be extendable to a global ordering $<$ of the edges. If it does, then 
 \be 
 X_e \prod_{e' <e} Z_{e'} = 
 X_e \prod_{e_1 \in E(f_1), e_1<_{f_1} e} Z_{e_1}\prod_{e_2 \in E(f_2), e_2<_{f_2} e} Z_{e_2}.
 \ee 
This is essentially related to the operator $\tilde{A}_{jk}$ in the Bravyi--Kitaev map on the dual lattice (see Appendix~\ref{app:BK}). On the other hand, if a local ordering of two edges of a polygon $f$ is incompatible with a global ordering, we can apply an extra controlled-$Z$ gate on these two edges that maps $X_{e_1} \to X_{e_1} Z_{e_2}$ and $X_{e_2} \to Z_{e_1}X_{e_2}$. Before the application, the unitary acts with respect to the global ordering as 
\begin{align*}
  X_{e_1} & \to X_{e_1} \prod_{e' \in E(e_1), e' < e_1} Z_{e'}, \\
 X_{e_2} &\to X_{e_2} \prod_{e' \in E(e_2), e' < e_2} Z_{e'}.
\end{align*} 
Suppose $e_1 <e_2$, then $Z_{e_1}$ shows up in the second product $\prod_{e' \in E(e_2), e' < e_2} Z_{e'}$ but $Z_{e_2}$ does not show up in the first product $\prod_{e' \in E(e_1), e' < e_1} Z_{e'}$. Now we apply the additional $CZ_{e_1 e_2}$, then only $X_{e_1}$ and $X_{e_2}$ are affected. Then $Z_{e_2}$ shows up in the first product $\prod_{e' \in E(e_1), e' < e_1} Z_{e'}$ but $Z_{e_1}$ is canceled in the first product $\prod_{e' \in E(e_2), e' < e_2} Z_{e'}$. Effectively, it reorders $e_1$ and $e_2$ within a face. In the end, it reduces to the case when local orderings are compatible with a global ordering on edges. In other words, we do not lose any generality by choosing local orderings compatible with a global order. The Bravyi--Kitaev bosonization method maps (ungauged) Majorana bilinears to spin operators. To make this final connection within our framework, we now consider the flatness condition on the fermionic side and the Gauss law on the spin side.

\subsection{Flatness condition and Gauss law}
\label{sec:flatness}
We have gauged the fermion parity of the fermionic system and mapped the gauged fermion system to a spin system. A bosonization scheme should map the (ungauged) fermionic system to a spin system. In 1D, it is well known that coupling a system to a background $\bbz_2$ gauge field is equivalent to twisting the boundary condition. Rendering the gauge field dynamical is equivalent to summing over all possible boundary conditions. For a fermionic symmetry, gauging parity, equivalently summing over different boundary conditions, amounts to summing over spin structures. Gauging the global $\bbz_2$ symmetry induces a dual $\bbz_2$ symmetry in the gauged system. Gauging this dual $\bbz_2$ symmetry can reproduce the original system with the original $\bbz_2$ global symmetry. No essential information of the system is lost in this duality transformation. 

In 2D, to obtain a duality transformation between the original fermion system and the parity-gauged fermion system, we need to impose the flatness condition on the gauge spins: 
\be 
\prod_{e \supset v} X_e =1
\ee 
for all $v$.
In a toric code,  
$\prod_{e \supset v} X_e$ is the vertex operator. The condition guarantees that fermions are only coupled to a flat gauge field with a vanishing ``magnetic" flux and that there is no extra Aharonov-Bohm phase difference when a fermion takes different local paths around a polygon to move from one point to another. In 1D, this is automatically satisfied because there is no curvature/flux, or equivalently, no alternative local paths. Under the flatness condition, gauging the parity of a fermionic system produces a dual 1-form symmetry in the parity-gauged fermionic system. The process can be reversed by gauging the dual 1-form symmetry \cite{gaiotto2016spin, bhardwaj2017state, thorngren2020anomalies, su$mathbbZ_2$2025a}. Thus, the flatness condition is required for the duality transformation. 

By the general expression in Eq.~(\ref{eq:general2}), the vertex operator is mapped to
\be 
 \prod_{e \supset v}  X_e  \to \left(\prod_{e \supset v} S_e \right) \cdot \prod_{e \supset v}\left[  X_e  \prod_{e' \in E(e), e' >e} Z_{e'} \right].
 \label{eq:vertex}
\ee
Here, we have assumed an ordering in the product. Also, $\prod_{e \supset v} S_e$ must be proportional to some product of $(-1)^{F_f}$ with $v\subset f$. Plugging in $(-1)^{F_f} =1$ yields a possibly nontrivial phase, depending on the surface graph and the assignment procedure. For the general case, if we assign all edges of each polygon to one type of Majorana fermions, no such $(-1)^{F_f}$  appears in this factor. Given a regular lattice,  such a factor can be easily evaluated. 

If we regard the parity-gauging and the disentangling unitary transformation as a duality transformation, then we obtain 
\be 
(-1)^{F_f} \to \prod_{e \subset f} Z_e
\label{eq:general3}
\ee 
and 
\be 
S_e  \to   X_e \prod_{e' \in E(e), e' <e} Z_{e'}.
\label{eq:general4}
\ee
A product of concatenated fermion bilinears on a closed loop around vertex $v$ is mapped to the Gauss law in the spin system up to a phase, as shown in Eq.~(\ref{eq:vertex}). These are the general bosonization maps on a simply connected surface graph, such as an infinite plane. Frequently, these maps can be written in a more compact form if more information about the surface graph is known. For example, if the surface graph is a triangulation and there is a branch structure. We will discuss this further in the following sections, we can make use of tools such as cup products in algebraic topology. 

If the surface graph is not simply connected, then there exist nontrivial loops/cycles in the space. Then the bosonization determined by the maps above fails to be an isomorphism of algebras. This is not surprising because even in the 1D case: when we map the periodic Majorana chain to the TFIM, a product of concatenated fermion bilinears over the entire chain $\prod_j (i \gamma_j \gamma_{j+1})$ is mapped to $\prod_j X_j$ which generates the dual global symmetry of the latter. Therefore, the 1D bosonization is only an isomorphism within certain charge  and twisted (with respect to the periodic boundary condition) sectors. This picture generalizes to 2D as well, except that now the dual symmetry is a 1-form symmetry. Indeed, a product of concatenated fermion bilinears over a nontrivial cycle of the space can be mapped to a product of $X_e$ over the cycle dressed by extra $Z_e$ factors determined by Eqs.~(\ref{eq:general3}) and (\ref{eq:general4}). We will encounter these 1-form symmetries again in Sec.~\ref{sec:noninv}.

\subsubsection{Example: Bosonization on square lattice}
\label{sec:bos_square}
\begin{figure}[bt]
    \centering \includegraphics[width=0.98\linewidth]{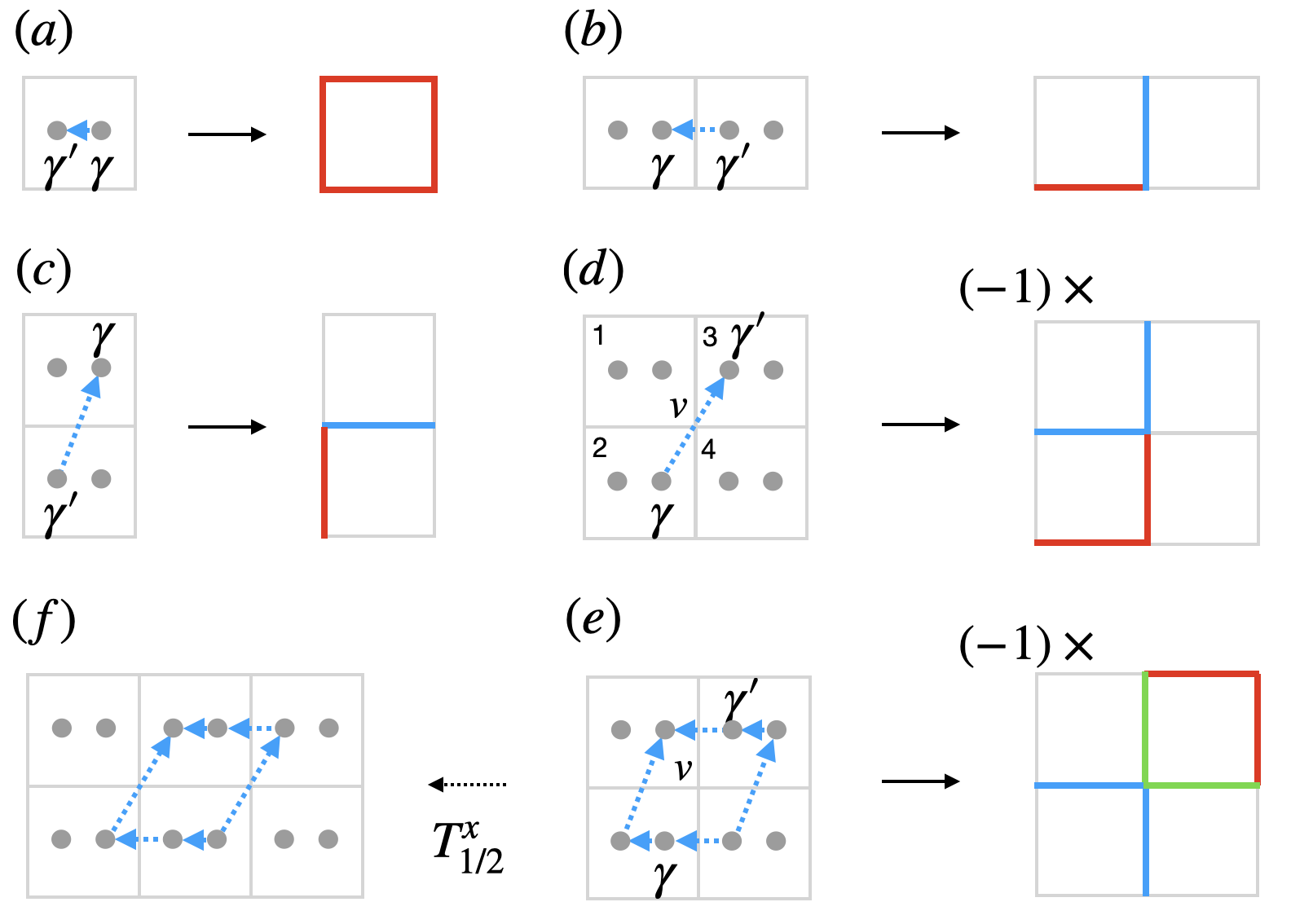}
    \caption{Bosonization of Majorana fermion bilinears on the square lattice.  Dashed blue arrows indicate the ordering of the two Majorana fermions in the bilinears. Corresponding spin operators are the products of Pauli operators on the colored edges: $Z$ (red), $X$ (blue), and $Y$ (green). If there is an extra minus sign, then it is placed at the upper left corner. In (e), a loop around vertex $v$ over which concatenated fermion bilinears are multiplied is shown;  the corresponding spin operator is the Gauss-law operator $G_v$. The loop in (f) is obtained from that in (e) by the minimal half-unit-cell translation $T^{1/2}_x$.}
    \label{fig_bos}
\end{figure}
Let us work out the bosonization map for a concrete example on the square lattice discussed in Sec.~\ref{sec:sq2}. Using Eq.~(\ref{eq:sq2}) and Eqs.~(\ref{eq:general3}) and (\ref{eq:general4}), we can obtain the bosonization maps from fermion bilinears to spin operators as shown in Fig.~\ref{fig_bos}(a-c). A red (blue) edge represents the Pauli operator $Z (X)$ on the edge. The dashed blue arrows fix the order within bilinears.  Let $e^f_{x/y}$ be the edge on the right/top of face $f$ and $e^f_{-x/y}$ be the edge on the left/bottom of face $f$.  Following Ref.~\cite{chen2018exact}, we define
\be W_f = \prod_{e \subset f} Z_e,\quad  V_{e^f_x} = X_{e^f_x} Z_{e^f_{-y}}, \quad V_{e^f_y} = X_{e^f_y} Z_{e^f_{-x}}.\ee 
Then, more explicitly, the bilinears are mapped to
\be 
i \gamma'_f \gamma_f  \to W_f,\quad i \gamma_{f} \gamma'_{f+a_x} \to V_{e^f_x},\quad i \gamma_{f} \gamma'_{f+a_y} \to V_{e^f_y},
\ee 
as shown in Fig.~\ref{fig_bos}(a-c). Maps for other bilinears can be generated by these maps. In particular, the bilinear in  Fig.~\ref{fig_bos}(d) is $i\gamma'_3 \gamma_2$, which can be written as
\be 
i  \gamma'_3 \gamma_2 = (i \gamma_1 \gamma'_3)  (i \gamma_1 \gamma'_2) (i \gamma'_2 \gamma_2), 
\label{eq:gamma23}
\ee 
and the corresponding spin operator is 
\be 
 V_{e^1_x}  V_{e^2_y} W_2  =  - X_{e^1_{x}}  X_{e^2_{y}} Z_{e^2_x} Z_{e^2_{-y}}, 
\ee
which is also shown in Fig.~\ref{fig_bos}(d). In general,
\be 
 i \gamma'_{f+a_x+a_y}\gamma_f \to   V'_{e^f_y}  \equiv - X_{e^{f+a_y}_{y}} X_{e^f_{y}}Z_{e^f_x} Z_{e^f_{-y}}.
\ee 
Note that there is another way to represent $i\gamma'_3 \gamma_2$:
\be 
i  \gamma'_3 \gamma_2 = (i \gamma'_3 \gamma_3)  (i \gamma_3 \gamma'_4) (i \gamma_2 \gamma'_4),
\ee 
then the corresponding spin operator is 
\be 
 W_{3} V_{e^4_y} V_{e^2_x} = -Z_{e^3_x} Z_{e^3_y} Z_{e^3_{-x}} Z_{e^2_{-y}} Y_{e^4_y} Y_{e^2_x} . 
\ee
For the two expressions for the spin operator to be consistent, we need  
\be 
 W_{3} V_{e^4_y} V_{e^3_x} W_2   V_{e^2_y}  V_{e^1_x} = \left( \prod_{e\subset v} X_{e}\right) W_3=1. 
\ee
This is also the image of the product of concatenated fermion bilinears along the closed loop shown in Fig.~\ref{fig_bos}(e). This is exactly the Gauss law \be 
G_{v} =\left( \prod_{e\subset v} X_{e}\right) W_{f_v} =1
\label{eq:gauss_modified}
\ee  at $v$ found in Ref.~\cite{chen2018exact}. Here, $W_{f_v}$ represents the plaquette operator on the face $f_v$ sitting on the northeast corner of $v$.  One can check this is consistent with the expression in Eq.~(\ref{eq:vertex}) for the global ordering we have chosen in Sec.~\ref{sec:sq2}. In the $\bbz_2$ lattice gauge theory associated with the conventional toric code, the Gauss law is $\prod_{e\subset v} X_{e} =1$. The Gauss law in Eq.~(\ref{eq:gauss_modified}) is modified by attaching a plaquette (flux) operator to the vertex operator. This modified Gauss law is associated with a twisted toric code \cite{chen2021disentangling}. Without setting $G_v$ to 1, $G_v$ can be viewed as a (1-form) symmetry operator, since it commutes with the Hamiltonian.  It gives rise to a 't Hooft anomaly and renders the basic excitations fermionic \cite{chen2018exact}. If the square lattice has a topology with nontrivial cycles, such as a torus, we need to consider a product of concatenated fermion bilinears over these nontrivial cycles. The bosonized operator is a loop of (possibly dressed) spin operators which generates a 1-form symmetry. As an example, consider the product of concatenated fermion bilinears along the zigzag path shown in Fig.~\ref{fig:uni1}. If the loop closes, say on a torus, the corresponding spin operator is a product of $Y$ operators over the zigzag path that generates a $\bbz_2$ 1-form symmetry.

\begin{figure}[bt]
    \centering
    \includegraphics[width=1.0\linewidth]{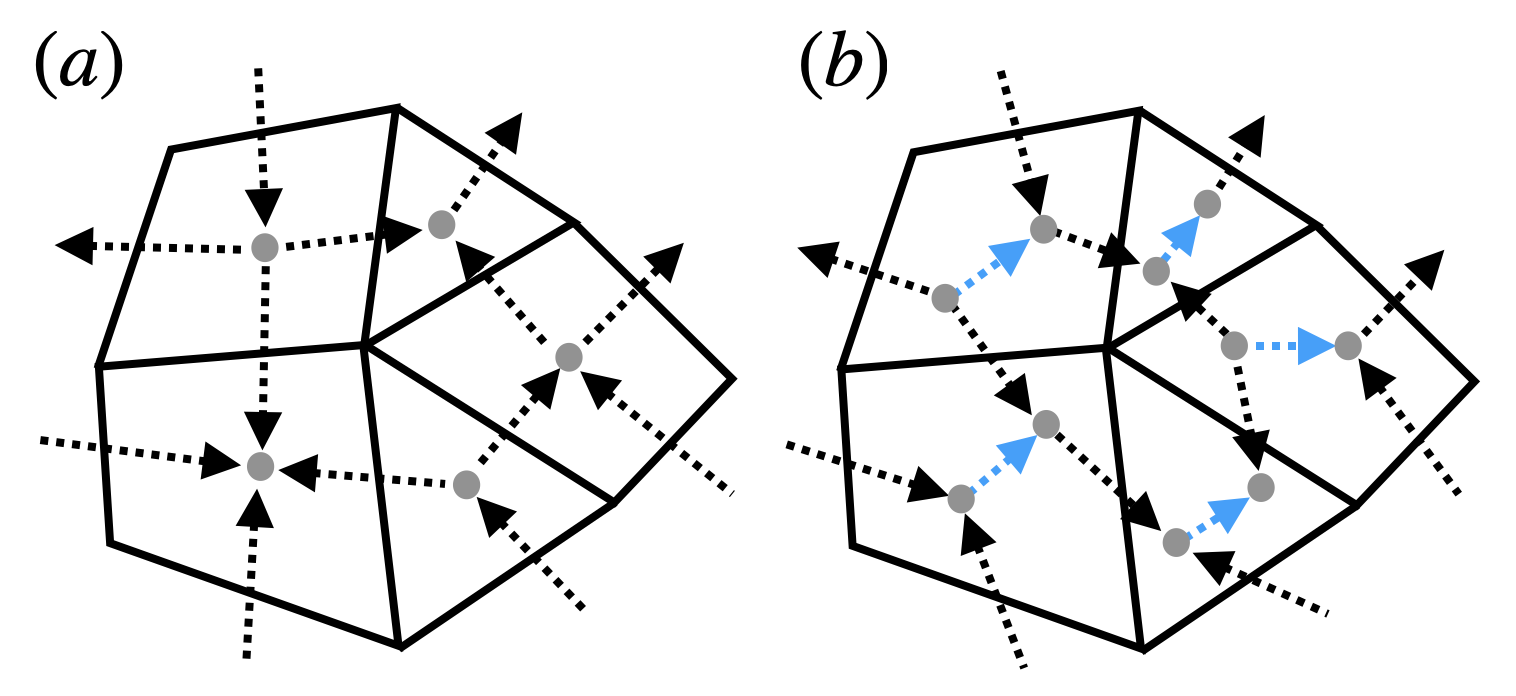}
    \caption{Surface graphs of edges with Majorana fermions on the each face. In the assignment procedure when all edges within a face are assigned to one type of the Majorana fermions $\gamma$ or $\gamma'$ on the face, these singled-out fermions form a dual surface graph as illustrated in (a). If both fermions on a face are assigned at least one edge, all fermions may participate in a dual surface graph as illustrated in (b). Intra-face bilinears (dashed blue) are distinguished from inter-face bilinears (dashed gray). The arrows indicate the ordering of the two Majorana fermions within a fermion bilinear: the fermion at the starting (ending) point is placed to the right (left). To illustrate Kasteleyn orientations, the orientations of the dual edges are chosen purposefully so that there are an odd number of clockwise-oriented edges for the dual face at the center.}
    \label{fig:dimer}
\end{figure}

\subsubsection{Dependence on Kasteleyn orientations}
\label{sec:Kasteleyn}
Let us now return to the general case and study the dependence of the Gauss law on the product of concatenated fermion bilinears over a closed loop. 

In the Sec.~\ref{sec:BK}, we assign all edges of each polygon to one type of Majorana fermions $\gamma_{v_2}$, then the product of concatenated fermion bilinears over a closed loop $\widetilde{\prod}_{e \supset v} S_e = \widetilde{\prod}_{e \supset v} (i \gamma_{v_2(e)}\gamma_{v_2(e)})$ is a phase. A schematic is shown in Fig.~\ref{fig:dimer}(a), where  dashed arrows between Majorana fermions $\gamma_{v_2}$ specify the ordering for each $S_e$. Here, we have used $\widetilde{\prod}$ to denote a product  such that  $\widetilde{\prod}_{e \supset v}  O_e = O_{e_{q_v}} O_{e_{q_v-1}}\cdots O_{e_{1}}$ where $e_1, e_2, \ldots, e_{q_v}$ is ordered counterclockwise and $q_v$ is the degree of the vertex $v$, i.e., the number of edges incident to it. Note that this ordering need not be compatible with the global ordering on all edges.

So far, we have left the orientations arbitrary between two Majorana fermions in $S_e$ and the disentangling unitary in Eq.~(\ref{eq:general_U}). To fix $\widetilde{\prod}_{e \supset v} S_e$,  let us consider the graph formed by the (dual) edges connecting the complex fermions, or equivalently, $\gamma_{v_2}$, as opposed to the graph formed by the edges where the gauge spins sit. For each face in the dual graph, if all edges are oriented counterclockwise, then the product is equal to $i^{q_v}$, where again $q_v$ is degree of the vertex $v$ dual to the face. We have used the definition that $\widetilde{\prod}$ is ordered counterclockwise and the fact that Majorana fermions square to the identity $\gamma^2=1$. Flipping the orientation of one edge also flips the sign of the product. Let $n_{\text{cw}}(v)$ be the number of clockwise edges, then 
\be 
\widetilde{\prod}_{e \supset v} S_e = i^{q_v} n_{\text{cw}}(v).
\ee  

If there are an even number of dual vertices, i.e., $\gamma_{v_2}$, then there is another way to simplify the expression. We now choose a Kasteleyn orientation on the dual graph such that for every dual face, the number of dual edges oriented clockwise around it is odd. Such a Kasteleyn orientation always exists on any surface graph with an even number of vertices \cite{cimasoni2007dimers}. In this case, 
\be \widetilde{\prod}_{e \supset v} S_e = i^{q_v} (-1) = i^{q_v-2},
\label{eq:fermionloop1}
\ee
Note that the expression remains invariant if the orientations of all the edges incident to any fermion $\gamma_{v_2}$ are flipped because for each face, the orientations of two edges are flipped. Since two Kasteleyn orientations are said to be equivalent if they are related by repeating the orientation-flip action on a subset of vertices, Eq.~(\ref{eq:fermionloop1}) only depends on the equivalence class of a Kasteleyn orientation.  Plugging Eq.~(\ref{eq:fermionloop1}) into Eq.~(\ref{eq:vertex}), we obtain
\be 
G_v \equiv i^{q_v-2} \widetilde{\prod}_{e \supset v}  \left[  X_e  \prod_{e' \in E(e), e' >e} Z_{e'}\right].
\label{eq:Gauss_general1}
\ee 
Alternatively, by Eq.~(\ref{eq:general2}),
\be 
 \widetilde{\prod}_{e \supset v} X_e S_e
 \to  \widetilde{\prod}_{e \supset v} \left[  X_e  \prod_{e' \in E(e), e' <e} Z_{e'} \right].
\ee 
and $ \widetilde{\prod}_{e \supset v} X_e =1 $
on the left-hand side, so we have 
\be
\begin{split}
 &(-i)^{q_v-2} 
\widetilde{\prod}_{e \supset v} S_e=1 \\
&\to G_v \equiv  (-i)^{q_v-2} \widetilde{\prod}_{e \supset v} \left[   X_e  \prod_{e' \in E(e), e' <e} Z_{e'}\right].
\end{split}
\label{eq:Gauss_general2}
\ee 
It is not hard to see that two expressions are equivalent. Thus, 
$G_v =1$ is the modified Gauss law for the spin system on a general 2D graph. This completes the final step to relate our bosonization scheme to the Bravyi--Kitaev method in Sec.~\ref{sec:BK}. Note that non-equivalent Kasteleyn orientations are in one-to-one correspondence with discrete spin structures. This is proved in Ref.~\cite{cimasoni2007dimers} by using the Kuperberg's construction to relate the number of clockwise-oriented (or counterclockwise-oriented) edges to vector fields with only singularities of even index in all faces and using the dimer configuration to eliminate the singularities of odd index on the vertices. We provide a slightly different proof in the next section. On a surface graph with genus $g$, there are $2^{2g}$ nonequivalent Kasteleyn orientations and spin structures on the surface. Thus, in this case, we have shown that the Gauss law implicitly depends on the spin structure. 

If the number of dual vertices is odd, there is at least a face violating the Kasteleyn orientation and the corresponding product of fermions in Eq.~(\ref{eq:fermionloop1}) carries an extra minus sign in this face. This is not surprising as the entire fermionic system is built from complex fermions, or equivalently, from twice as many Majorana fermions. To take into account all Majorana fermions, we need an assignment scheme such that none of the sets $E(v_1(f))$ and $E(v_2(f))$ are empty and then all vertices participate in the graph. 

More generally, consider such an assignment procedure that the Majorana fermions form a surface graph. This is the case when the edges in each subset are connected. See Fig.~\ref{fig:dimer}(b) for a schematic. Then there are an even number of vertices and the complex fermions form a natural dimer cover for the graph. Strictly speaking, this surface graph is not exactly the dual of the surface graph of gauge spins due to the extra local parity bilinear in each face, a fact that can be seen directly by comparing Fig.~\ref{fig:dimer}(b) with Fig.~\ref{fig:dimer}(a). A better terminology may be ``quasi-dual." However, for simplicity, we still refer to this type of surface graphs formed Majorana bilinears as the dual of the surface graphs formed by gauge spins.  On the dual surface graph, we can choose a Kasteleyn orientation so that there are an odd number of clockwise-oriented edges within any face. In this general case, the orientations of $(-1)^{F_f}$ also participate in the closed loop $\Gamma_v$ and 
\be 
\widetilde{\prod}_{e \supset \Gamma_v} S_e =i^{q_v-2}. 
\ee
Therefore, the left-hand side in Eq.~(\ref{eq:Gauss_general2}) is modified to 
\be 
 (-i)^{q_v-2} \left(
\widetilde{\prod}_{e \supset \Gamma_v} S_e \right)\left(  \widetilde{\prod}_{e \supset v} X_e \right)=1.
\ee 
Making use of Eqs.~(\ref{eq:general2}) and (\ref{eq:general3}), we can map this expression to the modified Gauss law. Compared to that in Eq.~(\ref{eq:Gauss_general2}), the general Gauss law has some insertions of the plaquette operators in the product $\widetilde{\prod}$ at the right place.  Once again, the expression is invariant under orientation-flip transformations at any Majorana vertex, so it only depends on the equivalent class of Kasteleyn orientations, and equivalently discrete spin structures.

\begin{figure}[bt]
    \centering
\includegraphics[width=1.0\linewidth]{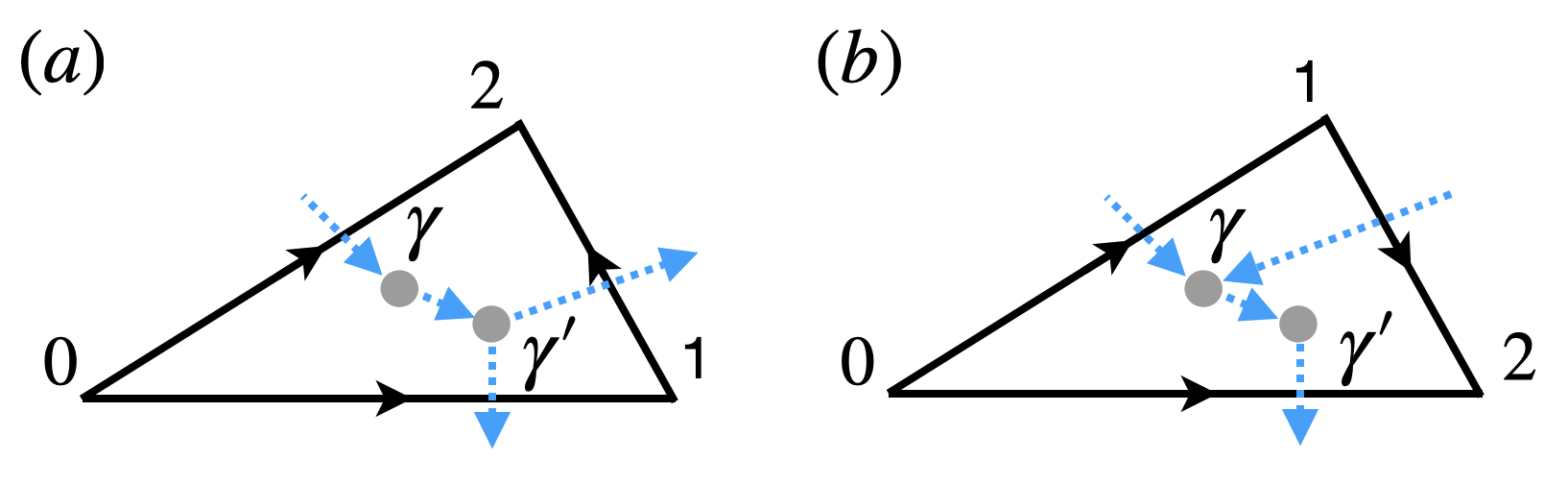}
    \caption{Branching structures and orientations on a triangle and its dual.  The triangle in (a) is called positively oriented while the triangle in (b) is called negatively oriented. The black arrows determine the ordering of vertices $0<1<2$, and also determine an assignment scheme: $\gamma$ ($\gamma'$) is assigned an edge (connected by a dashed blue arrow) if $\gamma$ ($\gamma'$) lies to the right (left) along the orientation of the edge. The dashed blue arrows also fix the ordering within each bilinear by pointing from the Majorana fermion on the right to the one on the left in Eq.~(\ref{eq:MajBilinears}).}
    \label{fig:branching}
\end{figure}

\subsection{Kasteleyn orientations and spin structures} 
We elaborate more on the relation between the Kasteleyn orientation and the second Stiefel-Whitney (SW) class (and discrete spin structures) in this section. The reader who is more interested in the application of the bosonization scheme to KW dualities in higher dimensions can skip this section.   
\label{sec:triangulation} 
\subsubsection{Triangulation}
The relation between the Kasteleyn orientation for products of concatenated fermion bilinears over closed loops and the discrete spin structure can be established when the spatial configuration of fermions induces a triangulated graph for the gauge spins. The triangulated graph can be endowed with a branching structure, which is a choice of an orientation for each edge such that there is no oriented loop on any face. The branching structure specifies the ordering of the vertices of a triangle: $i < j$ for the orientation from vertex $i$ to vertex $j$.  There are two types of branching structures on a triangle: positively oriented ($+$-oriented) as in Fig.~\ref{fig:branching}(a) and negatively oriented ($-$-oriented) as in Fig.~\ref{fig:branching}(b). The orientations of edges also determine a way to partition and assign the edges of any triangle $f$: we assign $\gamma'_f$ to an edge if $\gamma'_f$ lies to the left along its orientation and $\gamma_f$ to an edge if $\gamma_f$ lies to the right along its orientation. The assignment is represented by the red edges and dashed blue lines. We can specify the ordering within each bilinear such that  
\be 
(-1)^{F_f} = i \gamma'_f \gamma_f, \quad S_e = i \gamma_{R(e)}\gamma'_{L(e)},
\label{eq:MajBilinears}
\ee 
which also defines an orientation from the Majorana operator on the right to the one on the left. Note that two bilinears $S_e$ and $S_{e'}$ always commute unless $e$ and $e'$ are two concatenated edges with compatible orientations, i.e., not pointing toward each other. As we explain in Appendix~\ref{app:CK}, this property is properly captured by the cup product. 

For any vertex at the centered of Fig.~\ref{fig:dimer}, let us consider the orientations of the bilinears (on the dual edges) surrounding it. Using the convention shown in Fig.~\ref{fig:branching}, we can see that there are only two cases in which clockwise-oriented arrows surround it. The first is when the arrow is crossing an edge  that is pointing away from the vertex. The second is when the arrow is in a triangle such that the vertex is the smallest, i.e., 0. As defined in Appendix~\ref{app:algebraictopology}, these vertices are ``regular" in the corresponding edge or triangle in 2D. It turns out the number of clockwise-oriented arrows of the bilinears for each vertex $v$, $N_{\text{cw}}(v)$, is directly related to discrete spin structures. Indeed, an orientable $n$-dimensional manifold admits a spin structure if and only if the second SW class $w_2 \in H^2(M, \bbz_2)$ vanishes. By the Poincare duality, the second SW class $w_2 \in H^2(M, \bbz_2)$ is dual to a zeroth SW class in $H_0(M, \bbz_2)$ represented by a formal sum of vertices
\be 
\tilde{w}_0 = \sum_{v} N(v) v,
\label{eq:SW}
\ee 
where $v$ is any vertex ($0$-simplices),  and $N(v)$ is the number of simplices $\Delta_i$ such that $v$ is regular in $\Delta_i$ for any $i =0, 1, 2$ (see Appendix~\ref{app:algebraictopology}).  In particular, a vertex $v$ in 2D is regular in itself, regular in $\Delta_1$ if it is the smaller vertex, and regular in $\Delta_2$ if it is the smallest vertex. Thus, the number $N(v)$ is exactly one (due to itself) plus the number of clockwise-oriented arrows with respect to $v$:
\be 
N(v) = 1+ N_{\text{cw}}(v).
\ee 
In a Kasteleyn orientation, $N_{\text{cw}}(v)$ is odd, and thus $N(v)$ is even. As a result, all terms in Eq.~(\ref{eq:SW}) vanish, which is not unexpected since the dual second SW class is trivial. Flipping the orientation of one edge leads to the violation of the Kasteleyn condition in the neighboring faces, which is interpreted as a change in the representation of $\tilde{w}_0$ by a boundary. Since the chain $[\tilde{w}_0]$ is trivial, there exist 1-chains $S$ such that $\partial S =0$ using the representation in Eq.~(\ref{eq:SW}) for $[\tilde{w}_0]$. Different inequivalent choices of $S$ are then in one-to-one correspondence with the elements in $H_1(M, \bbz_2) \cong H^1(M, \bbz_2)$. This is compatible with our general discussion in the last section.  

It is worth noting that a similar geometric relation between the Kasteleyn orientation and the discrete spin structure is also discussed in Ref.~\cite{wang2018towards} where the authors use the relation to construct nontrivial fermionic symmetry-protected topological (SPT) phases by decorating the domain walls of the bosonic SPT phases with Majorana chains. There the Majorana lattice is different because there are more than two Majorana fermions within each triangle. See also Refs.~\cite{gaiotto2016spin, tarantino2016discrete, tata2023anomalies} for other types of Majorana lattices. In Refs.~\cite{chen2019bosonization, chen2020exact}, the dependence of the exact bosonization on the spin structure is also revealed using the cup product without referring to the Kasteleyn orientations. We show that the general framework presented in this work reproduces their results in Appendix~\ref{app:CK} with slight modifications.

\subsubsection{General graphs}
\label{sec:KO_general}
\begin{figure}[bt]
    \centering
\includegraphics[width=0.8\linewidth]{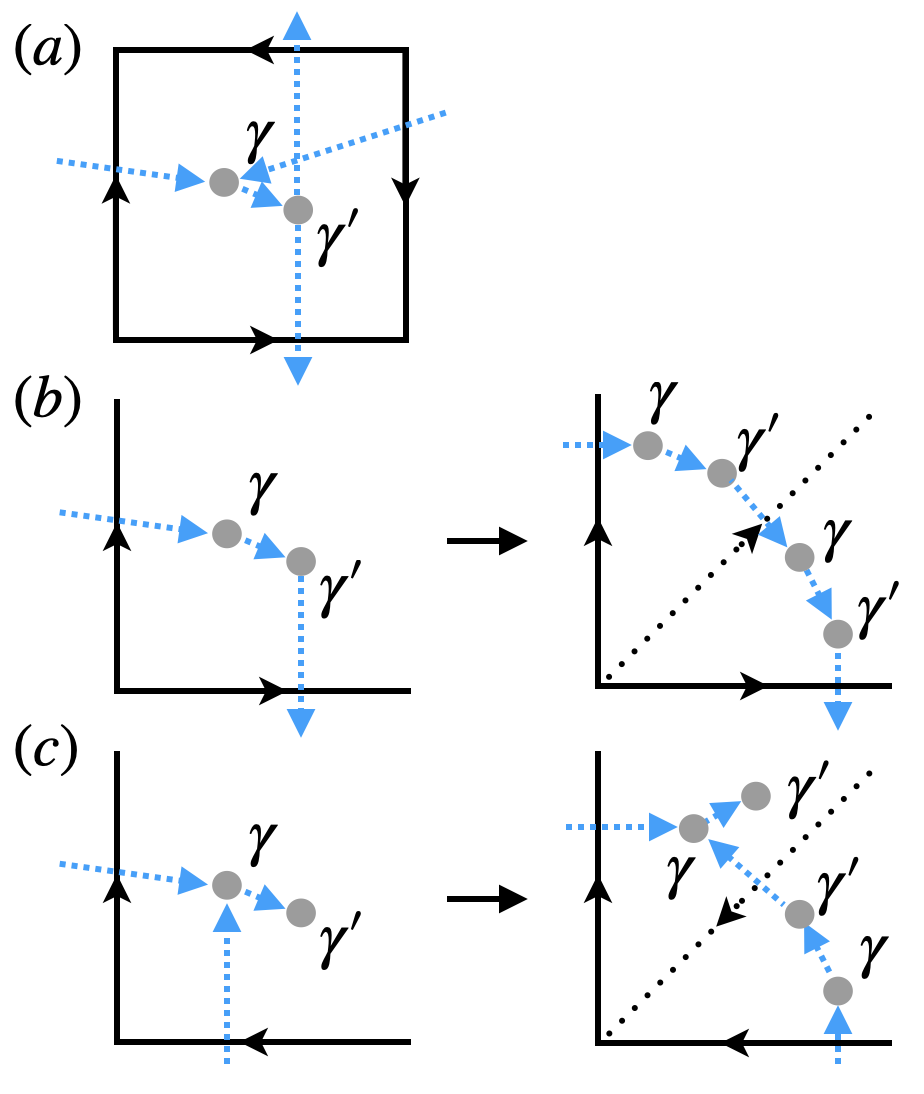}
    \caption{(a) Example of orientations of edges (black) and bilinears (dashed blue) that lead to a crossing of two dual edges. This case is not allowed for a (embedded) surface graph. (b,c) Two cases of adding a ``hidden" edge (dotted) to triangulate the general surface graph. $N_{\text{cw}}(v) \mod 2$ remains invariant. }
    \label{fig:Kas}
\end{figure}

Since the one-to-one correspondence between the equivalent Kasteleyn orientations and discrete spin structures in 2D does not require a triangulation with a branching structure, we expect the representation of the second SW class obtained above
\be \tilde{w}_0 = \sum_v [1+ N_{\text{cw}}(v)] v
\label{eq:Kas_general}
\ee 
to be valid for a general surface graph when the Majorana fermions also form a (dual) surface graph. Indeed, this can be proved inductively based on the triangulated case. 

Consider a general surface graph with a global ordering of the vertices. The global ordering induces an orientation on each edge and thereby  a branching structure of the surface graph. In each face there is a pair of Majorana fermions. As in the triangulated case, we fix the ordering within each bilinear and the corresponding orientation as in Eq.~(\ref{eq:MajBilinears}). Now we need to assume that the arrows formed by these bilinears form a dual surface graph as in Fig.~\ref{fig:dimer}(b). If this is not the case, then there are crossings of these dual edges, violating the (embedded) surface graph requirement. For example, in Fig.~\ref{fig:Kas}(a), $\gamma$ and $\gamma'$ are, respectively, assigned to two opposite edges in the quadrangle, leading to a crossing of two dual edges. This type of configuration should be ruled out. Once the arrows formed by these fermion bilinears form a dual surface graph, we can denote the number of clockwise-oriented arrows surrounding vertex $v$ by $N_{\text{cw}}(v)$. 

Let us now consider the representatives of $[\tilde{w}_0]$ for the general surface graph. There are many different ways to view the general graph (ignoring the fermions) as a triangulated graph with some ``hidden" edges. Pick one way by specifying the ``hidden" edges, and then a representative of $[\tilde{w}_0]$ for the general surface graph is $\tilde{w}_0 = \sum_v N(v) v$ as in the discussion above. With respect to this ``hidden" triangulation, we can relate $N(v)$ to the clockwise-oriented edges of the bilinears using the orientation rules discussed above in the triangulated case. Let $N_{\text{cw, tri}}(v)$ be the number of clockwise-oriented arrows in this ``secretly" triangulated graph with respect to vertex $v$. Then
$N(v) = 1+ N_{\text{cw, tri}}(v)$. We want to show that $N_{\text{cw, tri}}(v) =N_{\text{cw}}(v) \mod 2$. Since $N_{\text{cw}}(v)$ is fixed for the general surface graph, this equality also implies that any other way of specifying the ``hidden" edges yields the same $N_{\text{cw, tri}}(v)$. 

To prove $N_{\text{cw, tri}}(v) =N_{\text{cw}}(v) \mod 2$, we use induction on the consequence of adding a ``hidden" edge adjacent to $v$ in the general graph with an orientation compatible with the global order. All scenarios of adding one edge to $v$ can be enumerated. For example, for the two edges in the general graph in Fig.~\ref{fig:Kas}(b), there are three clockwise-oriented bilinears surrounding the vertex. Adding the ``hidden" edge represented by the dotted line increases the number of clockwise-oriented bilinears by exactly 2. Another example is shown in Fig.~\ref{fig:Kas}(c) where again adding the ``hidden" edge does not change the number of clockwise-oriented bilinears. Four other allowed cases can be analyzed in a similar way. Adding one ``hidden" edge for the triangulation preserves the value of $N_{\text{cw}}(v) \mod 2$. Obviously, this process can be repeated for all ``hidden" edges until the general graph is fully triangulated, leading to $N_{\text{cw, tri}}(v) =N_{\text{cw}}(v) \mod 2$. Therefore, we have shown that $\tilde{w}_0$ in Eq.~(\ref{eq:Kas_general}) is indeed a representation of the second SW class $w_2$ for the general case. Intuitively, it is the local branching structure of the surface graph that specifies the relation between $N_{\text{cw}}(v)$ and $N(v)$ and determines $w_2$. If a Kasteleyn orientation is chosen, then $N_{\text{cw}}(v) $ is odd for all $v$ and $\tilde{w}_0$ vanishes. Different spin structures $S$ can then be obtained, as in the triangulated case. We conclude our discussion of the relationship between Kasteleyn orientations and discrete spin structures.

\section{Kramers-Wannier dualities in 2D}
\label{sec:KW}
Equipped with a general bosonization framework, 
we now apply it to fermionic systems on regular lattices with translational invariance. 

\subsection{An intuitive look at KW dualities beyond 1D}
As reviewed in Section \ref{sec:review1d}, bosonization maps the minimal, i.e., half-unit-cell, translation of the free Majorana chain to the conventional KW duality of the TFIM: $Z_{j-1} Z_{j} \to X_j \to Z_j Z_{j+1}$. In physical terms, the Ising coupling term $Z_{j-1} Z_{j}$ measures domain walls of the model and $X_j$ measures spin polarizations with respect to the eigenbasis of $X$.
We want to generalize the duality to higher dimensions. 
One simple way to achieve it is to consider a square lattice and add terms that transform under the conventional KW duality. If we use $Z_{n, j}$ and $X_{n,j}$ to label spin operators in different rows, then we can add $(Z_{n, j-1}Z_{n, j}) (Z_{n+1, j-1}Z_{n+1, j})$ and $X_{n, j}X_{n+1, j}$ to the sum of the 1D TFIM in different row:
\begin{align}
H^{\text{2D}} =& - \sum_{n, j}t_1 Z_{n, j-1} Z_{n, j} +t_2  X_{n, j} \\
   + & t'_1 (Z_{n, j-1}Z_{n, j}) (Z_{n+1, j-1}Z_{n+1, j}) + t'_2 X_{n, j}X_{n+1, j}.   \nonumber  
\end{align}  
It is easy to see that the Hamiltonian has a KW duality along the rows, which is related to the $\bbz_2$ (subsystem) symmetry within each row. When some parameters in the Hamiltonian vanish, e.g. $t_1 = t'_2 =0$, the Hamiltonian can also possess a $\bbz_2$ subsystem symmetry along the columns. Composing these two $\bbz_2$ subsystem symmetries (and a basis rotational transformation $Z \leftrightarrow X$), we can show the Hamiltonian in this case is also dual \cite{xuStrongWeak2004, cobanera2011bond, caoSubsystem2023}.

In this section, we generalize the conventional KW duality to higher dimensions by bosonizing a higher dimensional Majorana lattice. Before we discuss the general scenario, let us gain some intuition by studying a limiting case. Imagine that we have infinitely many 1D identical Majorana chains placed on a 2D plane in parallel. Treating them as strictly 1D chains, each Majorana chain can be mapped independently to a 1D TFIM. The minimal translation along the chain direction is still mapped to the conventional KW duality. However, if we view them as a 2D system and want to apply the higher-dimensional bosonization scheme, we need to take into account the anticommutation relations of Majorana fermions in different chains. Based on our earlier discussion, gauging the fermion parity maps $i\gamma'_f \gamma_{f}$ to a plaquette operator $\prod_{e\subset f} Z_e$, generalizing the 1D case where $\prod_{e\subset f} Z_e$ consists of only two $Z_e$ operators instead of four. Consider the following Hamiltonian 
\begin{align}
 H_{\text{decoupled}} &=-\sum_f t_1 \prod_{e\subset f} Z_e + t_2 X_{e^f_x} \nonumber \\
 & =-\sum_f t_1 (Z_{e^f_y} Z_{-e^f_y}) Z_{e^f_x} Z_{-e^f_x} + t_2 X_{e^f_x}, 
 \label{eq:decoupled}
\end{align}
where $X$ only acts on the vertical edges. Therefore, $Z_{\pm e^f_y}$ on the horizontal edges commutes with the Hamiltonian and can be simultaneously diagonalized. $Z_{e^f_y} Z_{-e^f_y}$ can take values in $\pm 1$. Consequently, $ H_{\text{decoupled}}$ can be interpreted as many decoupled TFIMs with non-uniform couplings. In each sector parametrized by these eigenvalues, we can make a unitary transformation site-by-site on the spins on the vertical edges $\pm e^f_x$ to change all the coefficients multiplying $Z_{e^f_x} Z_{-e^f_x}$ to 1. Thus, even though the spectrum of $H_{\text{decoupled}}$ is infinitely degenerate, it is the same as that of a sum of decoupled TFIMs with uniform couplings in each sector. As a result, $H_{\text{decoupled}}$  still has a KW duality induced by that of decoupled TFIMs: $H_{\text{decoupled}}(t_1, t_2) \leftrightarrow H_{\text{decoupled}}(t_2, t_1)$. 

The Hamiltonian $H_{\text{decoupled}}$ above looks very artificial because there is no transverse coupling and the ground state is highly degenerate. To generalize the KW duality to more interesting spin systems, transverse couplings should be added and these spin systems need to be gauged.

\subsection{Free Majorana fermions on square lattice}
\begin{figure}
    \centering \includegraphics[width=0.8\linewidth]{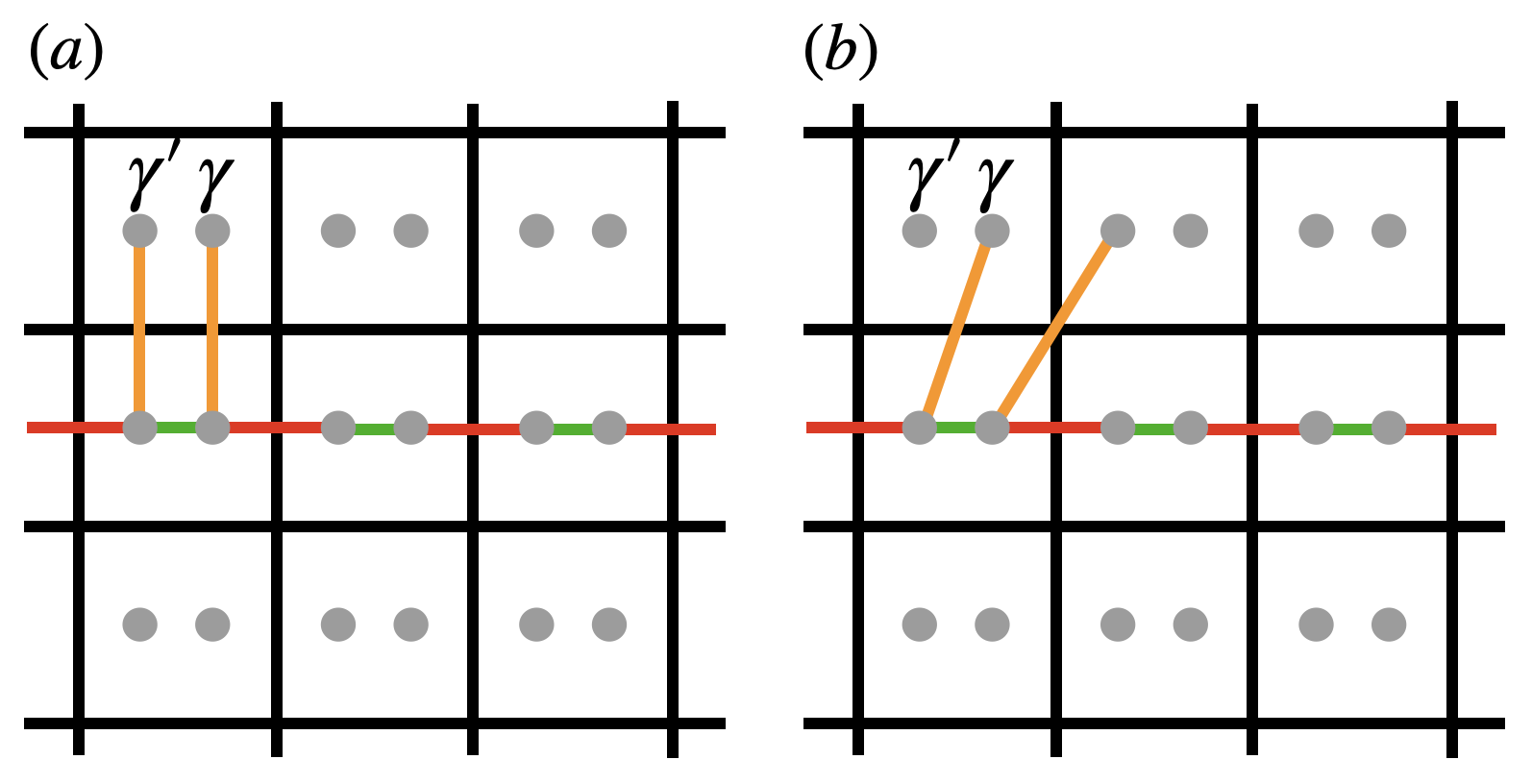}
    \caption{Free Majorana fermions on the square lattice with couplings $t_1$ (green) and    $t_2$ (red), and different transverse couplings $t'$ (orange).}
    \label{fig_lattice}
\end{figure}

In Sec.~\ref{sec:disentanling} and Sec.~\ref{sec:bos_square}, we have discussed the bosonization of a general fermionic system on a 2D square lattice. Let us now consider a concrete translationally invariant Hamiltonian motivated by the discussion above.

We place a pair of Majorana fermions $\gamma$ and $\gamma'$ on each face, and rewrite the 1D free Majorana Hamiltonian in Eq.~(\ref{eq:H_Maj1d}) as
\be 
H_{\text{Maj}} = -i\sum_{f} (t_1 \gamma'_{f} \gamma_{f} +t_2  \gamma_{f} \gamma'_{f+a_x} ),
\label{eq:H_Maj2d}
\ee 
but allow $f$ to span the entire 2D plane. $a_x$ and $a_y$ denote horizontal and vertical translations by one unit cell to the right and top, respectively. Next, we add couplings in the transverse direction of the chains. One concrete example is  
\be 
\begin{split}
 H =& -i\sum_{f} (t_1 \gamma'_{f} \gamma_{f} +t_2 \gamma_{f} \gamma'_{f+a_x}) \\
 & + (t'_1 \gamma_{f} \gamma_{f+a_y} +t'_2 \gamma'_{f} \gamma'_{f+a_y}).   
\end{split}
\label{eq:H_Maj2d2}
\ee   
This is a free Majorana system on the square lattice. See Fig.~\ref{fig_lattice}(a). For simplicity, we set $t' = t'_1 =t'_2$. The Hamiltonian is exactly solvable and the dispersion relation is given by
\be 
\varepsilon(k_x, k_y) = \pm \sqrt{t_1^2 +t_2^2 -2 t_1 t_2 \cos 2 k_x}  + 2 t' \sin k_y.
\label{eq:spectrum1}
\ee 
The first term corresponds to the dispersion relation of the 1D Majorana chain, and  the second term is due to couplings along the transverse direction to the neighboring chains. The dispersion relation remains invariant under $t_1 \leftrightarrow t_2$, although this does not necessarily indicate a phase transition. If $t_1 =t_2$, then the Hamiltonian in Eq.~(\ref{eq:H_Maj2d2}) is invariant under the horizontal minimal translation $T^x_{1/2}$, but not under the vertical minimal translation $T^y_{1/2}$. Instead, it is invariant under the one-unit-cell vertical translation $T^y_1$. We can reshape the unit cells and consider Hamiltonians invariant under $T^y_{1/2}$ but then the Hamiltonian is not invariant under $T^x_{1/2}$, which can also be viewed as the exchange $\gamma_f \leftrightarrow \gamma'_f$ followed by a one-unit-cell translation of fermions of one type:  $\gamma_f$ or $\gamma'_f$. Note that we can also choose other kinds of transverse couplings while preserving the invariance under $T_{1/2}^x$ at $t_1 =t_2$. For example,  consider transverse couplings shown in Fig.~\ref{fig_lattice}(b). In this case, the system can be regarded as complex fermions on a triangular lattice deformed from the square lattice by reshaping the unit cells.  

Let us now consider the bosonization of the Hamiltonian in Eq.~(\ref{eq:H_Maj2d2}) into a gauged spin system as discussed in Sec.~\ref{sec:bos_square}. The spin model is given by
\be 
H_b = - \sum_{f} (t_1 W_{f} +t_2 V_{e^f_x} -i t'  W_{f} V_{e^f_y} -i t'  V_{e^f_y} W_{f+a_y} ), 
\label{eq:H_bos1}
\ee
with the Gauss law
\be 
G_{v} =\left( \prod_{e\subset v} X_{e}\right) W_{f_v} =1,
\ee 
which is shown in Fig.~\ref{fig_bos}(e). Recall that $W_{f_v}$ is the plaquette operator on the face $f_v$ sitting on the northeast corner of $v$. On an infinite plane, these are the only basic constraints, from which all others can be generated. For example, consider the minimal translation $T_x^{1/2}$ of all Majorana fermions, the loop of Majorana fermions in  Fig.~\ref{fig_bos}(e)  becomes that in Fig.~\ref{fig_bos}(f). One can verify their equivalence using the identity in Eq.~(\ref{eq:gamma23}). Correspondingly, the Gauss law is invariant under the translation $T_x^{1/2}$. The gauged spin model is then also invariant under this (anisotropic) translation. Therefore, we can consider the corresponding KW duality in the spin model.

\begin{figure}[tb]
    \centering \includegraphics[width=0.9\linewidth]{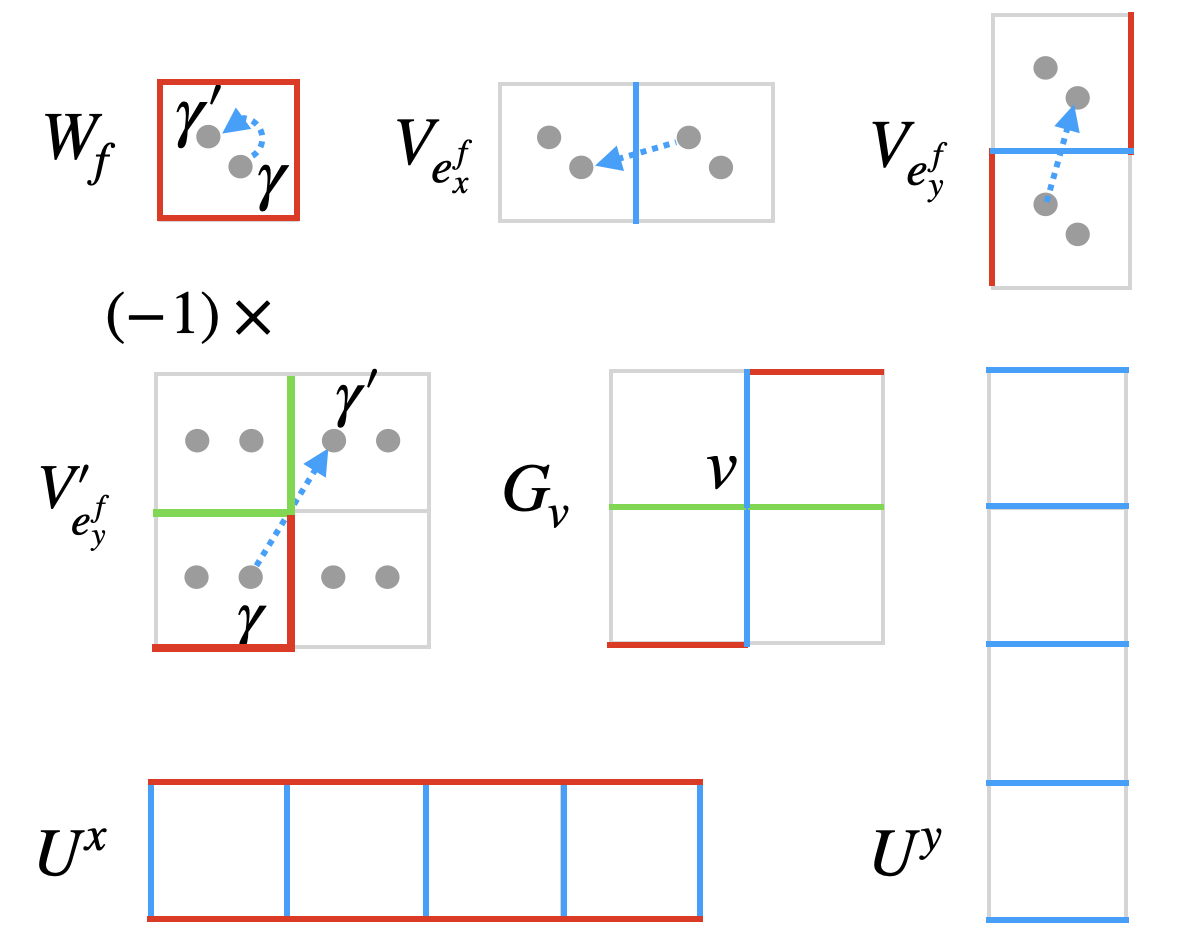}
    \caption{Bosonization on the square lattice in a different representation. Each Majorana bilinear is shown together with the corresponding spin operator. As in Fig.~\ref{fig_bos}, dashed blue arrows indicate the ordering of the two Majorana fermions in the bilinears. On a periodic lattice with the topology of a torus, there are nontrivial 1-form symmetries around the holes generated by $U^x$ and $U^y$.}
    \label{fig_KW}
\end{figure}

In fact, it is easy to see that the KW duality acts as 
\be 
W_f \to V_{e^f_x} \to  W_{f+a_x}
\ee 
and 
\be 
     W_{f+a_y} V_{e_y^f} \to   V_{e^f_y}W_f \to    W_{f+a_x+a_y} V_{e^{f+a_x}_y}.
\ee
The latter can also be represented equivalently as 
\be 
    V_{e^f_y} \to  V'_{e^f_y}      \to V_{e^{f+a_x}_y},
\ee
where $V'_{e^f_y} = V_{e^{f+a_y}_x}  V_{e^f_y} W_f$.
In Fig.~\ref{fig_bos}, this relates the spin operator in (c) to that in (d). 
The KW duality admits multiple equivalent representations due to the existence of many unitarily equivalent bosonization schemes. As an example, for each face $f$, we can apply a controlled-$Z$ transformation on the edge $e^f_x$ and the edge $e^f_{-y}$, then 
\be 
\begin{split}
 i\gamma'_f \gamma_f \to W_f, & \quad i\gamma_f \gamma'_{f+a_x} \to  V_{e^f_x} =X_{e^f_x}, \\
 i \gamma_{f+a_y}\gamma'_f &\to V_{e^f_y}= Z_{e^f_{-x}}X_{e^f_y} Z_{e^{f+a_y}_{x}},\\
 i \gamma'_{f+a_x+a_y}\gamma_f &\to  V'_{e^f_y}  = - Z_{e^f_{-y}}Z_{e^f_x}Y_{e^f_{y}}Y_{e^{f+a_y}_{y}}.
\end{split}
\label{eq:rep1}
\ee  
These maps are shown in Fig.~\ref{fig_KW}. Note that compared to Fig.~\ref{fig_bos}, we have condensed the bosonization map for each Majorana bilinear by placing together the bilinear and the corresponding spin operator. 
Along the $x$-direction, the spin operators $W_f$ and $X_{e^f_x}$ are the same as in Eq.~(\ref{eq:decoupled}). The KW duality transformations above can therefore be viewed as a direct generalizations of the 1D counterpart, as discussed around Eq.~(\ref{eq:decoupled}). Since $W_f$ measures $\bbz_2$ fluxes and $X_{e^f_x}$ measures the spin polarizations along vertical edges, the 2D duality generalizes the 1D duality between domain walls and spin polarizations to a duality between gauge fluxes and gauge spin polarizations, albeit only in one direction.  

Note that the Gauss-law constraint is crucial in generalizing the duality. Intuitively, we know that the KW duality comes from the different ways of choosing unit cells of two Majorana fermions. Gauging the fermion parity does not preserve this information if the gauge field is not flat. Therefore, the KW duality is also lost in the spin system without Gauss-law constraints, which is unitarily equivalent to a parity-gauged fermion system without the flatness condition.  

Algebraically, bosonization maps the generating set of the parity-even fermion algebra to the generating set, including $V_{e^f_x}$, $V_{e^f_y}$, and $W_f$, of the operator algebra with the gauge constraint $G_v =1$. On an infinite square lattice, this is an operator bijection (see Appendix~\ref{app:embedding}).
The KW duality between spin operators, which is isomorphic to the translation operation $T_x^{1/2}$ on the Majorana bilinears, is an operator algebra isomorphism among generators. These generators also form a bond algebra whose isomorphisms generally produce dualities \cite{cobanera2011bond}. See Appendix~\ref{app:bondalgebra} for a brief review. If the space manifold $M$ contains loops/cycles, then the bosonization map is only an isomorphism up to boundary conditions and charge sectors of 1-form symmetries. Nevertheless, we can still define the KW duality in the spin system without referring back to the fermionic system as in the 1D case.  

\subsection{KW duality as a noninvertible symmetry operator}
\label{sec:noninv}
As a direct generalization of the 1D case, we can write down an explicit expression of the KW duality transformation by bosonizing the minimal translations of the Majorana lattices. We loosely follow the procedure in Ref.~\cite{seiberg2024non}. 

Assuming an infinite square lattice, we can express the minimal Majorana translation along the (positive) $x$-direction up to a phase: 
\be 
T^{1/2}_x \sim  {\prod_f}' \frac{1-\gamma'_f \gamma_f}{\sqrt{2}} \frac{1-\gamma_f \gamma'_{f+a_x}}{\sqrt{2}}.
\ee 
Here, faces $f$ on the lattice are ordered from left to right and from bottom to top, and a larger (smaller) factor is placed to the right (left) in the product ${\prod}'$. It is easy to check that $T^{1/2}_x$ translates $\gamma'_f$ to $\gamma_f$ and $\gamma_f$ to $\gamma'_{f+a_x}$. By the bosonization method in the representation shown in Eq.~(\ref{eq:rep1}), the minimal Majorana translation is mapped to 
\be 
T^b_x \sim \sideset{}{'}\prod_f \frac{1+i W_f}{\sqrt{2}} \frac{1+ iX_{e^f_x}}{\sqrt{2}}.
\ee 
It maps $W_f \to X_{e^f_x} \to W_{f+a_x}$ and $ V_{e^f_y} \to  V'_{e^f_y}      \to V_{e^{f+a_x}_y}$. As in the 1D case, we need to account for the different charge and boundary sectors of the Hilbert space. Recall that $f = (m, n)$, then for $m$th row of the square lattice,
\be
\prod_n X_{e^{(m, n)}_x} \to \prod_n W_{(m, n)}.
\ee 
Here the products are taken over all squares on the row. The term on the left acts on the vertical edges, and the term on the right is a product of $Z$'s on all the horizontal edges of the row. The product of the two operators is exactly the $U^x$ operator shown in Fig.~\ref{fig_KW} that generates the $\bbz_2$ 1-form symmetry around a cycle along the $x$ direction if the lattice is periodic. For completeness, the generator $U^y$ operator of the $\bbz_2$ 1-form symmetry around a cycle along the $y$ direction is also shown. Let us denote $U^x$ on the $m$th row by $U^x_m$, then 
\be 
U^x_m =\prod_n X_{e^{(m, n)}_x}  \prod_n W_{(m, n)}.
\ee 
Multiplying $U^x_m$  by the product of the Gauss-law operators $G_v$ around this cycle can shift $U^x_m$ up or down by one row. Therefore, any gauge-invariant common eigenstate of $U^x_m$,  where $m$ runs over different rows, must have the same eigenvalues, 1 or $-1$, for all $U^x_m$. In addition,
\be 
\prod_f X_{e^f_x}= \prod_m U_m^x.
\ee

On a periodic lattice where we identify $L+1 \sim 1$, the minimal Majorana translation $T^{1/2}_x$ is slightly modified and one version of the bosonized operator is given by
\be
\begin{split}
T^{b}_x \sim & \prod_m \frac{1+ iX_{e^{(m, 1)}_{-x}}}{\sqrt{2}}  \frac{1+i W_{(m, 1)}}{\sqrt{2}} \frac{1+ iX_{e^{(m, 1)}_x}}{\sqrt{2}} \cdots \\
& \frac{1+i W_{(m, L)}}{\sqrt{2}} \frac{1+ iX_{e^{(m, L)}_x}}{\sqrt{2}}.  
\end{split}
\ee 
One can check explicitly that $W_f \to X_{e^f_x} \to W_{f+a_x}$ and $ V_{e^f_y} \to  V'_{e^f_y}      \to V_{e^{f+a_x}_y}$ are satisfied for all $f$ if $U^x_m =1$ in the projected sector of the Hilbert space. Thus, the KW duality operator $D_{KW}$ (up to a phase) can be written as   
\be
D_{KW} \sim  T^{b}_x \left(\prod_m \frac{U^x_m+1}{2} \right) .
\label{eq:noninv_KW}
\ee 
Under the duality transformation $D_{KW}$,   
\be
\prod_f X_{e^f_x} \to \prod_f W_{f+a_x}=1,
\ee 
and
\be 
\prod_f V_{e^f_y} =\prod_f X_{e^f_y} \to \prod_f V'_{e^f_y} =\prod_f X_{e^f_x} X_{e^f_y}.
\ee 
These transformations are compatible with the global condition:
\be
\prod_f X_{e^f_x} =1
\ee
in the subspace where the charge of the 1-form symmetry $U^x_m$ is 1.  $D_{KW}$ in Eq.~(\ref{eq:noninv_KW}) is obviously noninvertible, generalizing the 1D case.

\subsection{Two copies of free Majorana lattices: Bosonization and KW dualities}
\label{sec:twocopies}
In 1D, given a free Majorana Hamiltonian in Eq.~(\ref{eq:H_Maj1d}), there can be local conserved charge operators commuting with both sums $H_{\text{Maj}, 1}= -i\sum_j \gamma_{2j-1}\gamma_{2j}$ and $H_{\text{Maj}, 2}= -i\sum_j \gamma_{2j}\gamma_{2j+1}$. One such operator is $Q = -i\sum ( \gamma_{2j-1}\gamma_{2j+1}+\gamma_{2j}\gamma_{2j+2})$, which can be viewed as two decoupled copies of the free Majorana Hamiltonian, and it describes free (staggered) Dirac fermions when written in terms of complex fermions. If we view $Q$ as the Hamiltonian, then $H_{\text{Maj}, 1}$ and $H_{\text{Maj}, 2}$, related by a half-unit-cell translation, can be viewed as conserved charge operators. In fact, they can be interpreted as the vector charge and the axial charge of the Dirac fermionic system \cite{chatterjee2025quantized}. After bosonization, $H_{\text{Maj}, 1}$ and $H_{\text{Maj}, 2}$ are mapped to the Ising coupling terms and the transverse field in the TFIM, while $Q$ is mapped to the $XY -YX$ model or its cousins (up to boundary conditions)  which again can be viewed as a local conserved charge of the TFIM \cite{su2025z2}. These models have (self-)dualities derived from the KW duality of the TFIM. 

We now generalize the picture to 2D. We consider two decoupled copies of the free Majorana fermions on different lattices. As we will see in this section, it turns out that, similar to the 1D case, the system possesses two local conserved charges forming the Hamiltonian in Eq.~(\ref{eq:H_Maj2d}). The (self-)dualities  of the systems are also derived from the KW duality we discussed in the last section. Moreover, in the previous discussion, the KW duality is induced by the minimal translation along one direction in the fermionic system. For two decoupled copies of the free Majorana fermions, minimal translations in both directions lead to KW dualities in the gauged spin system in both directions.

\begin{figure}[tb]
    \centering \includegraphics[width=0.96\linewidth]{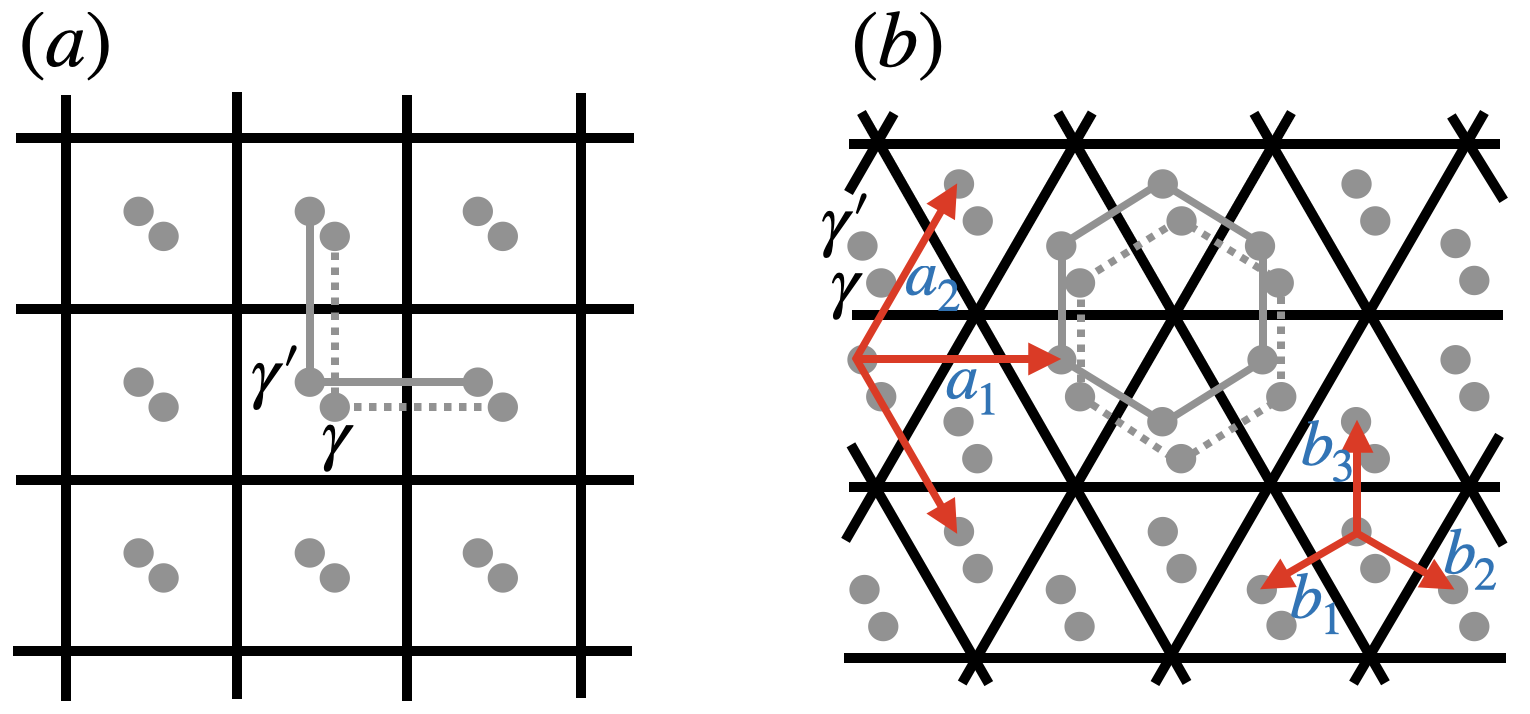}
    \caption{Couplings of two copies of free Majorana fermions on the square lattice (a) and the honeycomb lattice (b). Each face of the dual square lattice and the dual hexagon lattice contains two Majorana fermions, $\gamma$, $\gamma'$.  There are two types of fermionic bilinears: $i\gamma_f\gamma_{f'}$ (solid gray) and $i\gamma'_f\gamma'_{f'}$ (dashed gray) between nearest-neighboring faces $f$ and $f'$. In (b), three primitive vectors of the lattice $a_j$, $j =1, 2, 3$, are shown. Also shown are the three vectors $b_j$, $j =1, 2,3$, connecting a lattice point on a sublattice to its three nearest neighbors on the complementary sublattice. }
    \label{fig_lattice2}
\end{figure}

\subsubsection{Square lattice}
We first consider the case on the square lattice. Compared to Eq.~(\ref{eq:H_Maj2d}), we modify the bilinear couplings in the horizontal direction by changing nearest-neighbor couplings (as shown in Fig.~\ref{fig_lattice}) to next-nearest-neighbor couplings. To illustrate it, we redraw the schematic in Fig.~\ref{fig_lattice2}(a). Two Majorana fermions on each face are slightly displaced for a better view. Now there are no intracell couplings but only intercell couplings: $i\gamma_f\gamma_{f'}$ and $i\gamma'_f\gamma'_{f'}$ for nearest-neighboring faces $f$ and $f'$. The Hamiltonian is given by
\begin{align}
  H =&  - i\sum_{f}( t_1 \gamma_f \gamma_{f+a_x} + t_2 \gamma'_f \gamma'_{f+a_x}   \nonumber\\
  & \quad \quad +t'_1 \gamma_f \gamma_{f+a_y} + t'_2 \gamma'_f \gamma'_{f+a_y}). 
  \label{eq:square2copies}
\end{align}
Here $a_x$ and $a_y$ represent two primitive translation vectors. 
As expected, the Hamiltonian can be viewed as two decoupled copies of free Majorana lattices.  The spectrum consists of two copies of that in Eq.~(\ref{eq:spectrum1}): 
\be 
\varepsilon_{1, 2}(k_x, k_y) = \pm 2t_{1, 2} \sin k_x + 2 t'_{1,2} \sin k_y. 
\ee 
When $t = t_1 = t_2 = t'_1 = t'_2$, the Hamiltonian can be expressed as complex fermion hoppings
\begin{align}
 H &= 2t \sum_{\langle f, f'\rangle} c_f^{\dagger} c_{f'} +h.c. = - i t \sum_{\langle f,f'\rangle} \gamma_f \gamma_{f'} +\gamma'_f \gamma'_{f'} 
 \label{eq:H_comp}  \\
 &=-it \sum_{f}\gamma_f \gamma_{f+a_x} +\gamma_f \gamma_{f+a_y} +\gamma'_f \gamma'_{f+a_x}+\gamma'_f \gamma'_{f+a_y}, \nonumber
\end{align}  
where $\langle f,f'\rangle$ represent different pairs of nearest neighbors. Here, we have made use of the fact that the square lattice is bipartite and consists of two sublattices, and expressed complex fermions in terms of Majorana fermions on one of the sublattices:
\be 
c_f =\frac{1}{2} (\gamma_f - i\gamma'_f)
\ee 
and performed a phase rotation such that the new creation operator becomes $c_f \to -i c_f$ on the other sublattice. 

It is clear that the Hamiltonian in Eq.~(\ref{eq:H_comp}) has $U(1)$ symmetries. One is the conservation of complex fermion number,  generated by 
\be 
Q = i \sum_f   \gamma'_f \gamma_f.
\ee 
Other $U(1)$ symmetry generators are obtained by performing $T_{x/y}^{1/2}$: $\gamma \leftrightarrow \gamma'$ and a one-unit-cell translation for $\gamma'$ in the $x$-direction or the $y$-direction (which we still call a minimal translation for simplicity). They are 
\be
Q_{x/y} = i \sum_f  \gamma_f \gamma'_{f+a_{x/y}}.
\ee  
The $U(1)$ charges $Q$ and $Q_{x/y}$ can be viewed as direct generalizations of the vector and axial charges in the 1D case \cite{chatterjee2025quantized}. They are directly related to the terms in Eq.~(\ref{eq:H_Maj2d}). Note that they do not commute on the lattice level and form a nontrivial generalization of the Onsager algebra. 

\begin{figure}[tb]
    \centering \includegraphics[width=1.0\linewidth]{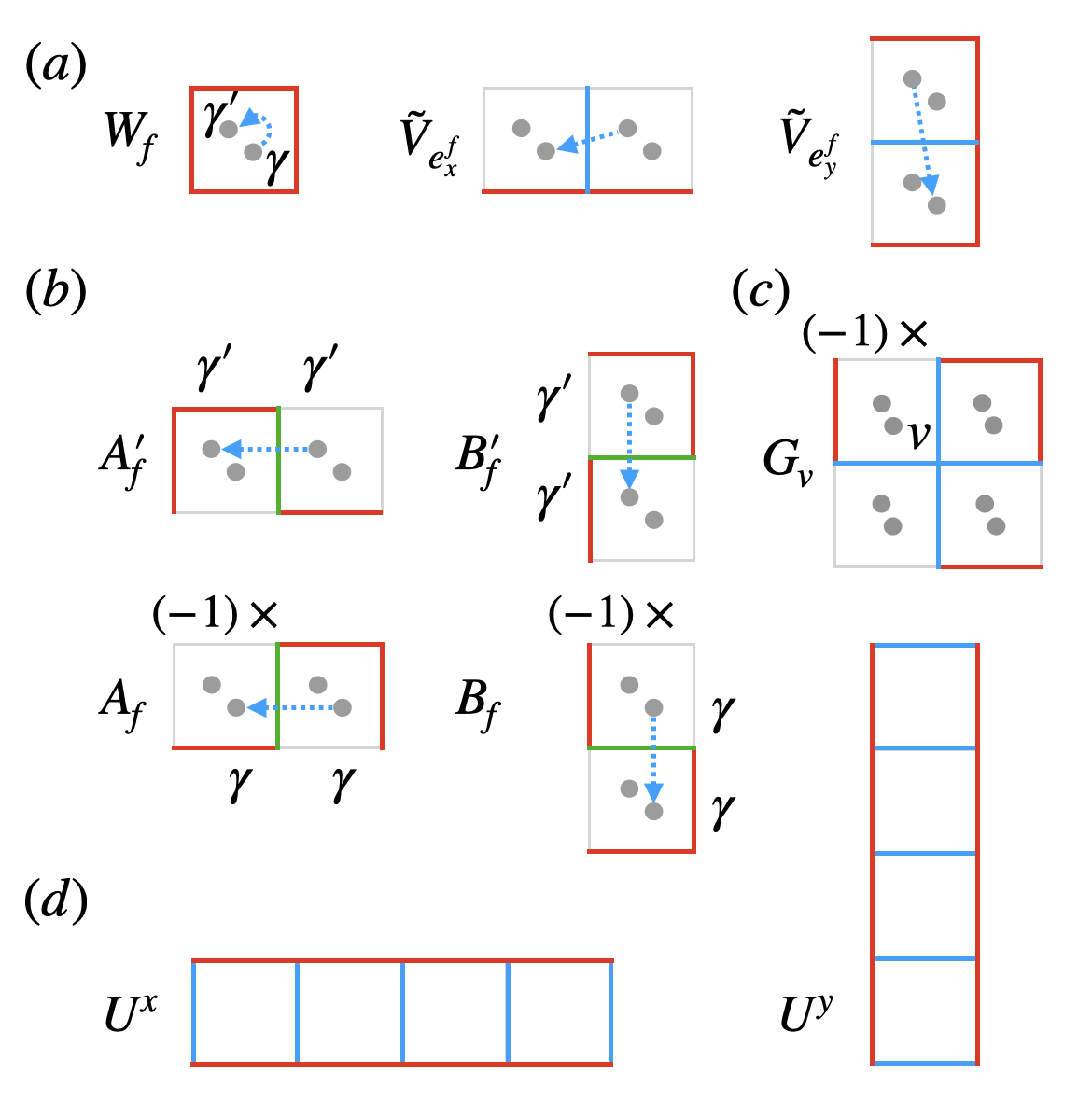}
    \caption{Bosonization for two copies of free Majorana fermions on the square lattice. Fermions bilinears in the charges ($Q$, $Q_x$, and $Q_y$) are shown in (a). Fermions bilinears in the Hamiltonian in Eq.~(\ref{eq:square2copies}) are shown in (b). The Gauss-law operator $G_v$ is shown in (c). Two 1-form symmetry generators $U_x$ and $U_y$ are shown in (d). Blue dashed arrows specify the ordering within fermion bilinears.  Corresponding spin operators are the products of Pauli operators on the colored edges: $Z$ (red), $X$ (blue), and $Y$ (green). If there is a extra minus sign, then it is placed at the upper left corner. }
    \label{fig_bos3}
\end{figure}

Now we apply bosonization to the system. The terms in the charges, $Q$, $Q_{x/y}$ are shown in Fig.~\ref{fig_bos3}(a), and the terms in the Hamiltonian are shown in Fig.~\ref{fig_bos3}(b). The Gauss-law operator $G_v$ in this representation is shown in Fig.~\ref{fig_bos3}(c).  Note that for each face $f$, we have applied $CZ$ transformations acting on $e^f_{-x}$ and $e^f_{-y}$ to the bosonized operators in Fig.~\ref{fig_bos}. This transformation renders the terms in the Hamiltonian more symmetrical in this representation. The Hamiltonian is now 
\be 
H_b =  - i\sum_{f}( t_1 A_{f} + t_2 A'_f   +t'_1 B_f + t'_2 B'_f). 
\ee 
Charge operators are  
\be 
Q_b = \sum_f W_f, \quad Q_{b, x} = \sum_f \tilde{V}_{e_x^f}, \quad Q_{b, y} = \sum_f \tilde{V}_{e_y^f}.
\ee

Now the KW duality transformation of $H_b$ is apparent. The Hamiltonian in Eq.~(\ref{eq:square2copies}) $H(t_1, t_2, t'_1, t'_2)$ becomes $H(t_2, t_1, t'_2, t'_1)$ under the transformation $T_x^{1/2}$ and $T_y^{1/2}$. Thus, there are also two types of KW duality transformations. In the horizontal direction, the duality $KW_x$ maps
\be 
A'_f \to A_f \to A'_{f+a_x},\quad B'_f \to B_f \to B'_{f+a_x}.
\ee 
In the vertical direction, the duality $KW_y$ maps
\be 
A'_f \to A_f \to A'_{f+a_y},\quad B'_f \to B_f \to B'_{f+a_y}.
\ee 
The Hamiltonian $H(t_1, t_2, t'_1, t'_2)$ is self-dual at $t_1 =t_2$ and $t'_1 =t'_2$. For the charges, $KW_x$ maps 
\be 
Q_b \to Q_{b, x}\to Q_b
\ee 
and $KW_y$ maps 
\be 
Q_b \to Q_{b, y}\to Q_b.
\ee
These are essentially the KW duality for one copy of free Majorana fermions discussed in the last section where these charges are terms in the Hamiltonian. Once again, the dualities hold as long as the Gauss-law constraints $G_v=1$ are imposed for all $v$. The 1-form symmetry generators along two different cycles on a periodic lattice $U_x$ and $U_y$ are shown in Fig.~\ref{fig_bos3}(d). In line with our expectation, the two generators are manifestly more symmetric in this representation. When we express $KW_x$ and $KW_y$ as noninvertible symmetry operators as in Eq.~(\ref{eq:noninv_KW}), $U_x$ and $U_y$ are used in the respective projection operators.     

As a side note, we remark that in Eq.~(\ref{eq:H_comp}), the couplings are uniform. Choosing nonuniform couplings along one direction can still preserve one of the minimal translations. For instance,  consider the Hamiltonian: $H = 2t\sum_f[ c^{\dagger}_f c_{f+a_x} +(-1)^mc^{\dagger}_f c_{f+a_y} +h.c.]$ where $f =(m, n)$. It describes free complex fermions on the square lattice with a transverse in-plane magnetic field. 
The dispersion is $
\varepsilon(k_x, k_y) = \pm 4t \sqrt{\cos^2 k_x + \cos^2 k_y} $ with $k_x \in [0, \pi)$ and $k_y \in [0, 2\pi)$, and the spectrum has two Dirac cones. It is invariant under $T^{1/2}_y$, and the corresponding spin model is self-dual under $KW_y$. This self-dual point can be viewed as a critical point.

\subsubsection{Honeycomb lattice}
\label{sec:honeycomb}
We can also consider the case on the honeycomb lattice as in Fig.~\ref{fig_lattice2}(b). In this case, the dual lattice is a triangular/hexagonal lattice. Each triangular face contains two Majorana fermions, and a unit cell contains two triangles. It is easy to see that the honeycomb lattice with nearest-neighbor couplings can be viewed as a deformed square lattice with period-2 zigzag couplings in the horizontal or vertical direction. The honeycomb lattice is bipartite with two sublattices $\Lambda =\Lambda_A \sqcup \Lambda_B$.  Since the analysis is similar to the square lattice case, we focus on the Hamiltonian that can be expressed in terms of free complex fermions. 
\begin{align}
 H &= 2t \sum_{\langle f, f'\rangle} c_f^{\dagger} c_{f'} +h.c. = - i t \sum_{\langle f,f'\rangle} \gamma_f \gamma_{f'} +\gamma'_f \gamma'_{f'} \label{eq:H_hc}  \\
 &=-it \sum_{f \in \Lambda_A, b_j }\gamma_f \gamma_{f+b_j} +\gamma'_f \gamma'_{f+b_j}  ,   \nonumber
\end{align} 
where $b_j$, $j=1, 2, 3$, denotes the three vectors connecting a lattice point on sublattice $\Lambda_A$ to its three nearest neighbors on the complementary sublattice $\Lambda_B$. See Fig.~\ref{fig_lattice2}(b). This case is interesting because the free (complex) fermionic system with nearest neighbor hoppings yields a dispersion relation for two Dirac fermions as in graphene.

\begin{figure}[tb]
    \centering \includegraphics[width=1.0\linewidth]{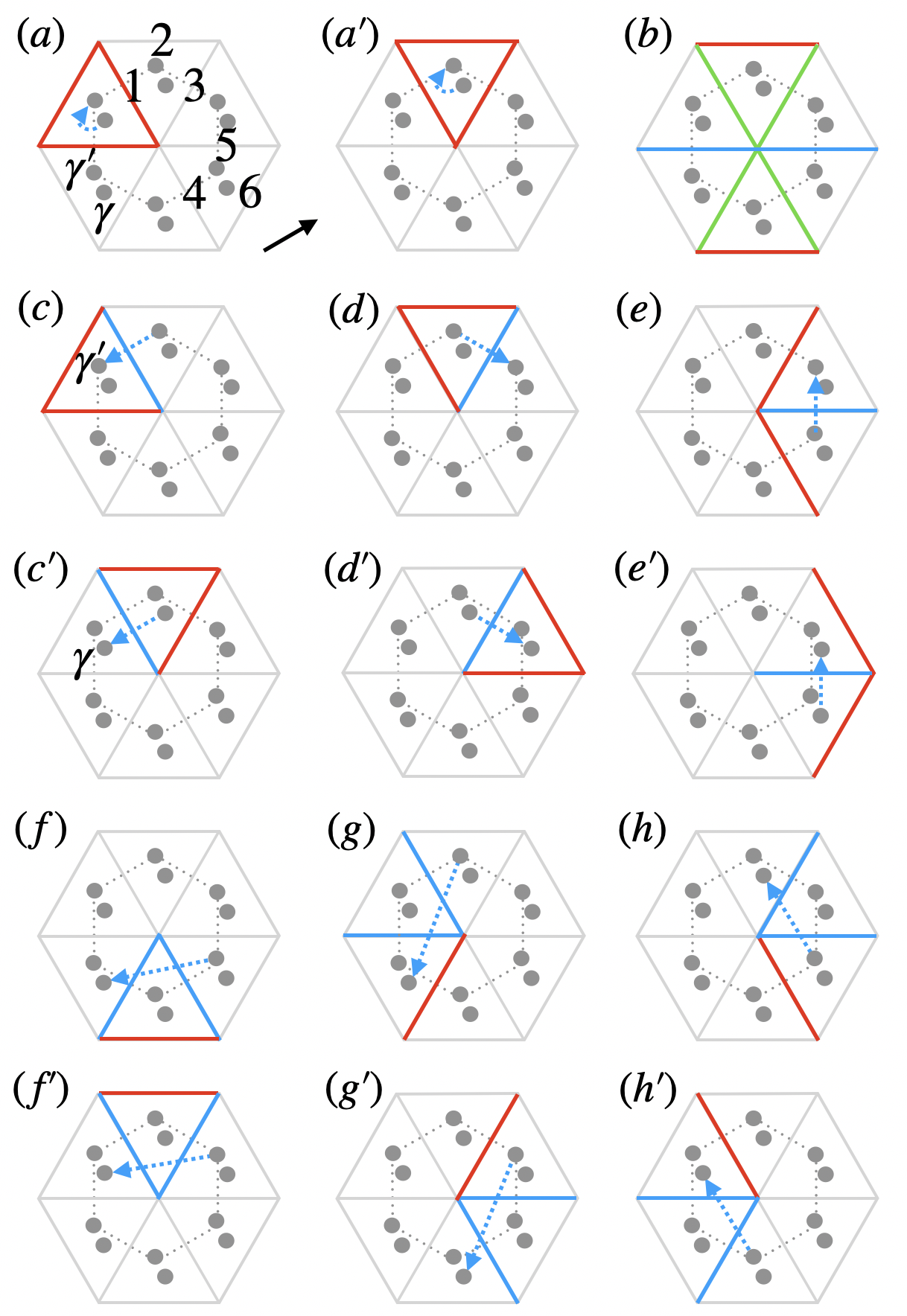}
    \caption{Bosonization for two copies of free Majorana fermions on the honeycomb lattice. Fermion bilinears are indicated by blue dashes with order-specifying orientations. Plaquette operators are shown in (a) and (a$'$). The global edge ordering is indicated by the numbers in (a) with a black arrow showing the increasing direction once edges along the diagonal direction have been ordered. The Gauss-law operator is shown in (b). Fermions bilinears in the Hamiltonian in Eq.~(\ref{eq:H_hc}) are shown in (c, d, e) and (c$'$, d$'$, e$'$).  Fermions bilinears in the charges ($Q$ and $Q^+_j$, $j=1, 2, 3$) are shown in (f-h) and (f$'$-h$'$). Corresponding spin operators are the products of Pauli operators on the colored edges: $Z$ (red), $X$ (blue), and $Y$ (green).}
    \label{fig_bos_honeycomb}
\end{figure}

To use the general formula Eq.~(\ref{eq:general2}), we order the edges as in Fig.~\ref{fig_bos_honeycomb}(a) and assign all three edges of a face to $\gamma'$ (instead of $\gamma$ as in Sec.~\ref{sec:BK}). The black arrow in Fig.~\ref{fig_bos_honeycomb}(a) represents the order-increasing direction once one diagonal ordering is finished. The plaquette operators for intracell bilinears are shown in (a) and (a$'$). The spin operators for intercell fermion bilinears appearing in Eq.~(\ref{eq:H_hc}) can be written down explicitly following Eq.~(\ref{eq:general2}), and are shown in (c-e) and (c$'$-e$'$). The Gauss-law operator $G_v$ for each vertex is shown in (b). The operator is a product of a vertex operator $\prod_{e \supset v} X_e$ associated with $v$ and two plaquette operators $\prod_{e \subset f} Z_e$ associated with the upper and lower triangles in the middle. 

The KW dualities are obtained from minimal translations $T^{\pm  1/2}_{j}$: $\gamma \leftrightarrow \gamma'$ followed by one-unit-cell translation of  all $\gamma'$ in the direction of the primitive vector $\pm a_j$ $(j =1,2,3)$. $a_1$, $a_2$, and $a_3 =a_2-a_1$ are shown in Fig.~\ref{fig_lattice2}(b).  In particular, under the KW dualities, the spin operators in (c-e) are mapped to those in (c$'$-e$'$), respectively; the spin operators in (c$'$-e$'$) are, respectively, mapped to those in (c-e) up to a primitive lattice translation associated with the KW duality. 

The Hamiltonian Eq.~(\ref{eq:H_hc}) has two types of $U(1)$ symmetries as well: total onsite charge
\be 
Q = i \sum_{f \in \Lambda}   \gamma'_f \gamma_f
\ee  
and non-onsite charge
\be 
 Q^{\pm}_{j} = i \sum_{f \in \Lambda}    \gamma_f \gamma'_{f\pm a_j},
\ee  
where $j =1, 2, 3$. $Q^{\pm}_{j}$ is obtained from $Q$ by a minimal translation $T^{\pm 1/2}_{ j}$.
Since there are two sublattices, each charge operator contains two types of fermion bilinears. In particular, the fermion bilinears in $Q^+_{j}$ for $j=1, 2, 3$ are displayed with their orientations in (f-h) and (f$'$-h$'$) where the corresponding spin operators in the bosonized charge operators $Q^{\pm}_{b, j}$ are also shown. Similar to the case on the square lattice, $Q$ and $Q^{\pm}_{j}$ do not commute in general. They form a generalized nonabelian Onsager algebra, an aspect explored recently in Ref.~\cite{pace2025parity}. 

Since the low energy theory is described by $(2{+}1)$D Dirac fermions with an $SU(2)$ flavor symmetry, it has a parity anomaly for the $SU(2)$ symmetry and orientation-reversing symmetries, such as time reversal $T$ and space reflection $P$ \cite{redlich2984gauge}. That is, if we couple the system to the background $SU(2)$ flavor gauge field and preserve $T$ or $P$ of the theory, the theory is not gauge invariant under some large gauge transformations. It can be viewed as a boundary theory of a $(3{+}1)$D fermionic SPT phase protected by $SU(2)$ and an orientation-reversal symmetry, $T$ or $P$. It is argued in Ref.~\cite{pace2025parity} that the $U(1)$ symmetries generated by $Q$ and $Q^{\pm}_{j}$ flow to the $SU(2)$ symmetry in the continuum and the time reversal and space reflections on the lattice also flow to the orientation-reversing symmetries. Therefore, the parity anomaly of the Dirac theory can be traced back to the microscopic symmetries in the lattice model. The gauged spin system described above then can possess a bosonic counterpart of the ultraviolet-infrared anomaly matching with a bosonic analog of the parity anomaly in the IR. In the Dirac theory, fermion parity $(-1)^F$ is a $\bbz_2$ center of the $SU(2)$ and $T^2 =(-1)^F$. If we gauge the fermion parity of the Dirac theory, $SU(2)$ reduces to $SO(3)$ and $T^2 =1$, which together form a $O(3)$ group. The bosonic parity anomaly can be captured by $(3{+}1)$D bosonic topological phases protected by $O(3)$, which are classified by cohomology groups $H^4(O(3), U(1)_T)$ \cite{chen2013} or more generally by cobordism group $\Omega_{SO}^4(BO(3), U(1)_T)$ \cite{kapustin2014symmetry} with twisted coefficients. We leave the verification of the bosonic parity anomaly in the gauged spin model, as well as its ultraviolet-infrared anomaly matching, for future work.

\section{Generalizations to higher dimensions}
\label{sec:general}

\subsection{Bosonization}

The discussion of the 2D case in the previous sections can be generalized to higher dimensions in a straightforward manner. The big picture is already presented in Fig.~\ref{fig_schematic}. We first gauge the fermion parity $(-1)^F$ of a (parity-even) fermionic system. Next, we perform a unitary transformation to map the gauged fermionic system to a spin system. If we impose the flatness condition on the gauge field, then the gauged fermionic system is dual to the (ungauged) fermionic system. The flatness condition on the gauge field is mapped to a Gauss law in the spin system by the unitary transformation. Then we can view the spin system with the Gauss law as the bosonized system of the original (ungauged) fermionic system. The crucial step is to construct the unitary transformation that disentangles the (fermionic) matter from the gauge field, but it is very similar to the 2D case.

\begin{figure}[bt]
    \centering
\includegraphics[width=0.8\linewidth]{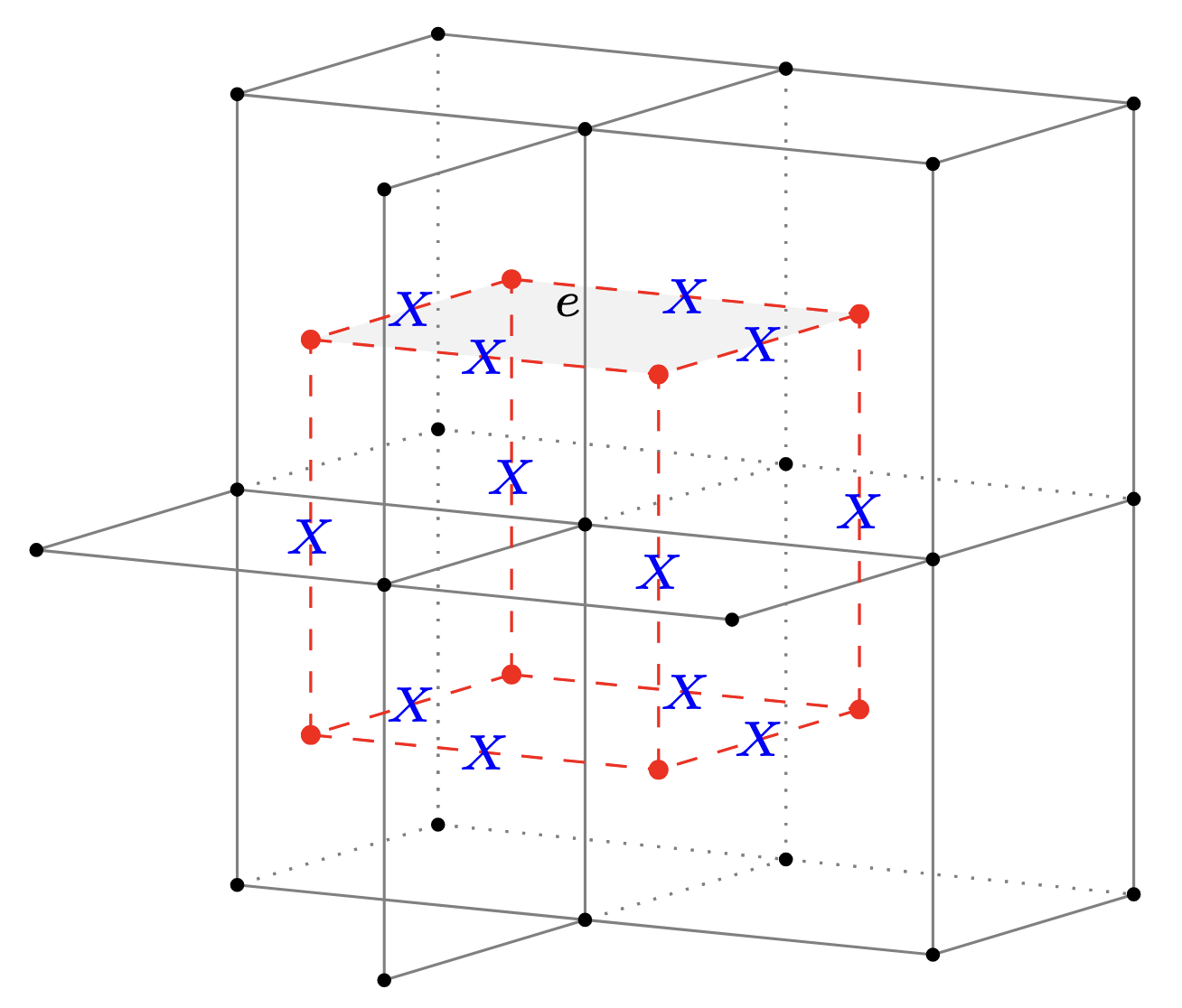}
    \caption{Schematic of a cubic lattice and its dual. The black vertices and black edges form the primal lattice. One complex fermion, or two Majorana fermions, is placed at the center of each cubic cell (red dots), which is a dual vertex of the dual lattice. One gauge spin is placed at the center of each face $f$, or equivalently, on the midpoint of a dual edge (dashed red). Four dual edges connecting four dual vertices form a loop encircling a dual face, as exemplified by the shaded area. This dual face intersects with an edge $e$ of the primal lattice. By duality, the four faces where the four gauge spins sit share the common edge $e$. }
    \label{fig_cubic}
\end{figure}

To construct such a disentangling unitary $U$, we first describe the gauging process in an $n$-dimensional space $M$. In principle, we can consider any polyhedral decomposition of $M$, generalizing the cellular decomposition in 2D where a general surface graph is embedded as a 1-skeleton, but we will illustrate the main ideas using a cubic decomposition in the 3D space. Generalizations to more general polyhedral decompositions in higher dimensions are very straightforward. A schematic of a cubic decomposition in 3D is shown in Fig.~\ref{fig_cubic}. The vertices and edges are drawn in black.  Now we place a complex fermion, or two Majorana fermions, at the center of each cube, colored in red. These fermions form the vertices of a dual cubic lattice whose edges connecting these vertices are drawn with dashed red lines. For each complex fermion, there are six nearby faces on the black primal lattice. Each face intersects with one edge on the dual lattice. At each intersection, we place one gauge spin, as exemplified by the Pauli $X$. To gauge fermion parity, we impose the following Gauss law 
\be 
G_c \equiv  (-1)^{F_c}  \prod_{f \subset c} Z_{f}=1.
\ee 
Here, the subscript $c$ exhausts all minimal cubes, and $f$ is one of the six faces of the cube. The local fermionic parity $(-1)^{F_c}$ can again be expressed as a Majorana bilinear $i\gamma_{v_1(c)} \gamma_{v_2(c)}$. As in the 2D case, the ordering with the bilinear implicitly assumes an orientation between these two Majorana fermions $\gamma_{v_1(c)} $ and $\gamma_{v_2(c)}$. The parity-even fermion algebra is generalized by these local fermion parity operators together with other Majorana bilinears connecting a Majorana fermion in a cube to another Majorana fermion in its neighboring cubes, which is represented by the dashed red lines in Fig.~(\ref{fig_cubic}). Each of these latter bilinears should be dressed with an extra $X_f$ at each intersecting point to preserve the Gauss law invariance as in any minimal coupling scheme. 

Next, we carry out the assignment procedure. For each cube $c$, we partition the set of its six faces $F(c)$ into two subsets and assign them to the two Majorana fermions $\gamma_{v_1(c)} $ and $\gamma_{v_2(c)}$ at the center such that $F(v_1(c)) \sqcup F(v_2(c)) = F(c)$. Then each face $f$ has two Majorana fermions in neighboring cubes $c_1(f)$ and $c_2(f)$ assigned to it. The set of edges assigned to these two Majorana fermions $\gamma_{c_1(f)}$ and $\gamma_{c_2(f)}$ are $E(f) \equiv E(c_1(f)) \cup E(c_2(f))$. If there is an orientation between $\gamma_{c_1(f)}$ and $\gamma_{c_2(f)}$, there is a unique bilinear associated with the face $f$: 
$S_f =i\gamma_{c_1(f)}\gamma_{c_2(f)}$. The corresponding minimally coupled bilinear is  $S_f X_f=i\gamma_{c_1(f)}\gamma_{c_2(f)} X_f$. 
Define a global ordering of the faces and $P_f^{\pm} = (1\pm Z_f)/2$, then the disentangling unitary is given by
\be 
U= \stackrel{\leftarrow}{\prod}_{f}[P_f^+ +P_f^- S_f].
\label{eq:general_U3d}
\ee 
Here, as in the 2D case, the terms on a smaller (larger) face are placed to the right (left) in the product. The unitary $U$ acts on the fermions as
\be 
\gamma_{v_1(c)} \to \prod_{f \in F(v_1 (c))}Z_f \gamma_{v_1(c)}, \  \gamma_{v_2(c)} \to \prod_{f \in F(v_2(c))}Z_f \gamma_{v_2(c)},
\ee 
\be 
(-1)^{F_c} \to \prod_{f \subset c} Z_f,
\ee 
and
\be 
S_f \to  S_f \prod_{f \in F(f)} Z_f.
\ee 
And the action on $X_f$ is 
\be 
 U X_f U^{-1}  = S_f X_f\prod_{f' \in F(f), f'>f} Z_{f'}, 
\ee  
and on $X_f S_f$ is 
\be 
X_f S_f    \to   X_f \prod_{f' \in F(f), f' <f} Z_{f'}.
\label{eq:xfsf_general}
\ee
The unitary maps 
\be 
G_c \to (-1)^{F_c} = 1.
\ee 
By gauge invariance, the image of all other fermionic operators can be written in terms of $(-1)^{F_c}$ for general $c$.  Thus setting $(-1)^{F_c} =1$ disentangles the fermions from the gauge spins. We have completed the construction of the unitary transformation from the parity-gauged fermionic system to the spin system. 

Now we need to consider the flatness condition on the gauge field when we gauge the fermion parity. On the primal cubic lattice as in Fig.~\ref{fig_cubic}, the fermions sit at the center of the cubes and the gauge spins on the faces. On the dual cubic lattice, the fermions sit on the (dual) vertices and the gauge spins on the (dual) edges. Four dual edges connecting four dual vertices can form a loop that encloses a dual face. See the shaded square in  Fig.~\ref{fig_cubic} for an example. The dual face intersects with an edge $e$ of the primal lattice. In the dual picture, the four faces $f$ in the primal lattice where the four gauge spins sit share the common edge $e$. The flatness condition on the gauge field is then  
\be \prod_{f \supset e} X_f =1.\ee
Physically, it means that there is no gauge flux passing through any dual square face enclosed by four dual edges where the gauge spins sit. This condition guarantees the duality between the original fermionic system and the parity-gauged fermionic system. Under the disentangling unitary, 
\be 
 \prod_{f \supset e}  X_f  \to \left(\prod_{f \supset e} S_f \right) \cdot \prod_{f \supset e}\left[  X_f  \prod_{f' \in F(f), f' >f} Z_{f'} \right]. 
\ee
The factor $\prod_{f \supset e} S_f$ can be simplified explicitly to a product of $(-1)^{F_c}$ once an orientation is chosen for each bilinear $S_f$ and $(-1)^{F_c}$.  Plugging in $(-1)^{F_c} =1$ yields a phase and the right-hand side becomes the Gauss law 
\be 
G_e =1
\label{eq:gauss3d}
\ee 
for the spin system. Thus, we have succeeded in encoding the flatness condition into the spin system. 

The composition of parity-gauging under the flatness condition and the disentangling unitary transformation produces a general bosonization scheme in this case:
\be 
(-1)^{F_c} \to \prod_{f \subset c} Z_f,
\label{eq:ff_general}
\ee 
and
\be 
S_f \to  X_f \prod_{f' \in F(f), f' <f} Z_{f'}.
\label{eq:sf_general}
\ee 
Mathematically, the modified Gauss law $G_e =1$ guarantees that a product of concatenated fermion bilinears surrounding edge $e$ equal to identity on the fermionic side is mapped to an identity on the spin side. If there is a nontrivial cycle in the total space $M$, then the  product of concatenated fermion bilinears along that cycle is also mapped to a (dressed) spin operator supported along that cycle. This new spin operator is a generator of the dual 2-form symmetry after gauging the (0-form) fermion parity.    Physically, the Gauss law guarantees that the dual 2-form symmetries have a nontrivial 't Hooft anomaly and the local gauge flux excitations have a nontrivial fermionic statistics \cite{gaiotto2016spin}. Apparently, the big picture highlighted in Fig.~\ref{fig_schematic} generalizes easily to more general polyhedral  decompositions than a cubic decomposition and to arbitrary dimensions. 

\subsection{Triangulation and Kasteleyn orientations}
\label{sec:Kasteleyn3d}

Inspired by the 2D case, we may want to generalize the Kasteleyn orientation to higher dimensions in order to show the explicit dependence of the Gauss law in Eq.~(\ref{eq:gauss3d}) and thus the bosonization scheme on the discrete spin structure. Generalizations to arbitrary regular decompositions are lacking, but can be accomplished when the total space is triangulated with complex fermions sitting in the top-dimensional simplicies and gauge spins residing on the faces.   

As illustrated in Fig.~\ref{fig_cubic}, for each edge, there is a loop of fermions surrounding it. This loop can be viewed as the boundary of the 2-cell dual to the edge $e$. Recall that the 
dual is only loosely defined here because of the extra local parity bilinear between two Majorana fermions within the same simplex. Given an orientation for each generating fermionic bilinear, the Majorana bilinears form a loop with links pointing along or opposite to the designated direction of the loop. The orientations of these links encode the information similar to Kasteleyn orientations in 2D.        

Let us consider a triangulated 3D space $M$ with a branching structure. The branching structure induces a natural orientation for each edge, which also fixes a positive orientation of the loop of bilinears circling it by the right-hand rule. Each edge $e$ is the boundary of many triangular faces and the branching structure also determines a canonical positive normal direction for each triangular face. These normal directions specify the ordering within bilinears $S_f = i\gamma_{t_+(f)}\gamma'_{t_-(f)}$ where $t_{\pm}(f)$ are the two neighboring 3-simplices located in the positive or negative normal direction of $f$. As in 2D, the orientation of $S_f$ points from $\gamma'_{t_-(f)}$ to $\gamma_{t_+(f)}$, and that of $(-1)^{F_t} = i \gamma'_{t}\gamma_t$ for each tetrahedron $t$ points from $\gamma_t$ to $\gamma'_t$. Hence, the orientation of each link associated with a bilinear on the loop surrounding $e$ is determined. Then we repeat the story in 2D to show that the number of clockwise-oriented arrows of the bilinears for each edge $e$, $N_{\text{cw}}(e)$, is directly related to a representative of the second SW class $w_2 \in H^2(M, \bbz_2)$.

\begin{figure}[bt]
    \centering
\includegraphics[width=1.0\linewidth]{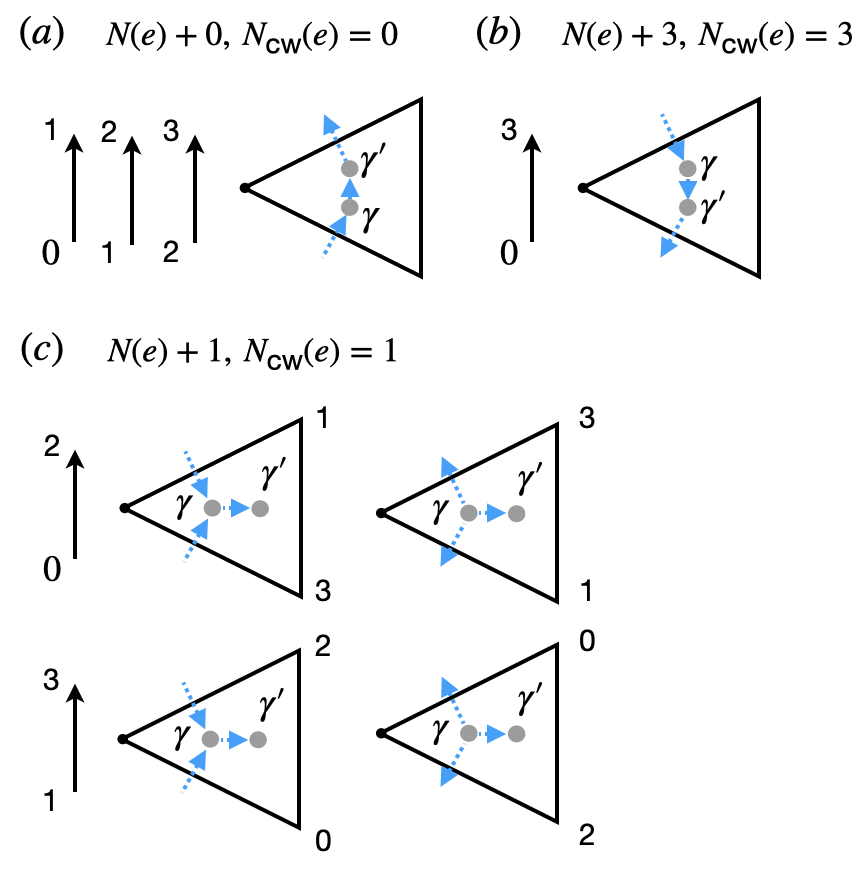}
    \caption{Configurations of orientations of bilinears within a tetrahedron circling the edge (black dot) on the left. The indices of the possible edge is indicated to the left. Each triangle represents a top view of the tetrahedron from the larger vertex of the edge. $N(e)+s$ means the contribution of the 2-simplices and the 3-simplex, i.e., the tetrahedron, to the number of regularity $N(e)$ is $s$. $N_{\text{cw}}(e)$ is the number of clockwise-oriented arrows.}
    \label{fig:Kas3d}
\end{figure}

In 3D, the Poincare dual to the second SW class $w_2 $ is a first class in $H_1(M, \bbz_2)$ which can be represented by a formal sum of edges
\be 
\tilde{w}_1 = \sum_{e} N(e) e,
\label{eq:SW3d}
\ee 
where $N(e)$ is the number of simplices in which $e$ is regular. In 3D, an edge $e$ is regular in itself, and regular in $\Delta_2$ and/or $\Delta_3$ if it connects the smallest vertex to the largest vertex in that simplex. See Appendix~\ref{app:algebraictopology} for more information. In Fig.~\ref{fig:Kas3d}, we list all the configurations of the arrows of bilinears circling an edge of a tetrahedron. Here, what is presented is the top view of tetrahedrons with respect to the edge on the left, represented by the black dot. The indices of the edge within the tetrahedron are indicated by arrows to the left. In Fig.~\ref{fig:Kas3d}(a), three different edges correspond to the same configuration of bilinears. For example, for the edge pointing from vertex 0 to vertex 1, which is represented by single black dot in the triangle, the other two vertices can be either 2 or 3. Then the upper edge and the lower edge represent two triangular faces in 3D with the canonical normal directions determined by the right-hand rule. It is easy to see that the corresponding orientations of the bilinears are exactly represented by the dashed blue arrows. Note that, by definition, the edge $\langle 01\rangle$ is regular in neither the two triangles nor the tetrahedron. If we denote the contribution from the triangles and the tetrahedron to the number $N(e)$ by $s$, then $s =0$ in this case. Other cases can be enumerated similarly except that in the case presented in Fig.~\ref{fig:Kas3d}(c), for the edges indicated to the left, there are two different possible configurations for the bilinears depending on the order of the other two indices within the tetrahedron. In all six cases, we can see that each contribution $s$ from triangles and tetrahedrons to the number $N(e)$ is in one-to-one correspondence with a clockwise-oriented arrow. Taking into account the fact that each edge is regular in itself, we conclude 
\be 
N(e) =1+ N_{\text{cw}}(e).
\ee 
The analysis above is very similar to that in Ref.~\cite{wang2018towards} except that they construct a resolved dual lattice with more Majorana fermions in each tetrahedron. As a direct consequence, when $N_{\text{cw}}(e)$ is odd for any edge $e$ under the decomposition, i.e., a generalized Kasteleyn orientation is chosen, the representative $\tilde{w}_1$ vanishes. A 2D surface $S$ satisfying $\partial S =0$ represents a discrete spin structure in $H_2(M, \bbz_2) \cong H^1(M,\bbz_2)$. 

One may ask whether we can generalize the above analysis to arbitrary polyhedral decompositions. In 2D, as we have shown in Sec.~\ref{sec:KO_general}, we are able to generalize it to general surface graphs based on the triangulated case. However, the canonical way to determine the normal direction based on the branching structure is invalid for a general face. Therefore, the connection between the Kasteleyn orientation and the discrete spin structure is unclear in the general case. In $n$-dimension, the second SW class $w_2$ can be represented by an $(n-2)$-chain that is a formal sum of regular or polyhedral $(n-2)$-cells with $\bbz_2$ coefficients. The dual of each $(n-2)$-cell is a regular or polyhedral dual 2-cell with a loop boundary on which fermion bilinears can reside. As a characteristic class, the second SW class $w_2$ represents the obstruction to extending a nonvanishing section (in the $SO(n)$-frame bundle) from 1-cells to 2-cells (or more rigorously, from a 1-chain to a 2-chain), assuming that the total space $M$ is an $n$-dimensional CW complex \cite{hatcher2002algebraic}. In 2D, one can use Kuperberg's construction to relate the Kasteleyn orientation to the winding number of a section extended from the 1-cell to the 2-cell, and make use of the dimer configuration to eliminate the singularities of the vector field with odd indices on the vertices \cite{cimasoni2007dimers}. Such a construction is lacking in higher dimensions so far. It is yet to be shown whether generalized Kasteleyn orientations on the boundary of the dual 2-cell can still determine $w_2$ and represent different discrete spin structures (if $w_2$ vanishes).

\subsection{KW dualities}
Analogous to the 2D case, the bosonization in general dimensions can be exploited in translationally invariant fermionic systems to construct KW dualities in general dimensions. The fundamental example is free Majorana fermions on a (dual) hypercubic lattice. Again, we place a complex fermion, or two Majorana fermions $\gamma$ and $\gamma'$, within each hypercube, and consider the minimal translation defined by an exchange of $\gamma \leftrightarrow \gamma'$ followed by a translation of $\gamma'$ fermions along one direction, say the (positive) $x$-direction. The couplings/hoppings of the Majorana fermions must ensure that the Hamiltonian is invariant under this minimal translation. Then we apply the bosonization scheme to the fermionic model to obtain a gauged spin model. Two Majorana bilinears related by the minimal translation (along the $x$-direction) are thus mapped to two different spin operators related by the generalized KW duality. In particular, there is one representation where 
\be (-1)^{F_f} \to W_c \equiv \prod_{f\subset c} Z_f\ee 
and 
\be S_{f^c_{x}} \to X_{f^c_{x}}.\ee
Here, ${f^c_{x}}$ represents the face in the $x$-direction of cube $c$. This representation can be obtained from Eqs.~(\ref{eq:ff_general}) and (\ref{eq:sf_general}) as follows. We partition the six faces of cube $c$ into $\{f_x^c, f_y^c, f_z^c\}$ and $\{f_{-x}^c, f_{-y}^c, f_{-z}^c\}$, and assign the former set to $\gamma$ and the latter to $\gamma'$. We define $S_{f^c_{x}} \equiv i \gamma_{c}\gamma'_{c+a_x}$, where $c+a_x$ denotes the neighboring cube sitting in the $x$-direction of cube $c$. When choosing an ordering of all the faces of the cubic lattice,  we first order faces perpendicular to the $x$-axis before ordering faces perpendicular to the $y$- and $z$-axis. In this way, for any  ${f^c_{x}}$, there is no $f' <f^c_x$ in  $E(f^c_{x})$ and $S_{f^c_{x}}$ is mapped to $X_{f^c_{x}}$.

In this representation, the KW duality maps 
\be W_c \to X_{f^c_x} \to W_{c+a_x}.
\ee  
This is a direct generalization of the 1D KW duality which maps a domain wall measured by an Ising coupling $ZZ$ to a spin polarization measured by $X$. In general, it maps a gauge flux measured by $W_c$ to a gauge spin polarization measured by $X_{f^c_x}$. On a periodic lattice, the symmetry generators associated with topologically nontrivial loops give rise to 2-form symmetries. Projections onto their eigenspaces with eigenvalue one appear in the spin representation of the noninvertible KW duality operator.

If we consider two copies of decoupled free Majorana fermions on a regular hyperlattice, then the minimal translation is not restricted to one direction but generalizes to any translationally invariant direction. The KW duality then also exists in those directions. This is a direction generalization of the discussion in Sec~\ref{sec:twocopies}.  Continuous $U(1)$ symmetries again exist at the self-dual point, and KW dualities interchange these symmetries.

\section{Discussion}
In our work, we presented a general, systematic, yet  intuitive framework to extend bosonization to arbitrary dimensions. It consists of three steps: gauging fermion parity via minimal coupling, imposing the flatness condition, and constructing the disentangling unitary. It is a direct generalization of the bosonization of the Majorana chain to the 1D TFIM by gauging fermion parity. The construction of the disentangling unitary based on a general assignment procedure works for any polyhedral decomposition in general dimensions. In particular, fermions are not required to reside in top-dimensional simplices of a triangulated space. The dependence of the Gauss law of the spin system on the Kasteleyn orientations (and spin structures) of an arbitrary surface graph in 2D and of triangulation with a branching structure in higher dimensions is manifest, and so is the emergence of dual higher form symmetries on a non-simply-connected space. Moreover, the bosonization framework subsumes previous bosonization methods in cases where these methods are applicable. While we have only demonstrated the derivation of the Bravyi--Kitaev and the Chen--Kapustin approaches, other methods, such as that in Ref.~\cite{verstraete2005mapping}, follow as well, based on the equivalence between them \cite{chen2023equivalence}.  Our construction imposes neither freeness of the system nor restrictions on the form of interactions. Although we have focused on spinless fermions, the framework extends straightforwardly to multiple fermion species, e.g., with different spins.
Hence, our bosonization scheme applies to general fermionic lattice systems, and the analysis in our work substantiates the general picture advocated in Fig.~\ref{fig_schematic}.

As an application to translationally invariant Majorana systems on regular lattices, we generalized the KW (self-)duality to general dimensions. The minimal (half-unit-cell) translation defined as an exchange of $\gamma \leftrightarrow \gamma'$ followed by a translation of $\gamma$ is mapped to KW dualities in the gauged spin system. For one copy of the free Majorana lattice, the (self-)duality is unidirectional in the spin model. It maps a gauge flux to a gauge spin polarization. For two decoupled copies of free Majorana lattices, the duality in the spin model is more ``isotropic" and exists in multiple directions along the primitive lattice vectors. In particular, the self-dual transformation interchanges the $U(1)$ symmetry charges at the self-dual point. The spin models also possess higher-form symmetries, and projections onto the eigenspaces with eigenvalue one of these higher-form symmetries give rise to the noninvertibility of the KW (self-)dualities. Although we have used free systems as illustrative examples, KW dualities readily generalize to interacting systems.  Moreover, we do not need to restrict to minimal translations. We can define minimal rotations: exchanging $\gamma \leftrightarrow \gamma'$ followed by a rotation of the $\gamma$ or $\gamma'$ Majorana fermions. Other more general crystalline symmetries can also be considered. The invariance under these ``minimal crystalline symmetries" in a fermionic system is mapped to generalized dualities of the gauged spin system. We leave this generalization to future work.

One of the key open questions is the relation between the KW (self-)dualities constructed in this work and those obtained by gauging generalized symmetries. In particular, the basic KW dualities arising from minimal translations of a single copy of the coupled Majorana square lattice cannot be conventional, since the self-dualities are incompatible with the required degree of the higher-form global symmetry, at least in even spatial dimensions. The directional dependence of the KW dualities further suggests that the conventional construction of a KW duality by gauging half of space \cite{choi2022noninvertible} is insufficient, and different inequivalent cycles of the spacetime manifold must be taken into account. In the accompanying paper \cite{su$mathbbZ_2$2025a}, we show that how the self-dualities studied in this work are different from conventional gauging. Furthermore, the gauge-flux-particle (self-)dualities warrant additional investigation. In 1D, the KW duality can be viewed as the boundary manifestation of the electromagnetic duality in the topological order of the bulk \cite{freedTopological2022}. This raises the natural question of whether an analogous bulk-boundary correspondence exists in higher dimensions. One may also ask whether the 2D KW self-duality is related to the particle-vortex duality in 2D.

In our work, we have focused on the KW dualities in the Hamiltonian formalism. We have not studied the dependence of the dualities on the boundary conditions in detail as in the 1D case \cite{seiberg2024majorana, aksoyLiebSchultzMattis2024}. In higher dimensions, the projections in the KW duality operators are associated with higher form symmetries, as exemplified in Eq.~(\ref{eq:noninv_KW}). We expect a similar analysis as in 1D. Furthermore, similar to the 1D case, there can be a strong dependence on the total system size \cite{cheng2023lieb, su2025z2}. It is not hard to envision that the self-dualities have some generalized 't Hooft anomalies and rule out the existence of trivially gapped ground states \cite{seiberg2024non}.  Also, it would be helpful to extend them to the spacetime picture and examine the Euclidean partition function \cite{chen2019bosonization, chen2020exact}.  We should analyze the topological defects associated with the KW dualities directly in the partition function. It would be interesting to study the continuum limit of the KW self-dualities. In 1D, the defects do not form a category because of the translation involved \cite{seiberg2024non} but flow to a fusion category in the infrared. By analogy, they do not form higher categories in higher dimensions either but are likely to flow to higher fusion categories in the low-energy theory. The manifestation of the KW dualities in the field theory, especially at the self-dual points, is very intriguing. 

As regards bosonization, there are several statements to be analyzed more carefully. First, we have discussed the dependence of the bosonization scheme on Kasteleyn orientations over the loop of concatenated fermionic bilinears and discrete spin structures, but we are yet to prove (or disprove) the relation between Kasteleyn orientations and discrete spin structures for more general decompositions than triangulation in dimensions higher than 2D. Second, we are yet to rigorously determine the fate of the bosonic version of the parity anomaly in 2D mentioned in Sec.~\ref{sec:honeycomb} and its generalizations. For example, we can bosonize the Hamiltonian models on a 3D cubic lattice with a chiral global symmetry constructed in Ref.~\cite{gioia2025exact}, and determine the fate of the bosonized $SU(2)$ anomaly. More broadly, we should ask how the exact bosonization framework for a lattice flows to the bosonization of a field theory in the infrared.  

Lastly, in the path integral formalism, bosonization is expressed as the summation over different spin structures of the partition function. The bosonized theory has emergent fermionic excitations and possesses an anomalous higher-form symmetry generated by the worldline of an emergent fermion. We have only briefly discussed this higher-form symmetry in our framework, but there are generalizations, called higher bosonization in Ref.~\cite{thorngren2020anomalies},  based on higher-group symmetries. It would be interesting to study the connection within our framework in detail. On the other hand, the reverse direction, also known as fermionization, can be carried out \cite{su$mathbbZ_2$2025a}, providing a two-way bridge between fermionic SPT phases and bosonic SPT phases. Our work provides a natural yet overarching framework to carry out the transformations exactly on general lattices. As evidenced by the relation between fermionic and bosonic systems via gauging fermion parity, the symmetries and anomalies of fermionic and bosonic systems are often entangled in a richly orchestrated and irresistibly captivating tango.

\section{Acknowledgment} 
We thank Aashish Clerk, Michael Levin, and Zohar Nussinov for helpful discussions. This work was supported by the US Department of Energy, Office of Science, Basic Energy Sciences, Materials Sciences and Engineering Division.

\appendix

\section{Operator algebras}
\label{app:algebra}
\subsection{Operator isomorphism}
\label{app:embedding}
In the main text, we discussed how bosonization maps the parity-even fermionic algebra to the algebra generated by the corresponding spin operators modulo the Gauss law. From an algebraic point of view, this is an isomorphism of operator algebras when the space $M$ is simply connected. To have a better sense of the isomorphism, let us work on an infinite square lattice. 

\begin{figure}[tb]
    \centering \includegraphics[width=0.8\linewidth]{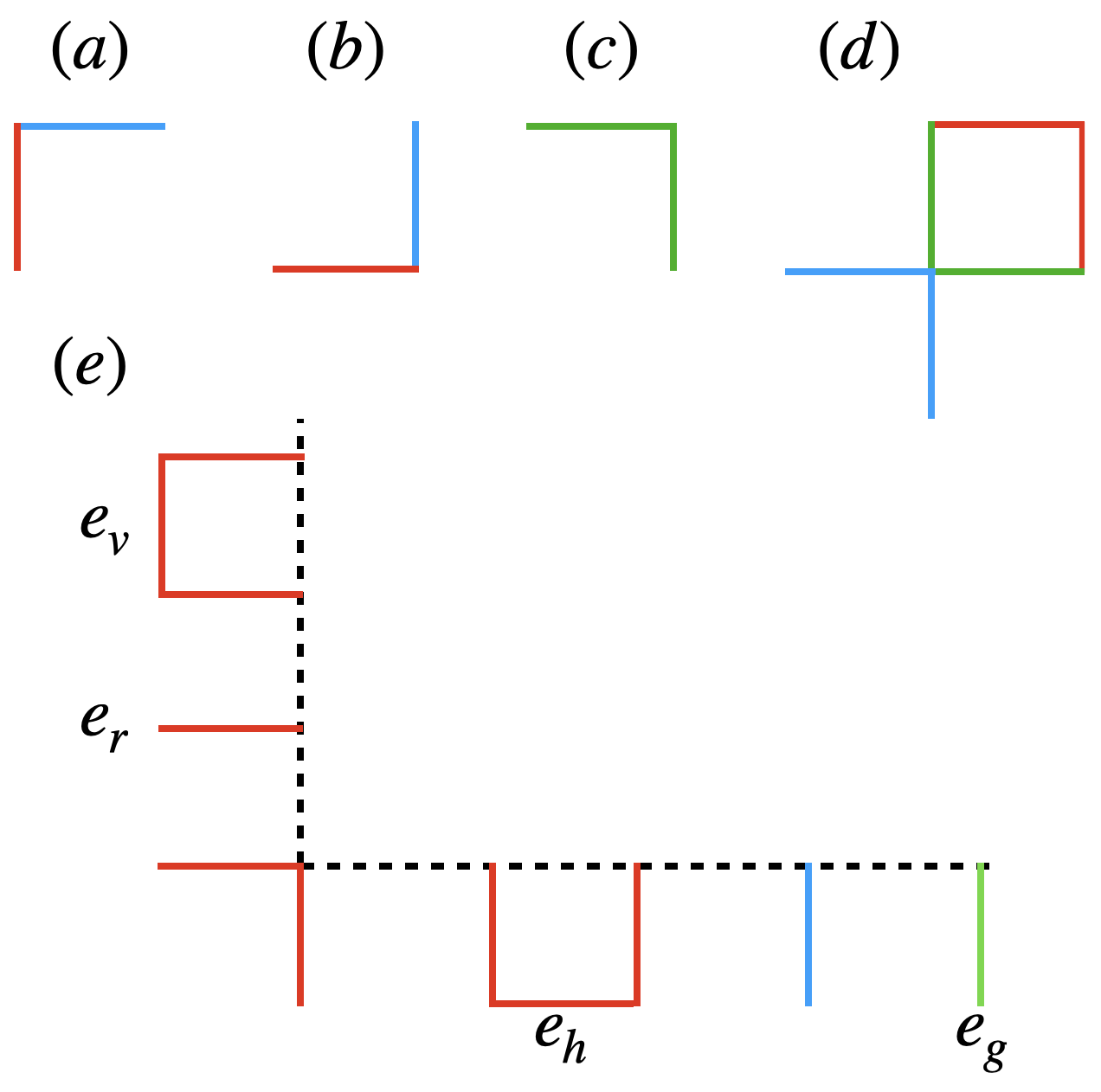}
    \caption{(a-c) Generators of the spin algebra (d) the Gauss-law operator (up to a minus sign). Each operator is the product of Pauli operators on the colored edges: $Z$ (red), $X$ (blue), and $Y$ (green). (e) Schematic showing a bounded operator on the bottom row and left column.}
    \label{fig_iso}
\end{figure}

The parity-even fermionic algebra is generated by $S_e$ and $(-1)^{F_f}$ modulo a subalgebra generated by products of concatenated Majorana bilinears along closed loops. By construction, the bosonization maps define an embedding of the fermionic algebra into the algebra of the gauged spin system generated by $V_{e^f_x}$, $V_{e^f_y}$, and $W_f$, or equivalently the three generators shown in Fig.~\ref{fig_iso}(a-c) modulo the subalgebra generated by the Gauss-law operator shown in (d). We can check that the fermionic and the bosonic Hilbert space dimensions match. The Hilbert space of fermions for each unit cell is 2. There are two independent spins associated with each unit cell, so the total dimension is 4. However, there is one Gauss-law constraint for each unit cell, reducing the dimension of the physical Hilbert space to $4/2 =2$. 

Furthermore, we can show explicitly that any gauge-invariant spin operator $O$, including sums of such operators, is generated by the three spin generators, at least when $O$ has a bounded support. Since it is bounded in range, we can assume that the operator is a product of operators whose range is covered by a finite rectangle consisting of $w \times h$ squares with $w$ being the width and $h$ the height. In the following, we reduce the operator $O$ by the three generators to the identity. Without loss of generality, we focus on spin components of $O$ on the bottom or the left side of the rectangle. 
Since $O$ is bounded in support, we can always identify the lowest spin components of $O$. They are either on horizontal edges as $e_h$, or on vertical edges as $e_g$ shown in (e). Since $O$ commutes with the Gauss-law operator in (d) at any vertex, the spin operator on horizontal edges of the lowest row must be $Z$. To reduce $O$, we can eliminate these $Z$'s by multiplying $O$ with the generator in (b).  Spins on the vertical edges can be removed by multiplying $O$ with the generator in (a) and/or the generator in (c). By checking the spins on the leftmost or the rightmost vertical edge, we can easily see that this process reduces the height of the operator $O$ by one row without increasing the width. Similarly,  we can perform reductions on the left side of the rectangle to reduce the width of the operator without increasing its height. 
Repeating these steps at most $w\times h$ times reduces $O$ to a simple operator acting on the square. It is straightforward to see that the three generators generate the operators on a square that commute with the Gauss law. This completes the proof.

\subsection{Bond algebras}
\label{app:bondalgebra}
There is a close relation between isomorphisms of bond algebras and dualities \cite{cobanera2011bond}.

Let $\Ha$ be a Hilbert space, and let $B(\Ha)$ denote the algebra of bounded operators on $\Ha$. If ${\cal{S}} \subset B(\Ha)$ is an arbitrary subset, its commutant (or centralizer) ${\cal{S}}' \in B(\Ha)$ is the subalgebra defined by 
\be {\cal{S}}' = \{O \in B(\Ha)| \forall A \in {\cal{S}}, [O, A]=0\}.\ee  A subalgebra $\Aa \subset B(\Ha)$ is a von Neumann algebra if it satisfies the following three conditions: 
\begin{enumerate}
    \item Unital: $\Aa$ contains the identity operator, $\1 \in \Aa$.
    \item Self-adjoint:  $\Aa$ is closed under the adjoint operation, $O \mapsto O^{\dagger}$.
    \item Double commutant property: $\Aa$ is equal to its bicommutant, $\Aa = \Aa''$.
\end{enumerate}
A map $\Phi: \Aa_1 \to \Aa_2$ between von Neumann algebras is a homomorphism if it preserves algebraic operations and the adjoint, i.e., $\Phi(O^{\dagger}) = \Phi(O)^{\dagger}$. If $\Phi$ is a bijection, then it is an isomorphism.  

We then have the following theorem:

Let $\Aa_i$ be von Neumann algebras over Hilbert spaces $\Ha_i$ for $i =1,2$. If $\Phi: \Aa_1 \to \Aa_2$ is an isomorphism, then there exists an auxiliary Hilbert space $\Ha'$ such that for any $O \in \Aa_1$, there is a unitary transformation $U: \Ha_1\otimes \Ha' \to \Ha_2 \otimes \Ha'$ that satisfies $\Phi(O) \otimes \1 = U(O\otimes \1 )U^{\dagger}$. 

Two Hamiltonians are dual to each other if the isomorphism maps one to another. 
The theorem implies identical spectra up to degeneracies and boundary conditions. If a set of local operators (called ``bonds") generates a von Neumann algebra, then it is called a bond algebra. Thus, dualities can be viewed as isomorphisms of bond algebras.  

We can show that in the parity-gauged fermionic system, the algebra generated by gauge-invariant Majorana bilinears is a von Neumann algebra. In particular, $\prod_{e \supset v} X_e$ generates the commutant (with finite support) of the algebra. Modifying the argument in the last section, we can show that the algebra generated by $V_{e^f_x}$, $V_{e^f_y}$, and $W_f$ is also a von Neumann algebra and that the commutant (with finite support) of the algebra is generated by the Gauss-law operator. The bosonization map is an isomorphism of von Neumann algebras that maps the two commutants to each other. Thus, the theorem above applies, with the unitary  given in the main text.

\section{Mathematical basics}
In this Appendix, we briefly introduce the minimal mathematical ingredients in simplicial algebraic topology to facilitate the presentation in the main text. We also briefly introduce the concept of spin structures. The reader who wants to know more can always refer to standard textbooks, e.g., Refs.~\cite{hatcher2002algebraic, nakahara2018geometry}. 

\subsection{Homology and cohomology groups}
\label{app:algebraictopology}
A free abelian group (or a formal sum) generated by the $p$-simplices of the triangulated $n$-dimensional manifold $M$, with coefficients in $\bbz_2$, is called a $\bbz_2$ (simplicial) $p$-chain group $C_p(M,\bbz_2)$. Here, 0-simplices are vertices, 1-simplices are edges, and 2-simplices are triangles, and higher-dimensional simplices follow similarly. We can similarly define a $\bbz_2$-valued (simplicial) $p$-cochain group $C^p(M,\bbz_2)$: it is a formal sum of  $\bbz_2$-valued functions on the $p$-simplices. We use $\bm{\Delta_p}$ in boldface to denote a function that maps the $p$-simplex $\Delta_p$ to 1 and all other simplices to 0, then, by definition, it is a $p$-cochain. We use $\int_{c} \alpha_p = \sum_{\Delta_p \in c}\alpha(\Delta_p)$ to denote the evaluation of $p$-cochain $\alpha_p$ on $p$-chain $c$. If $c$ is not specified, the evaluation is over the top-dimensional chains $\int \alpha_p = \int_M \alpha_p$. 

We can define a boundary map $\partial: C_p(M, \bbz_2) \to C_{p-1}(M, \bbz_2)$ that maps each $p$-simplex $\Delta_p$ to (a formal sum of) the $(p-1)$-simplices of its boundaries. Here, we are using $\bbz_2$-coefficients, so the sign $\pm 1$ in front of each boundary term in the formal sum is not important. In general, the signs need to be specified. It is easy to see that $\partial^2 =0$ because the boundary of the boundary (of a manifold) is empty. In this case, the $p$-chains form a so-called (simplicial) chain-complex. Chains in the kernel of $\partial: C_p(M, \bbz_2) \to C_{p-1}(M, \bbz_2)$ are called $p$-cycles, and those in the image of $\partial: C_{p+1}(M, \bbz_2) \to C_{p}(M, \bbz_2)$are called $p$-boundaries. The $p$th homology group $H_p(M, \bbz_2)$ is defined to be the quotient of $p$-cycles by $p$-boundaries. Two $p$-cycles that differ by a boundary are identified in the group.  

Similarly, we can define a coboundary map $\delta: C^p(M, \bbz_2) \to C^{p+1}(M, \bbz_2)$ that maps $p$-cochain $\bm{\Delta_p}$ to (a formal sum of) the $(p{+}1)$-cochains $\bm{\Delta_{p+1}}$ such that the corresponding simplex $\Delta_p$ is a boundary of the simplices $\Delta_{p+1}$. For example, for a vertex $v$, $\delta \bm{v}$ is a sum of $\bm{e}$, where $e$ is any edge incident to $v$. By abuse of notation, we also write $\delta v = \sum e$. It is easy to show that $\delta^2=0$. Therefore, cochains $C^p(M, \bbz_2)$ together with $\delta$ form a cochain complex. Cochains in the kernel $\delta: C^p(M, \bbz_2) \to C^{p+1}(M, \bbz_2)$ are called $p$-cocycles and those in the image $\delta: C^{p-1}(M, \bbz_2) \to C^{p}(M, \bbz_2)$ are called $p$-coboundaries. The $p$th cohomology group $H^p(M, \bbz_2)$ is defined to be the quotient of $p$-cocycles by $p$-coboundaries. 

What we defined above are simplicial homology and cohomology groups. There are other homology and cohomology theories. For the purpose of this work, we assume that they are identical since we consider either a triangulated manifold or a manifold with a polyhedral decomposition. For any orientable closed (compact and without boundary) $n$-dimensional manifold $M$, the Poincar\'e duality theorem states that 
\be 
H^{i}(M, G) \cong H_{n-i}(M, G)
\ee 
for any coefficient group $G$. If $G =\bbz_2$, which is our focus, then $M$ need not be orientable. Since the only noncompact case is when $M$ is infinite and contractible and  thus topologically trivial, we do not need to refine the Poincar\'e duality theorem to include this case. 

To derive the Chen--Kapustin method in the next Appendix, we can introduce the cup product of two cochains. Endowing the triangulated space with a branching structure fixes the ordering of vertices. Each $p$-simplex can be denoted by $\langle v_0, v_1, \ldots, v_p\rangle$ or simply $\langle 0, 1, \ldots, p\rangle$ for a  specific $p$-simplex. The cup product of a $p$-cochain $\alpha_p$ and a $q$-cochain $\beta_q$ on a $(p+q)$-simplex denoted by $\langle 0, 1, ..., p+q\rangle$ is a $(p+q)$-cochain defined as
\be
\begin{split}
  \alpha_p \cup \beta_q &(\langle 0, 1, ..., p+q\rangle) = \\ & \alpha_p(\langle 0, 1, ..., p-1\rangle) \beta_q (\langle p, p+1, ..., p+q\rangle) . 
\end{split}
\ee 
For example, 
\be
\bm{e} \cup \bm{e}' (\langle 0, 1, 2\rangle) = \bm{e}(\langle 0, 1\rangle) \bm{e}'(\langle 1, 2\rangle).
\ee 
The ordering in $\langle 0, 1, 2\rangle$ is determined by the branching structure. Therefore, the evaluation $\bm{e}\cup \bm{e}' + \bm{e}'\cup \bm{e}$ yields 1 only if $e$ and $e'$ are concatenated with a compatible orientation, or in other words, they share vertex 1 within a triangle. The cup product will play a role in deriving the Chen--Kapustin method in 2D in the next Appendix. To relate our approach to theirs in three or higher dimensions, higher cup products can be used. Since this is not our focus in this work, we will omit them.

\subsection{Spin structures}
Consider the tangent bundle of a (smooth) $n$-dimensional orientable manifold $M$. The structure group can be taken to be $SO(n)$. There is an $SO(n)$ frame bundle associated with the tangent bundle.  Let $\{U_i\}$ be an open cover of $M$. The transition functions $t_{ij}: U_i \cap U_j \to SO(n)$ satisfy the consistency conditions
 \be 
t_{ii} =I, \quad t_{ij} t_{ji} = I,\quad t_{ij} t_{jk} t_{ki} =I. 
 \ee 
 Consider another principal $Spin(n)$-bundle with transition functions $\tilde{t}_{ij} \in Spin(n)$ satisfying $\varphi(\tilde{t}_{ij})=t_{ij}$ and
\be 
\varphi(\tilde{t}_{ii}) =I, \quad \varphi(\tilde{t}_{ij})\varphi(\tilde{t}_{ji}) = I,\quad \varphi(\tilde{t}_{ij}) \varphi(\tilde{t}_{jk}) \varphi(\tilde{t}_{ki}) =I, 
 \ee  
where $\varphi: Spin(n)\to SO(n)$ is a double cover, i.e.,
\be 
0 \to \bbz_2 \to Spin(n) \to SO(n) \to 0
\ee 
is a short exact sequence of group homomorphisms. In particular, $Spin(2)\cong U(1)$ and $Spin(3) \cong SU(2)$. The set of $\tilde{t}_{ij}$, if it exists, defines a spin bundle on $M$ and $M$ is said to admit a spin structure. Physically, a spin structure determines how spinors are transported around a loop (i.e., 1-cycle) -- in particular, whether they return to themselves or change sign. In other words, it tells us whether fermions are periodic or anti-periodic around each cycle. In our main text, it is related to the phases of the products of concatenated fermion bilinears over each cycle.

Define a C\v{e}ch $p$-cochain to be a totally symmetric $\bbz_2$-valued function $f(i_0, i_1,\ldots, i_p)$ on $U_{i_0}\cap U_{i_1}\cap \cdots \cap U_{i_p} \neq \O$. Such $p$-cochains generate an additive cochain group $C^p(M, \bbz_2)$. The coboundary map $\delta: C^p(M, \bbz_2) \to C^{p+1}(M, \bbz_2)$ is defined by 
\be 
(\delta f)(i_0, i_1, \ldots, i_{p+1}) = \sum_{j=0}^{p+1} f(i_0, \ldots, \hat{i}_j, \ldots, i_{p+1}),
\ee 
where the variable with a hat is omitted. One can check that $\delta^2 f =0$, i.e., it defines a C\v{e}ch cochain complex for C\v{e}ch cochains. Cocycles, coboundaries, and cohomology groups can be defined analogously to the simplicial ones. The C\v{e}ch cohomology group is isomorphic to the simplicial cohomology group for the class of manifolds considered in this work. 

Now let us return to the transition functions $\tilde{t}_{ij}$  to determine the existence of a spin structure. Define the C\v{e}ch 2-cochain $f: U_i\cap U_j \cap U_k \to \bbz_2$ by 
\be 
\varphi(\tilde{t}_{ij}) \varphi(\tilde{t}_{jk}) \varphi(\tilde{t}_{ki}) =(-1)^{f(i, j, k)}I. 
\ee 
$f$ is symmetric in all three variables and $\delta f =0$. Thus, it defines a class in the  C\v{e}ch cohomology group $H^2(M, \bbz_2)$, called the second Stiefel--Whitney (SW) class. It is easy to show that $M$ admits a spin structure if and only if the second SW class $w_2 \in H^2(M, \bbz_2)$ vanishes. If $w_2=0$, then changing $\tilde{t}_{ij}$ can change the spin structure. Different classes of $\tilde{t}_{ij}$ determined by $H^1(M, \bbz_2)$ are therefore in one-to-one correspondence with inequivalent spin structures as a set. Denote the set of spin structures by $S(M)$. Then it is easy to show that $H^1(M, \bbz_2)$ acts freely and transitively on $S(M)$. In other words, $S(M)$ is a principal homogeneous space or torsor for the group $H^1(M, \bbz_2)$. In fact, the elements in $H^1(M, \bbz_2)$ can be viewed as $\bbz_2$ gauge fields which can twist the boundary conditions of fermions along a loop/cycle.

By the Poincar\'e duality, $w_2$ is dual to a class in $H_{n-2}(M, \bbz_2)$. On a triangulated space with a branching structure, an explicit representative of the class is given in Ref.~\cite{goldstein1976formula} as follows. 
Let $s =\langle v_0, v_1, \ldots, v_p\rangle$ and $\Delta_i$ be an $i$-simplex such that $s\subset \Delta_i$. Define the sets $B_k$ for $-1\le k\le p$ 
\be 
B_k =
\begin{cases} \{v \subset \Delta_i| v<v_0\},\  k =-1 \\
\{v \subset \Delta_i| v_k < v<v_{k+1}\},\  0\le k < p\\
\{v \subset \Delta_i| v>v_p\},\ k=p.
\end{cases}
\ee 
If the number of elements in the set $\#(B_k) =0$ for all odd $k$, $s$ is said to be regular in $\Delta_i$. In particular, a vertex $v$ in 2D is regular in itself, regular in $\Delta_1$ if it is the smaller vertex, and regular in $\Delta_2$ if it is the smallest vertex. In 3D, an edge $e$ is regular in itself, and regular in $\Delta_2$ and/or $\Delta_3$ if it connects the smallest vertex to the largest vertex. For a general $q$-dimensional simplex, $\Delta_{q}$, let $N(\Delta_{q})$ be the number of simplices with respect to which $\Delta_{q}$ is regular. Then, the class in the $(n-2)$th homology group dual to the second SW class $w_2$ is represented by
\be 
\tilde{w}_{n-2} = \sum_{\Delta_{n-2}} N(\Delta_{n-2}) \Delta_{n-2}.
\label{eq:SW_general}
\ee 
If $\partial S = \tilde{w}_{n-2}$ for an $(n-1)$-chain $S \in C_{n-1}(M,\bbz_2)$, then $\tilde{w}_{n-2}=0$ and $S$ corresponds to a spin structure. By the Poincar\'e duality, $S$ represents an element in $H^1(M, \bbz_2) \cong H_{n-1}(M, \bbz_2)$. 

On a 2D orientable Riemann surface, $w_2$ is always trivial. Hence, spin structures can always be defined in this case. On the one hand, it can be shown that the set of spin structures $S(M)$ determines a quadratic form on $H^1(M, \bbz_2)$ that is a quadratic refinement of the intersection form on the surface. One can then use the quadratic form to define the Arf invariant for the spin structure, which serves as a topological invariant of the 2D topological quantum field theory on a spin manifold. On the other hand,  a Kasteleyn orientation on a surface graph exists if and only if the number of vertices is even, as in the case of Majorana fermion pairs discussed in the main text.
A Kasteleyn orientation and a dimer configuration can be used to construct a vector field with only even index singularities. The latter determines a spin structure in $S(M)$. Therefore, inequivalent Kasteleyn orientations are in one-to-one correspondence with inequivalent spin structures \cite{cimasoni2007dimers}.

\section{Bravyi--Kitaev method}
\label{app:BK}
In this Appendix, we summarize the Bravyi--Kitaev superfast transformation \cite{bravyi2002fermionic}.  Consider a graph where the vertices are labeled by $k$ and the edges are labeled by $(j, k)$. We associate two Majorana fermions to vertex $k$ and define
$B_k = - i\gamma_{2k} \gamma_{2k+1}$ and $A_{jk} = -i \gamma_{2j}\gamma_{2k}$. These terms generate the parity-even fermionic algebra.  We choose an orientation for each edge of the graph such that $\epsilon_{kj} = - \epsilon_{jk} \in \{\pm 1\}$ and order the edges incident on $k$ (denoted by $<_k$).  To represent them in terms of spin operators, we put one qubit on each edge and map $B_k$ and $A_{jk}$, respectively, to 
\be 
\begin{split}
\tilde{B}_k &= \prod_{(j, k)} Z_{jk},   \\
 \tilde{A}_{jk} &= \epsilon_{jk} X_{jk} \prod_{(l,j) <_j (k,j)} Z_{lj} \prod_{(s, k)<_k (j,k)} Z_{sk}.
\end{split}
\ee 
The gauge constraint (Gauss-law) operator is 
\be 
\tilde{C}_{j_0,\ldots, j_p} =i^p \tilde{A}_{j_0, j_1}\tilde{A}_{j_1, j_2}\cdots \tilde{A}_{j_{p-1}, j_0}
\ee 
for any closed loop $(j_0, j_1, \ldots, j_{p-1})$. It can be verified that the mappings preserve the operator algebra. In the main text, we consider the dual graph where Majorana fermions reside on the faces and the qubits reside on the dual edges. 

\section{Deriving Chen--Kapustin method}
\label{app:CK}
In this Appendix, we derive the Chen--Kapustin method \cite{chen2019bosonization, chen2020exact} in the case of a triangulated space with a branching structure using our framework. We will focus on 2D, but the generalization to higher dimensions is very similar.

We follow the assignment procedure in Sec.~\ref{sec:triangulation}. Unlike the convention used there, we fix the ordering within each bilinear such that  
\be 
(-1)^{F_f} = i \gamma'_f \gamma_f, \quad S_e = i \gamma_{L(e)}\gamma'_{R(e)}.
\label{eq:convention}
\ee 
In this new convention, left and right are switched in $S_e$. Note that the two bilinears $S_e$ and $S_{e'}$ still always commute unless $e$ and $e'$ are two concatenated edges with compatible orientations. That is, they are not pointing toward each other. This property is exactly captured by the cup product, i.e.,
\be 
S_e S_{e'} = (-1)^{\int \bm{e}\cup \bm{e}' + \bm{e}'\cup \bm{e}} S_{e'} S_e.
\label{eq:Se_commutation}
\ee 
For convenience, we also define a product of bilinears over a 2-cochain 
\be 
S_{\beta} = (-1)^{\sum_{e<e'\in \beta} \bm{e}\cup \bm{e}'} \prod_{e\in \beta} S_e,  
\ee
where $\beta =\{e_1, e_2, \ldots, e_n\}$ is an order set of edges and 
\be 
\prod_{e\in \beta} S_e = S_{e_n} S_{e_{n-1}}\cdots S_{e_1}.
\ee 
Using the property in Eq.~(\ref{eq:Se_commutation}), we can see that $S_{\beta}$ does not depend on the order of elements in $\beta$. We also identify $\beta = e_1+e_2+\cdots + e_n$.

We now consider the disentangling unitary [Eq.~(\ref{eq:general_U})] expressed as
\be 
U = \stackrel{\leftarrow}{\prod}_e (P^+_{e} + P^-_{e}S_e).
\label{eq:U_triangulation}
\ee 
Once again, all the edges have been ordered.
Its action on $X_e$ is
\begin{align}
     X_e &\to  X_e  S_e \prod_{e'>e}[P^+_{e'} + (-1)^{ \int \bm{e}\cup \bm{e}' + \bm{e}'\cup \bm{e} }P^-_{e'}] \nonumber \\
     & = X_e  S_e \prod_{e'>e}Z_{e'}^{ \int \bm{e}\cup \bm{e}' + \bm{e}'\cup \bm{e} }.
\end{align}
Here we have used the property of $\bm{e}\cup \bm{e}' + \bm{e}'\cup \bm{e}$ to avoid explicitly specifying the set  $E(e)$, which consists of the edge $e$ and all other edges such that $\int \bm{e}\cup \bm{e}' + \bm{e}'\cup \bm{e}$ is nontrivial. Now we can assume that the global ordering is compatible with the local ordering $\langle 0, 1\rangle <\langle 1, 2\rangle$ within any 2-simplex. If it is not, we can use the reordering trick explained in the main text and compose $U$ with an extra $CZ$ on these two edges to effectively swap the order. In this case, the expression above can be further simplified:
\be 
     X_e\to  X_e  S_e \prod_{e'}Z_{e'}^{ \int \bm{e}\cup \bm{e}'}.
\ee
Since under the transformation, 
\be S_e \to S_e \prod_{e'} Z_{e'}^{ \int \bm{e}\cup \bm{e}' + \bm{e}'\cup \bm{e} },
\ee 
we obtain 
\be 
     X_e S_e \to  X_e  \prod_{e'}Z_{e'}^{ \int \bm{e}'\cup \bm{e}}.
\ee

Let us now consider how a closed loop of concatenated fermion bilinears is  mapped to the modified Gauss law. Again, this is where the spin structure plays a role in  Refs.~\cite{chen2019bosonization, chen2020exact}. We relate it to the Kasteleyn orientation in the main text, but they present it in a slightly different way. Using the orientation convention in Eq.~(\ref{eq:convention}), one can show that 
\begin{align}
  S_{\delta v} &= (-1)^{\sum_{e<e'\in \delta v}\int \bm{e}\cup \bm{e}'} 
\prod_{e\in \delta v} S_e \nonumber \\
&= (-1)^{\int_{w_2} \bm{v}}  \prod_f P_f^{\int \bm{v} \cup \bm{f} + \bm{f}\cup \bm{v}},  
\end{align} 
where $v$ is a vertex, $f$ is a 2-simplex, and $P_f = (-1)^{F_f}$. $ \delta v=\{e_1, e_2,\ldots, e_q\}$, where $q$ is the degree of $v$, is considered to be ordered. The exponent $\int \bm{v} \cup \bm{f} + \bm{f}\cup \bm{v}$ is nonzero only if $v$ is either vertex 0 or vertex 2 of face $f$.  $w_2 \in C_0(M, \bbz_2)$ represents the Poincare dual to the second SW class of $M$.  More explicitly
\be 
w_2 = \sum_v v+ \sum_{-\text{-oriented\ } f}  \langle 1 \rangle_f.
\ee 
The first term is a sum over all 0-simplices, i.e., vertices, of the graph. In the second term, vertex 1 in $-$-oriented face $f$ as in Fig.~\ref{fig:branching}(b) is singled out. The factor $(-1)^{\int_{w_2} \bm{v}}$ compensates the phase of the product of concatenated Majorana bilinears $S_e$ and $(-1)^{F_f}$ over a closed loop. If $M$ is a spin manifold, the second SW class vanishes and $w_2 = \partial E$ for some $E \in C_1(M, \bbz_2)$, which can be interpreted as a spin structure on $M$.

Applying the disentangling unitary, we have
\begin{align}
    \prod_{e \in \delta v} X_e 
    &\to  \prod_{e \in \delta v} X_e  S_e \prod_{e'}Z_{e'}^{ \int \bm{e}\cup \bm{e}' } \nonumber \\
    & =  \prod_{e \supset v} X_e  \left(\prod_{e'}Z_{e'}^{ \int \delta \bm{v} \cup \bm{e}' }\right) S_{\delta v}  \nonumber \\
    & =\prod_{e \supset v} X_e  \left(\prod_{e'}Z_{e'}^{ \int \delta \bm{v} \cup \bm{e}' }\right) \nonumber\\
    &\quad \times (-1)^{\int_{w_2} \bm{v}}  \prod_f P_f^{\int \bm{v} \cup \bm{f} + \bm{f}\cup \bm{v}}. 
\end{align} 
Plugging in $P_f =1$ for all $f$, we obtain
\be
G_v \equiv  (-1)^{\int_{w_2} \bm{v}}\prod_{e \supset v} X_e  \left(\prod_{e'}Z_{e'}^{ \int \delta \bm{v} \cup \bm{e}' }\right) =1 . 
\ee 
To get rid of the phase $(-1)^{\int_{w_2} \bm{v}}$, we can modify the unitary in Eq.~(\ref{eq:U_triangulation}) to 
\be 
U = \stackrel{\leftarrow}{\prod}_e [P^+_{e} + P^-_{e} (-1)^{\int_E \bm{e}} S_e]. 
\ee 
Then 
\be 
(-1)^{\int_E \bm{e}} S_e \to (-1)^{\int_E \bm{e}} X_e S_e \to  X_e  \prod_{e'}Z_{e'}^{ \int \bm{e}'\cup \bm{e}}.
\ee 
Making use of the relation $ \int_{w_2} \bm{v} = \int_E \delta \bm{v}$, we obtain 
\be
G_v = \prod_{e \supset v} X_e  \left(\prod_{e'}Z_{e'}^{ \int \delta \bm{v} \cup \bm{e}' }\right) =1 .  
\ee 
This is the modified Gauss law in 2D using the Chen--Kapustin bosonization method \cite{chen2020exact}. To derive the Gauss law in higher dimensions, one must invoke higher cup products; however, the derivation remains straightforward.

\bibliography{Bosonization} 

\end{document}